\documentclass[12pt]{article}
\pdfoutput=1
\usepackage{jheppub}
\usepackage{mathtools,amssymb,amsthm,amsmath}
\usepackage[utf8]{inputenc}
\usepackage{subcaption}
\usepackage{mathtools,slashed}
\usepackage{cancel}
\usepackage[normalem]{ulem}
\usepackage{parselines} 
 \usepackage[verbose]{placeins}

\def\be{\begin{equation}}
\def\ee{\end{equation}}
\def\ba#1\ea{\begin{align}#1\end{align}}
\def\bg#1\eg{\begin{gather}#1\end{gather}}
\def\bm#1\em{\begin{multline}#1\end{multline}}
\def\bmd#1\emd{\begin{multlined}#1\end{multlined}}

\def\a{\alpha}
\def\b{\beta}

\def\e{\epsilon}

\def\G{\Gamma}

\def\r{\rho}
\def\s{\sigma}

\def\z{\zeta}

\def\la{\label}

\def\re{\ref}
\def\er{\eqref}

\def\fr{\frac}

\def\pa{\partial}

\def\nn{\nonumber}

\def\qqu{\qquad}

\def\({\left(}
\def\){\right)}
\def\[{\left[}
\def\]{\right]}
\def\<{\langle}
\def\>{\rangle}

\def \bea {\begin{eqnarray}}
\def \eea {\end{eqnarray}}
\def \bf {\textbf}

\newcommand{\Tr}{\operatorname{Tr}}

\begin{document}

\title{\boldmath The Expressivity of Classical and Quantum Neural Networks on Entanglement Entropy }


\author[a]{Chih-Hung Wu}
\author[b]{and Ching-Che Yen}


\affiliation[a]{Department of Physics, University of California, Santa Barbara, CA 93106, USA}
\affiliation[b]{MediaTek Inc. }

\emailAdd{chih-hungwu@physics,ucsb,edu}
\emailAdd{johnson.yan@mediatek.com}

\abstract{Analytically continuing the von Neumann entropy from R\'enyi entropies is a challenging task in quantum field theory. While the $n$-th R\'enyi entropy can be computed using the replica method in the path integral representation of quantum field theory, the analytic continuation can only be achieved for some simple systems on a case-by-case basis. In this work, we propose a general framework to tackle this problem using classical and quantum neural networks with supervised learning. We begin by studying several examples with known von Neumann entropy, where the input data is generated by representing $\Tr \rho_A^n$ with a generating function. We adopt KerasTuner to determine the optimal network architecture and hyperparameters with limited data. In addition, we frame a similar problem in terms of quantum machine learning models, where the expressivity of the quantum models for the entanglement entropy as a partial Fourier series is established. Our proposed methods can accurately predict the von Neumann and R\'enyi entropies numerically, highlighting the potential of deep learning techniques for solving problems in quantum information theory.}

\maketitle
\flushbottom

\section{Introduction} \la{s1}

The \textit{von Neumann entropy} is widely regarded as an effective measure of quantum entanglement, and is often referred to as \textit{entanglement entropy}. The study of entanglement entropy has yielded valuable applications, particularly in the context of quantum information and quantum gravity (see \cite{Faulkner:2022mlp, Bousso:2022ntt} for a review). However, the analytic continuation from the \textit{R\'enyi entropies} to von Neumann entropy remains a challenge in quantum field theory for general systems. We tackle this problem using both classical and quantum neural networks to examine their expressive power on entanglement entropy and the potential for simpler reconstruction of the von Neumann entropy from R\'enyi entropies.

Quantum field theory (QFT) provides an efficient method to compute the $n$-th R\'enyi entropy with integer $n >1$, which is defined as \cite{Renyi:1961}
\be
S_n (\rho_A) \equiv \frac{1}{1-n} \ln{\Tr (\rho_A^n)}.
\ee
The computation is done by replicating the path integral representation of the reduced density matrix $\rho_A$ by $n$ times. This step is non-trivial; however, we will be mainly looking at examples where explicit analytic expressions of the R\'enyi entropies are available, especially in two-dimensional conformal field theories (CFT$_2$) \cite{Calabrese:2004eu, Calabrese:2009ez, Calabrese:2009qy, Calabrese:2010he}. Then upon analytic continuation of $n \to 1$, we have the von Neumann entropy
\be
S(\rho_A)=\lim_{n \to 1} S_n (\rho_A).
\ee
The continuation can be viewed as an independent problem from computing the $n$-th R\'enyi entropy. Although the uniqueness of $S(\rho_A)$ from the continuation is guaranteed by Carlson's theorem, analytic expressions in closed forms are currently unknown for most cases.

Furthermore, while $S_n (\rho_A)$ are well-defined in both integer and non-integer $n$, determining it for a set of integer values $n>1$ is not sufficient. To obtain the von Neumann entropy, we must also take the limit $n\rightarrow 1$ through a \textit{space} of real $n>1$. The relationship between the R\'enyi entropies and the von Neumann entropy is therefore complex, and the required value of $n$ for a precise numerical approximation of $S(\rho_A)$ is not clear.

Along this line, we are motivated to adopt an alternative method proposed in \cite{DHoker:2020bcv}, which would allow us to study the connection between higher R\'enyi entropies and von Neumann entropy "accumulatively." This method relies on defining a generating function that manifests as a Taylor series
\be \la{gen2}
G(w;\rho_A)=\sum_{k=1}^\infty \frac{\tilde{f}(k)}{k} w^k, \quad \tilde{f}(k)=\Tr [\rho_A (1-\rho_A)^k].
\ee
Summing over $k$ explicitly yields an absolutely convergent series that approximates the von Neumann entropy with increasing accuracy as $w \to 1$. This method has both numerical and analytical advantages, where we refer to \cite{DHoker:2020bcv} for explicit examples. Note that the accuracy we can achieve in approximating the von Neumann entropy depends on the truncation of the partial sum in $k$, which is case-dependent and can be difficult to evaluate. It becomes particularly challenging when evaluating the higher-order Riemann-Siegel theta function in the general two-interval case of CFT$_2$ \cite{DHoker:2020bcv}, which remains an open problem.

On the other hand, deep learning techniques have emerged as powerful tools for tackling the analytic continuation problem \cite{yoon2018analytic, Fournier_2020, xie2019analytic, song2020analytic, huang2022learned, sun2023neural}, thanks to their universal approximation property. The universal approximation theorem states that artificial neural networks can approximate any continuous function under mild assumptions \cite{citeulike:3561150}, where the von Neumann entropy is no exception. A neural network is trained on a dataset of known function values, with the objective of learning a latent manifold that can approximate the original function within the known parameter space. Once trained, the model can be used to make predictions outside the space by extrapolating the trained network. The goal is to minimize the prediction errors between the model's outputs and the actual function values. In our study, we frame the supervised learning task in two distinct ways: the first approach involves using densely connected neural networks to predict von Neumann entropy, while the second utilizes sequential learning models to extract higher R\'enyi entropies.

Instead of using a static "define-and-run" scheme, where the model structure is defined beforehand and remains fixed throughout training, we have opted for a dynamic "define-by-run" approach. Our goal is to determine the optimal model complexity and hyperparameters based on the input validation data automatically. To achieve this, we employ KerasTuner \cite{omalley2019kerastuner} with Bayesian optimization, which efficiently explores the hyperparameter space by training and evaluating different neural network configurations using cross-validation. KerasTuner uses the results to update a probabilistic model of the hyperparameter space, which is then used to suggest the next set of hyperparameters to evaluate, aiming to maximize expected performance improvement.

A similar question can be explicitly framed in terms of quantum machine learning, where a trainable quantum circuit can be used to emulate neural networks by encoding both the data inputs and the trainable weights using quantum gates. This approach bears many different names \cite{McClean_2016, romero2021variational, Mitarai_2018, farhi2018classification, McClean_2018,Benedetti_2019}, but we will call it a \textit{quantum neural network}. Unlike classical neural networks, quantum neural networks are defined through a series of well-defined unitary operations, rather than by numerically optimizing the weights for the non-linear mapping between targets and data. This raises a fundamental question for quantum computing practitioners: can \textit{any unitary operation} be realized, or is there a particular characterization for the \textit{learnable function class}? In other words, is the quantum model universal in its ability to express any function with the given data input?  Answering these questions will not only aid in designing future algorithms, but also provide deeper insights into how quantum models achieve universal approximation \cite{perez2021one, goto2021universal}.

Recent progress in quantum neural networks has shown that data-encoding strategies play a crucial role in their expressive power. The problem of data encoding has been the subject of extensive theoretical and numerical studies \cite{gan2022fock, chen2021expressibility, shin2022exponential, caro2021encoding}. In this work, we build on the idea introduced in \cite{gil2020input, Schuld_2021}, which demonstrated the expressivity of quantum models as partial Fourier series. By rewriting the generating function for the von Neumann entropy in terms of a Fourier series, we can similarly establish the expressivity using quantum neural networks. However, the Gibbs phenomenon in the Fourier series poses a challenge in recovering the von Neumann entropy. To overcome this, we reconstruct the entropy by expanding the Fourier series into a basis of Gegenbauer polynomials.

The structure of this paper is as follows. In Sec.~\ref{s2}, we provide a brief overview for the analytic continuation of the von Neumann entropy from R\'enyi entropies within the framework of QFT. In addition, we introduce the generating function method that we use throughout the paper. In Sec.~\ref{s3}, we use densely connected neural networks with KerasTuner to extract the von Neumann entropy for several examples where analytic expressions are known. In Sec.~\ref{s4}, we employ sequential learning models for extracting higher R\'enyi entropies. Sec.~\ref{s5} is dedicated to studying the expressive power of quantum neural networks in approximating the von Neumann entropy. In Sec.~\ref{s6}, we summarize our findings and discuss possible applications of our approach. Appendix.~\ref{sA} is devoted to the details of rewriting the generating function as a partial Fourier series, while Appendix.~\ref{sB} addresses the Gibbs phenomenon using Gegenbauer polynomials.

\section{Analytic continuation of von Neumann entropy from R\'enyi entropies} \la{s2}

Let us discuss how to calculate the von Neumann entropy in QFTs \cite{Sorkin:1984kjy, Bombelli:1986rw, Srednicki:1993im, Holzhey:1994we}. Suppose we start with a QFT on a $d$-dimensional Minkowski spacetime with its Hilbert space specified on a Cauchy slice $\Sigma$ of the spacetime. Without loss of generality, we can divide $\Sigma$ into two disjoint sub-regions $\Sigma=A \cup A^c$. Here $A^c$ denotes the complement sub-region of $A$. Therefore, the Hilbert space also factorizes into the tensor product $\mathcal{H}_\Sigma= \mathcal{H}_A \otimes \mathcal{H}_{A^c}$. We then define a reduced density matrix $\rho_A$ from a pure state on $\Sigma$, which is therefore mixed, to capture the entanglement between the two regions. The von Neumann entropy $S(\rho_A)$ allows us to quantify this entanglement 
\be
S(\rho_A)\equiv-\Tr (\rho_A \ln{\rho_A})=\frac{\text{Area}(\pa A)}{\epsilon^{d-2}}+\cdots.
\ee
Along with several nice properties, such as the invariance under unitary operations, complementarity for pure states, and a smooth interpolation between pure and maximally mixed states, it is therefore a fine-grained measure for the amount of entanglement between $A$ and $A^c$. The second equality holds for field theory, where we require a length scale $\epsilon$ to regulate the UV divergence encoded in the short-distance correlations. The leading-order divergence is captured by the area of the entangling surface $\partial A$, a universal feature of QFTs \cite{Witten:2018zxz}.\footnote{While in CFT$_2$, the leading divergence for a single interval $A$ of length $\ell$ in the vacuum
state on an infinite line is a logarithmic function of the length, this is the simplest example we will consider later.}

There have been efforts to better understand the structure of the entanglement in QFTs, including free theory \cite{Casini:2009sr}, heat kernels \cite{Solodukhin:2008dh, Hertzberg:2010uv}, CFT techniques \cite{Rosenhaus:2014woa} and holographic methods based on AdS/CFT \cite{Myers:2010tj, Liu:2012eea}. But operationally, computing the von Neumann entropy analytically or numerically is still a daunting challenge for generic interacting QFTs. For a review, see \cite{Faulkner:2022mlp}.

Path integral provides a general method to access $S(\rho_A)$. The method starts with the R\'enyi entropies \cite{Renyi:1961}
\be
S_n(\rho_A)=\frac{1}{1-n} \ln \Tr \rho^n_A,
\ee
for real $n > 1$. As previously mentioned, obtaining the von Neumann entropy via analytic continuation in $n$ with $n \to 1$ requires two crucial steps. An analytic form for the $n$-th R\'enyi entropy must be derived from the underlying field theory in the first place, and then we need to perform analytic continuation toward $n \to 1$. These two steps are independent problems and often require different techniques. We will briefly comment on the two steps below.

Computing $\Tr{\rho^n_A}$ is not easy; therefore, the replica method enters. The early form of the replica method was developed in \cite{Holzhey:1994we}, and was later used to compute various examples in CFT$_2$ \cite{Calabrese:2004eu, Calabrese:2009ez, Calabrese:2009qy, Calabrese:2010he}, which can be compared with holographic ones \cite{Faulkner:2013yia}. The idea behind the replica method is to consider an orbifold of $n$ copies of the field theory to compute $\Tr{\rho^n_A}$ for positive integers $n$. The computation reduces to evaluating the partition function on a $n$-sheeted Riemann surface, which can be alternatively computed by correlation functions of twist operators in the $n$ copies. For more details on the construction in CFTs, see \cite{Calabrese:2004eu, Calabrese:2009ez, Calabrese:2009qy, Calabrese:2010he}. If we are able to compute $\Tr \rho^n_A$ for any positive integer $n \geq 1$, we have
\be
S(\rho_A)=\lim_{n \to 1} S_n(\rho_A)=-\lim_{n \to 1} \frac{\pa }{\pa n} \Tr \rho^n_A.
\ee
This is computable for special states and regions, such as ball-shaped regions for the vacuum of the CFT$_d$. However, in CFT$_2$, due to its infinite-dimensional symmetry being sufficient to fix lower points correlation functions, we are able to compute $\Tr \rho^n_A$ for several instances.

The analytic continuation in $n \to 1$ is more subtle. Ensuring the existence of a unique analytic extension away from integer $n$ typically requires the application of the Carlson's theorem. This theorem guarantees the uniqueness of the analytic continuation from R\'enyi entropies to the von Neumann entropy, provided that we can find some locally holomorphic function $\mathcal{S}_\nu$ with $\nu \in \mathbb{C}$ such that $\mathcal{S}_n=S_n(\rho)$ for all integers $n > 1$ with appropriate asymptotic behaviors in $\nu \to \infty$. Then we have unique $S_\nu (\rho)=\mathcal{S}_\nu$ \cite{Boas1954, Witten_2019}. Carlson's theorem addresses not only the problem of unique analytic continuation but also the issue of continuing across non-integer values of the R\'enyi entropies.

There are other methods to evaluate $S(\rho_A)$ in the context of string theory and AdS/CFT; see for examples \cite{Dabholkar:1994ai, Witten:2018xfj,Agon:2013iva, Lewkowycz:2013nqa, Akers:2020pmf, Dong:2021clv}. In this work, we would like to focus on an effective method outlined in \cite{DHoker:2020bcv} that is suitable for numerical considerations.  In \cite{DHoker:2020bcv}, the following generating function is used for the analytic continuation in $n$ with a variable $z$
\be \la{genz}
G(z;\rho_A) \equiv -\Tr\bigg(\rho_A \ln{\frac{1-z \rho_A}{1-z}} \bigg)=\sum_{n=1}^\infty \frac{z^k}{k} \bigg( \Tr(\rho_A^{k+1})-1 \bigg).
\ee
This manifest Taylor series is absolutely convergent in the unit disc with $|z| <1$. We can analytically continue the function from the unit disc to a holomorphic function in $\mathbb{C} \setminus [1,\infty)$ by choosing the branch cut of the logarithm to be along the positive real axis. The limit $z \to -\infty$ is within the domain of holomorphicity and is exactly where we obtain the von Neumann entropy
\be
S(\rho_A)= \lim_{z \to -\infty} G(z;\rho_A).
\ee
However, a more useful form can be obtained by performing a M\"obius transformation to a new variable $w$ 
\be \la{Taylor}
G(w;\rho_A)=-\Tr \bigg(\rho_A \ln{\{1-w(1- \rho_A)\}} \bigg), \quad w =\frac{z}{z-1}.
\ee
It again manifests as a Taylor series
\be
G(w ; \rho_A)=\sum_{k=1}^\infty \frac{\tilde{f}(k)}{k} w^k,
\ee
where
\be
\tilde{f}(k)=\Tr[\rho_A(1- \rho_A)^k]=\sum_{m=0}^k \frac{(-1)^m k!}{m! (k-m)!} \Tr{(\rho_A^{m+1})}.
\ee
We again have a series written in terms of $\Tr \rho_A^n$, and it is absolutely convergent in the unit disc $|w|<1$. The convenience of using $w$ is that by taking $w \to 1$, we have the von Neumann entropy
\be \la{gen1}
S(\rho_A)=\lim_{w \to 1} G(w; \rho_A)=\sum_{k=1}^\infty \frac{\tilde{f}(k)}{k}.
\ee
This provides an exact expression of $S(\rho_A)$ starting from a known expression of $\Tr \rho_A^n$. Numerically, we can obtain an accurate value of $S(\rho_A)$ by computing a partial sum in $k$. The method guarantees that by summing to sufficiently large $k$, we approach the von Neumann entropy with increasing accuracy.

However, a difficulty is that we need to sum up $k \sim 10^3$ terms to achieve precision within $10^{-3}$ in general \cite{DHoker:2020bcv}. It will be computationally costly for certain cases with complicated $\Tr \rho_A^n$. Therefore, one advantage the neural network framework offers is the ability to give accurate predictions with only a limited amount of data, making it a more efficient method. 

In this paper, we focus on various examples from CFT$_2$ with known analytic expressions of $\Tr \rho_A^n$ \cite{Calabrese:2009qy}, and we use the generating function $G(w; \rho_A)$ to generate the required training datasets for the neural networks.

\section{Deep learning von Neumann entropy} \la{s3}

This section aims to utilize deep neural networks to predict the von Neumann entropy via a supervised learning approach. By leveraging the gradient-based learning principle of the networks, we expect to find a non-linear mapping between the input data and the output targets. In the analytic continuation problem from the $n$-th R\'enyi entropy to the von Neumann entropy, such a non-linear mapping naturally arises. Accordingly, we consider $S_n(\rho_A)$ (equivalently $\Tr \rho_A^n$ and the generating function) as our input data and $S(\rho_A)$ as the target function for the training process. As supervised learning, we will consider examples where analytic expressions of both sides are available. Ultimately, we will employ the trained models to predict the von Neumann entropy across various physical parameter regimes, demonstrating the efficacy and robustness of the approach.

The major advantage of using deep neural networks lies in that they improve the accuracy of the generating function for computing the von Neumann entropy. As we mentioned, the accuracy of this method depends on where we truncate the partial sum, and it often requires summing up a large $k$ in \er{gen1}, which is numerically difficult. In a sense, it requires knowing much more information, such as those of the higher R\'enyi entropies indicated by $\Tr \rho_A^n$ in the series. Trained neural networks are able to predict the von Neumann entropy more accurately given much fewer terms in the input data. We can even predict the von Neumann entropy for other parameter spaces without resorting to any data from the generating function. 

Furthermore, the non-linear mappings the deep neural networks uncover can be useful for investigating the expressive power of neural networks on the von Neumann entropy. Additionally, they can be applied to study cases where analytic continuations are unknown and other entanglement measures that require analytic continuations.

In the following subsections, we will give more details on our data preparation and training strategies, then we turn to explicit examples as demonstrations.

\subsection{Model architectures and training strategies} \la{s3.1}

Generating suitable training datasets and designing flexible deep learning models are empirically driven. In this subsection, we outline our strategies for both aspects.\\

\textbf{Data preparation}

To prepare the training datasets, we consider several examples with known $S(\rho_A)$. We use the generating function $G(w;\rho)$, which can be computed from $\Tr \rho^n_A$ for each example. This is equivalent to computing the higher R\'enyi entropies with different choices of physical parameters since the "information" available is always $\Tr \rho_A^n$. However, note that all the higher R\'enyi entropies are distinct information. Therefore, adopting the generating function is preferable to using $S_n(\rho_A)$ itself, as it approaches the von Neumann entropy with increasing accuracy, making the comparison more transparent.

We generate $N=10000$ input datasets for a fixed range of physical parameters, where each set contains $k_{\text{max}}=50$ terms in \er{gen1}; their corresponding von Neumann entropies will be the targets. We limit the amount of data to mimic the computational cost of using the generating function. We shuffle the input datasets randomly and then split the data into 80$\%$ for training, 10$\%$ for validation, and 10$\%$ as the test datasets. Additionally, we use the trained neural networks to make predictions on another set of $10000$ test datasets with a different physical parameter regime and compare them with the correct values as a non-trivial test for each example.\\

\textbf{Model design}

To prevent overfitting and enhance the generalizability of our model, we have employed a combination of techniques in the design of neural networks. ReLU activation function is used throughout the section. We adopt Adam optimizer \cite{kingma2014adam} in the training process with mean square error (MSE) as the loss function.

\begin{figure}[hbt!]
\centering
\includegraphics[width=0.70\textwidth]{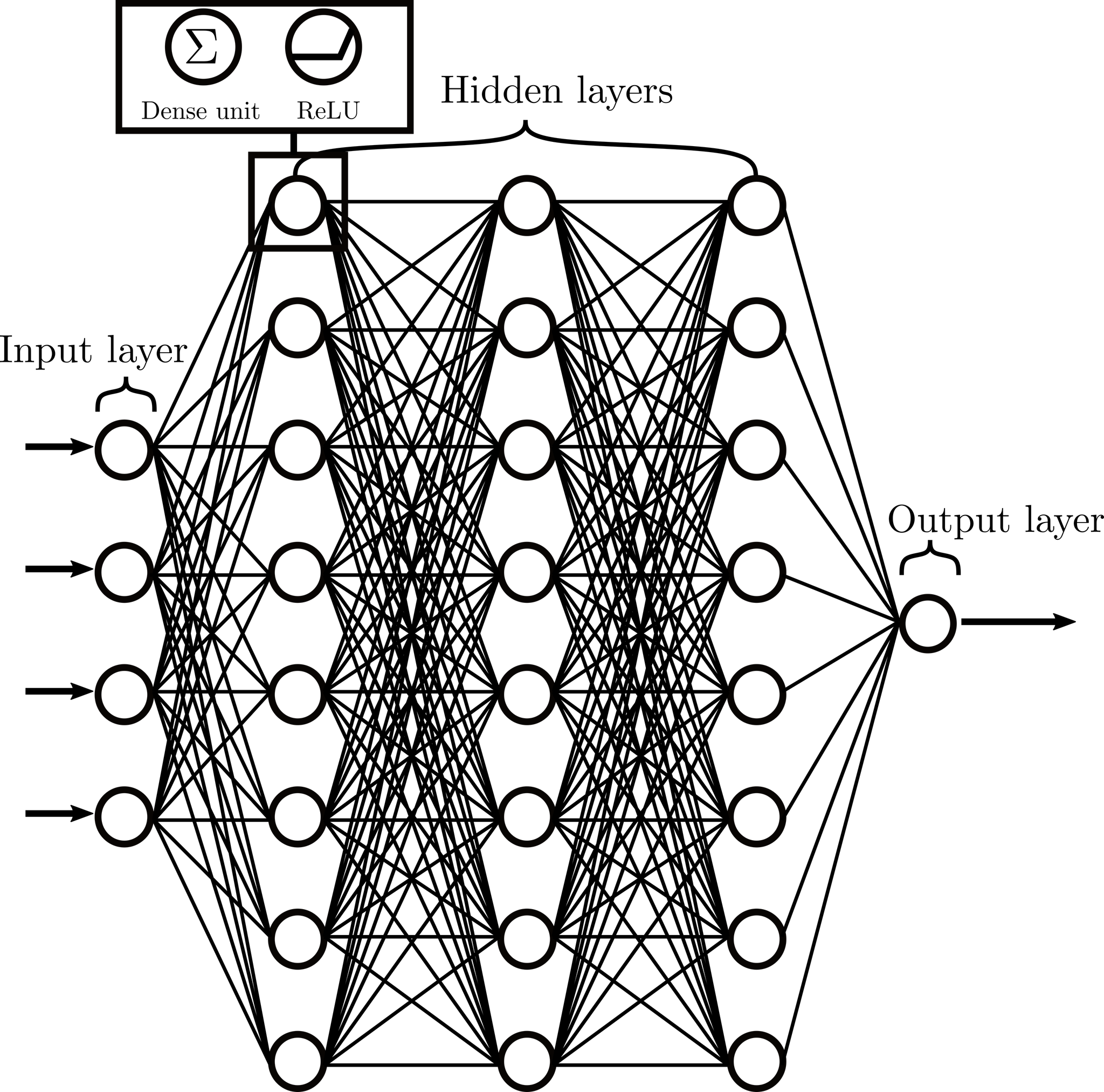}
\caption{An architecture of 3 densely connected layers, where each layer has $8$ units. The final output layer is a single Dense unit with a unique output corresponding to the von Neumann entropy.}
\label{DenseU}
\end{figure}

We consider a neural network consisting of a few hidden Dense layers with varying numbers of units in TensorFlow-Keras \cite{tensorflow2015-whitepaper, chollet2015keras}. In this case, each neuron in a layer receives input from all the neurons in the previous layer. The Dense connection allows the model to find non-linear relations between the input and output, which is the case for analytic continuation. The final layer is a Dense layer with a single unit that outputs a unique value for each training dataset, which is expected to correspond to the von Neumann entropy. As an example, we show a neural network with 3 hidden Dense layers, each with 8 units, in Figure~\ref{DenseU}.

\FloatBarrier

\begin{figure}[hbt!]
\centering
\includegraphics[width=1.00\textwidth]{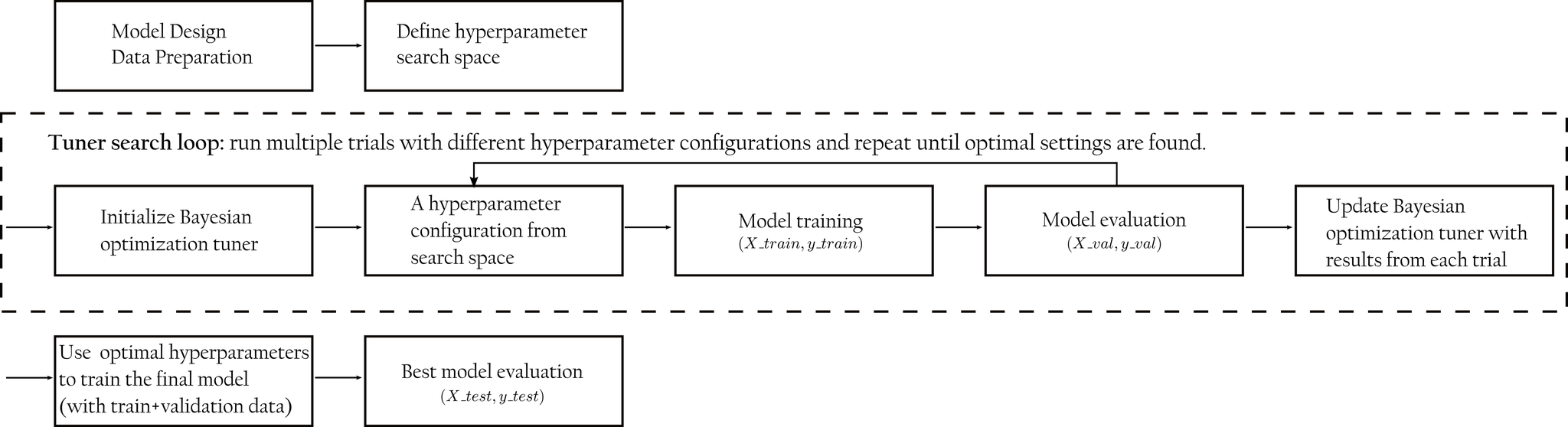}
\caption{Flowchart illustrating the steps of KerasTuner with Bayesian optimization. Bayesian optimization is a method for finding the optimal set of designs and hyperparameters for a given dataset, by iteratively constructing a probabilistic model from a prior distribution for the objective function and using it to guide the search. Once the tuner search loop is complete, we extract the best model in the final training phase by including both the training and validation data.}
\label{KerasTuner}
\end{figure}
\FloatBarrier

To determine the optimal setting of our neural networks, we employ KerasTuner \cite{omalley2019kerastuner}, a powerful tool that allows us to explore different combinations of model complexity, depth, and hyperparameters for a given task. An illustration of the KerasTuner process can be found in Figure~\ref{KerasTuner}. We use Bayesian optimization, and adjust the following designs and hyperparameters:

\begin{itemize}
    \item We allow a maximum of 4 Dense layers. For each layer, we allow variable units in the range of $16$ to $128$ with a step size of $16$. The number of units for each layer will be independent of each other.
    \item We allow BatchNormalization layers after the Dense layers as a Boolean choice to improve generalization and act as a regularization.
    \item A final dropout with log sampling of a dropout rate in the range of $0.1$ to $0.5$ is added as a Boolean choice.
    \item In the Adam optimizer, we only adjust the learning rate with log sampling from the range of 3$\times$10$^{-3}$ to 9$\times$10$^{-3}$. All other parameters are taken as default values in TensorFlow-Keras. We also use the AMSGrad \cite{reddi2019convergence} variant of this algorithm as a Boolean choice.
\end{itemize}

We deploy the KerasTuner for 100 trials with 2 executions per trial and monitor the validation loss with EarlyStopping of patience $8$. Once the training is complete, since we will not be making any further hyperparameter changes, we no longer evaluate performance on the validation data. A common practice is to initialize new models using the best model designs found by KerasTuner while also including the validation data as part of the training data. Indeed, we select the top $5$ best designs and train each one 20 times with EarlyStopping of patience $8$. We pick the one with the smallest relative errors from the targets among the $5 \times 20$ models as our final model. We set the batch size in both the KerasTuner and the final training to be 512.

In the following two subsections, we will examine examples from CFT$_2$ with $\Tr{\rho^n_A}$ and their corresponding von Neumann entropies $S(\rho_A)$ \cite{Calabrese:2004eu, Calabrese:2009ez, Calabrese:2009qy, Calabrese:2010he, DHoker:2020bcv}. These instances are distinct and worth studying for several reasons. They have different mathematical structures and lack common patterns in their derivation from the field theory side, despite involving the evaluation of certain partition functions. Moreover, the analytic continuation for each case is intricate, providing strong evidence for the necessity of independent model designs.

\subsection{Entanglement entropy of a single interval} \la{s3.2}

Throughout the following, we will only present the analytic expression of $\Tr \rho^n_A$ since it is the only input of the generating function. We will also keep the UV cut-off $\epsilon$ explicit in the formula.  \\

\noindent \bf{Single interval}

The simplest example corresponds to a single interval $A$ of length $\ell$ in the vacuum state of a CFT$_2$ on an infinite line. In this case, both the analytic forms of $\Tr{\rho^n_A}$ and $S(\rho_A)$ are known \cite{Calabrese:2004eu}, where $S(\rho_A)$ reduces to a simple logarithmic function that depends on $\ell$. We have the following analytic form with a central charge $c$
\be \la{single}
\Tr{\rho_A^n}= \bigg(\frac{\ell}{\epsilon} \bigg)^{\frac{c}{6}(\frac{1}{n}-n)},
\ee
that defines $G(w;\rho_A)$. The corresponding von Neumann entropy is given by
\be \la{sin}
S(\rho_A)=\frac{c}{3} \ln \frac{\ell}{\epsilon}.
\ee

\begin{figure}[hbt!]
\centering
\includegraphics[width=0.45\textwidth]{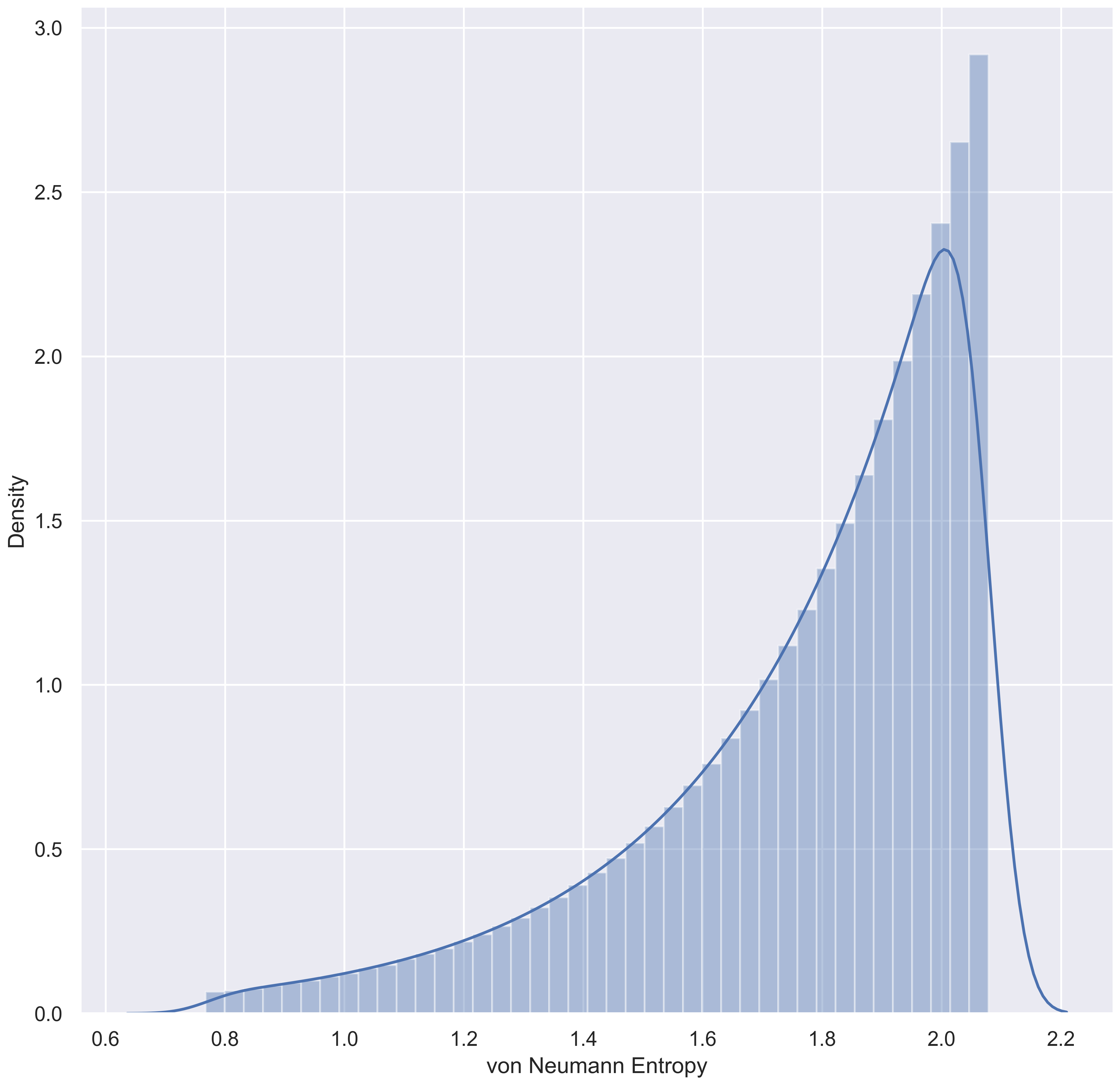}
\includegraphics[width=0.45\textwidth]{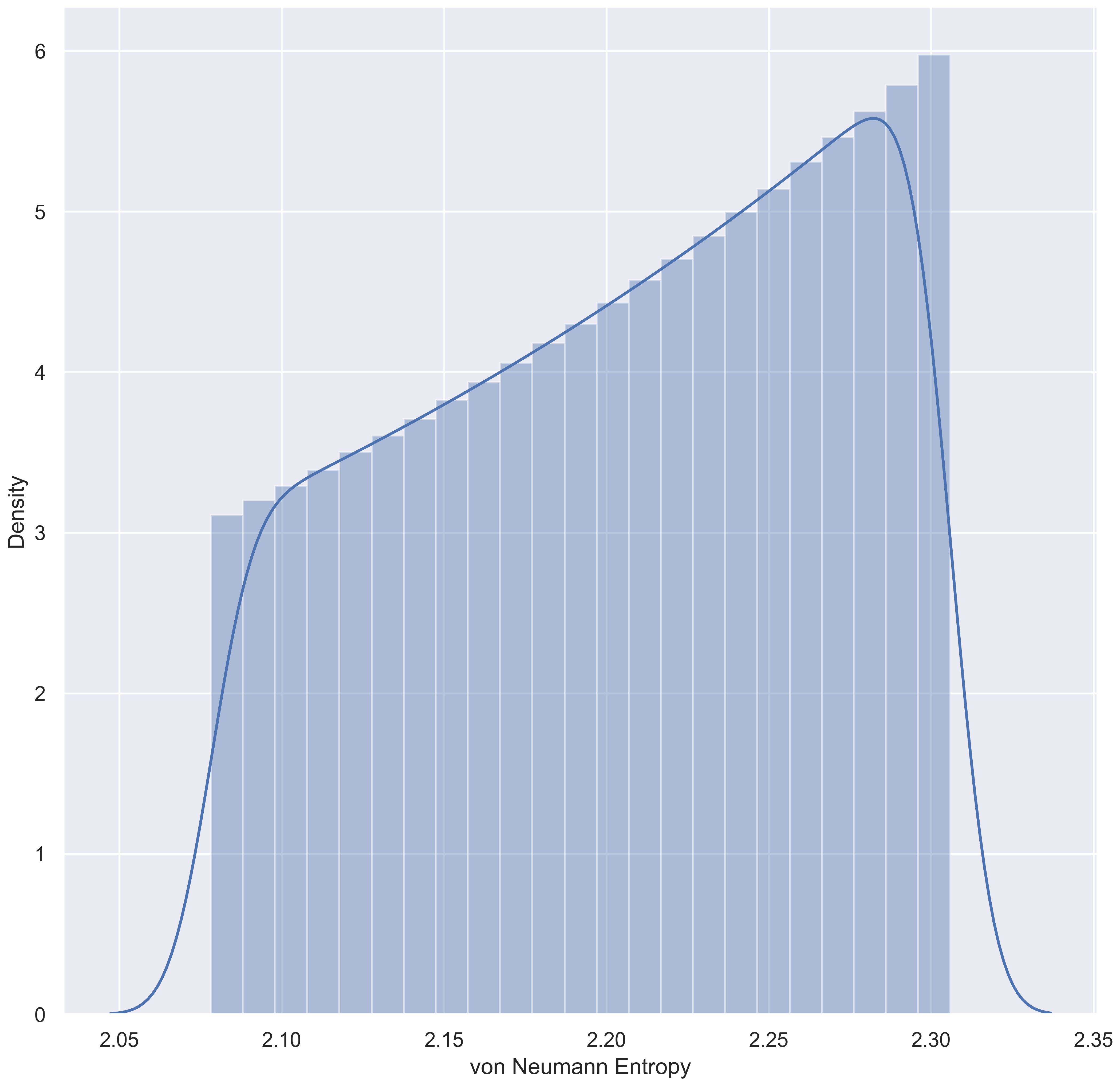}
\caption{ The distribution of the data for
the case of a single interval, where we plot density as a function of the
von Neumann entropy computed by \er{sin} with varying $\ell$. The left plot represents the 10000 datasets for the train-validation-test split, while the right
plot corresponds to the additional 10000 test datasets with a different physical parameter regime.}
\label{1Data}
\end{figure}

We fixed the central charge $c=1$ and the UV cutoff $\epsilon=0.1$ when preparing the datasets. We generated $10000$ sets of data for the train-validation-test split from $\ell =1$ to $50$, with an increment of $\Delta \ell = 5 \times 10^{-3}$ between each step up to $k=50$ in $G(w;\rho_A)$. To further validate our model, we generated an additional $10000$ test datasets for the following physical parameters: $\ell=51$ to $100$ with $\Delta \ell= 5 \times 10^{-3}$. For a density plot of the data distribution with respect to the target von Neumann entropy, see Figure~\ref {1Data}.

\FloatBarrier 

\begin{figure}[hbt!]
\centering
\includegraphics[height=3.8cm, width=7.9cm]{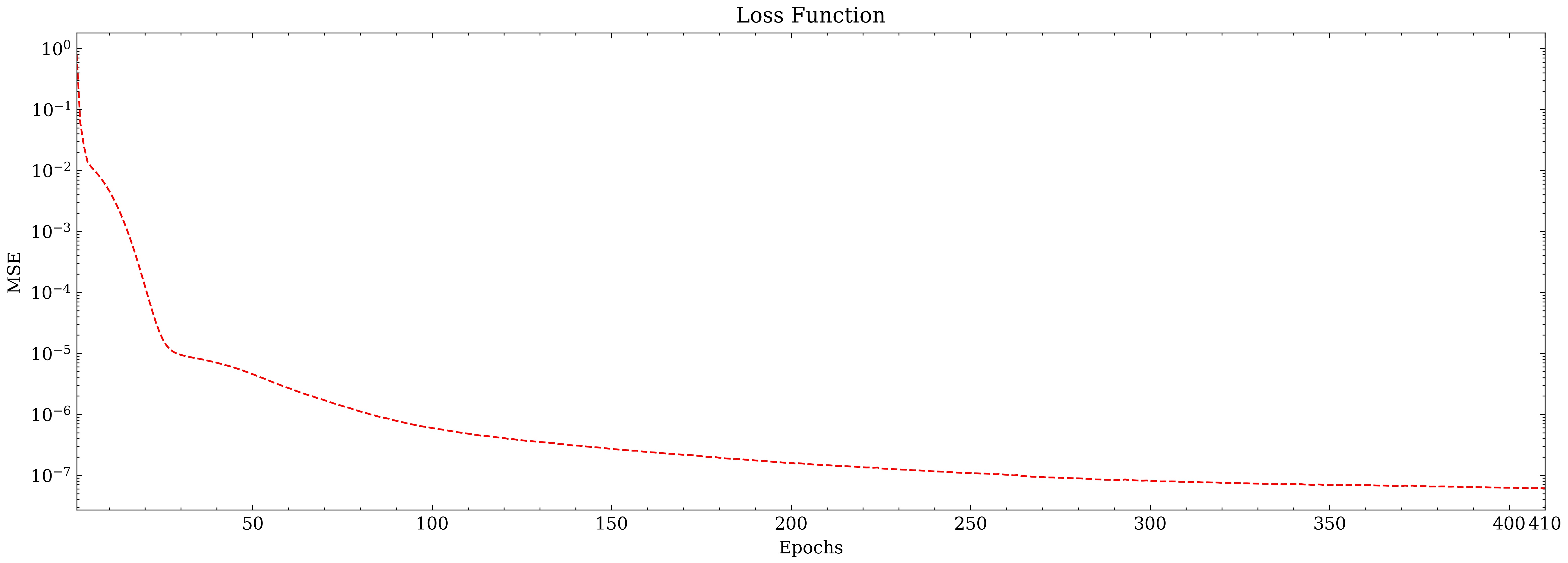}
\includegraphics[width=0.45\textwidth]{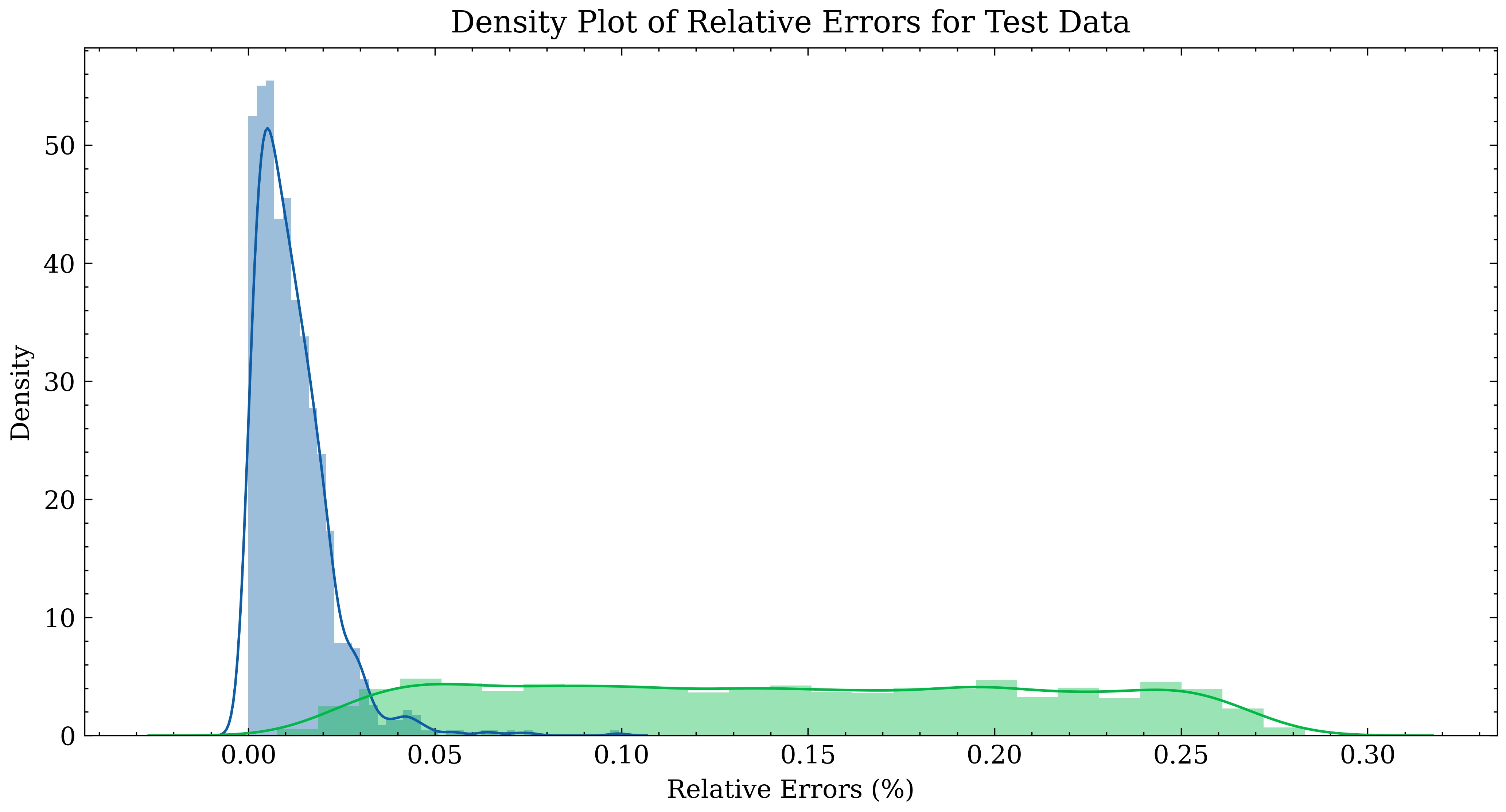}
\caption{Left: The MSE loss function as a function of epochs. We monitor the loss function with
EarlyStopping, where the minimum loss is achieved at epoch 410 with loss $\approx 10^{-7}$ for this instance. Right: The density plot of relative errors between the model predictions and targets. Note that the blue color corresponds to the test datasets from the initial train-validation-test split, while the green color is for the additional test datasets. We can see clearly that for both datasets, we have achieved high accuracy with relative errors $\lesssim 0.30 \%$.}
\label{1Loss}
\end{figure}
\FloatBarrier 

\begin{figure}[hbt!] 
\centering
\includegraphics[width=0.45\textwidth]{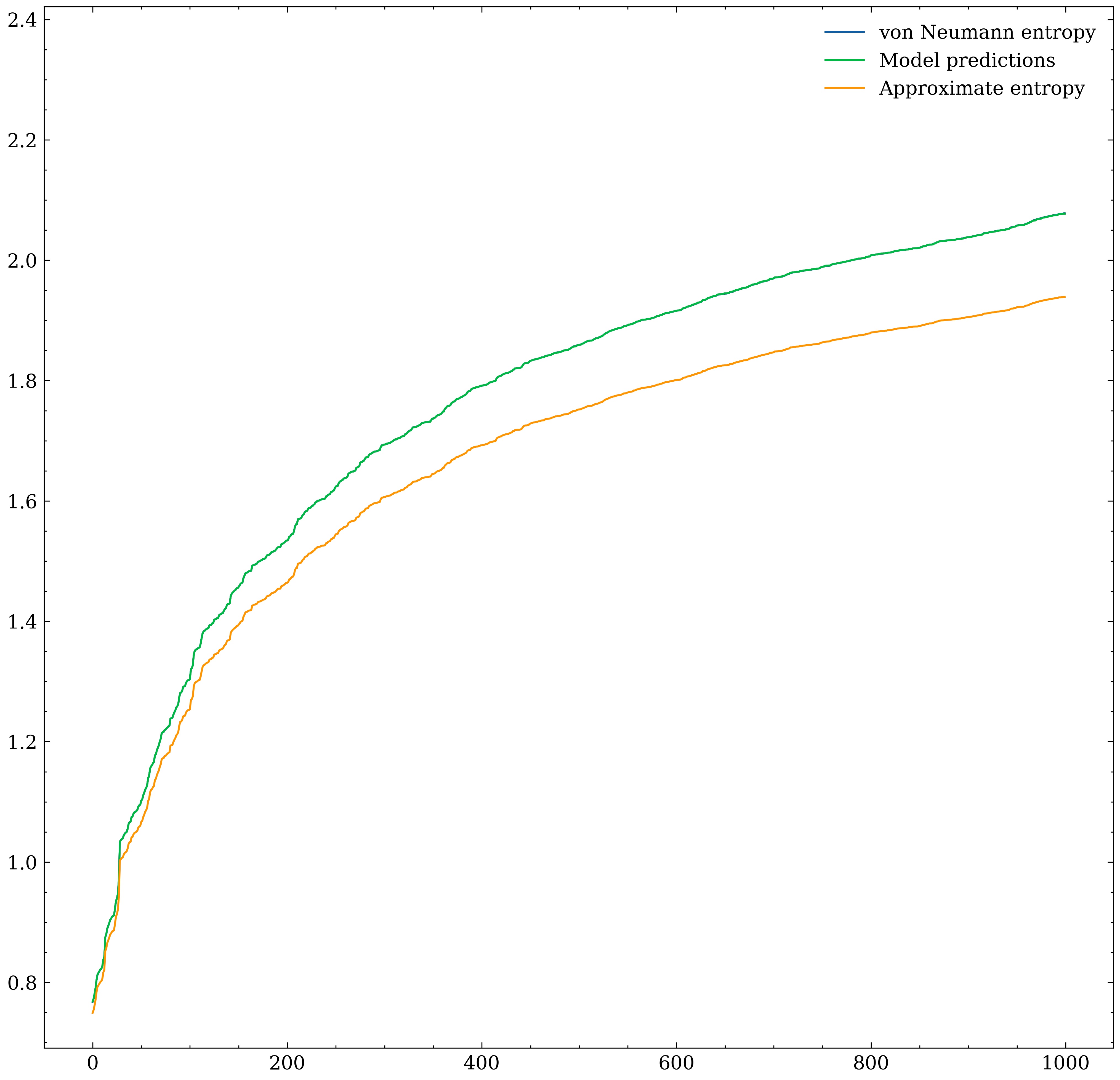}
\includegraphics[width=0.45\textwidth]{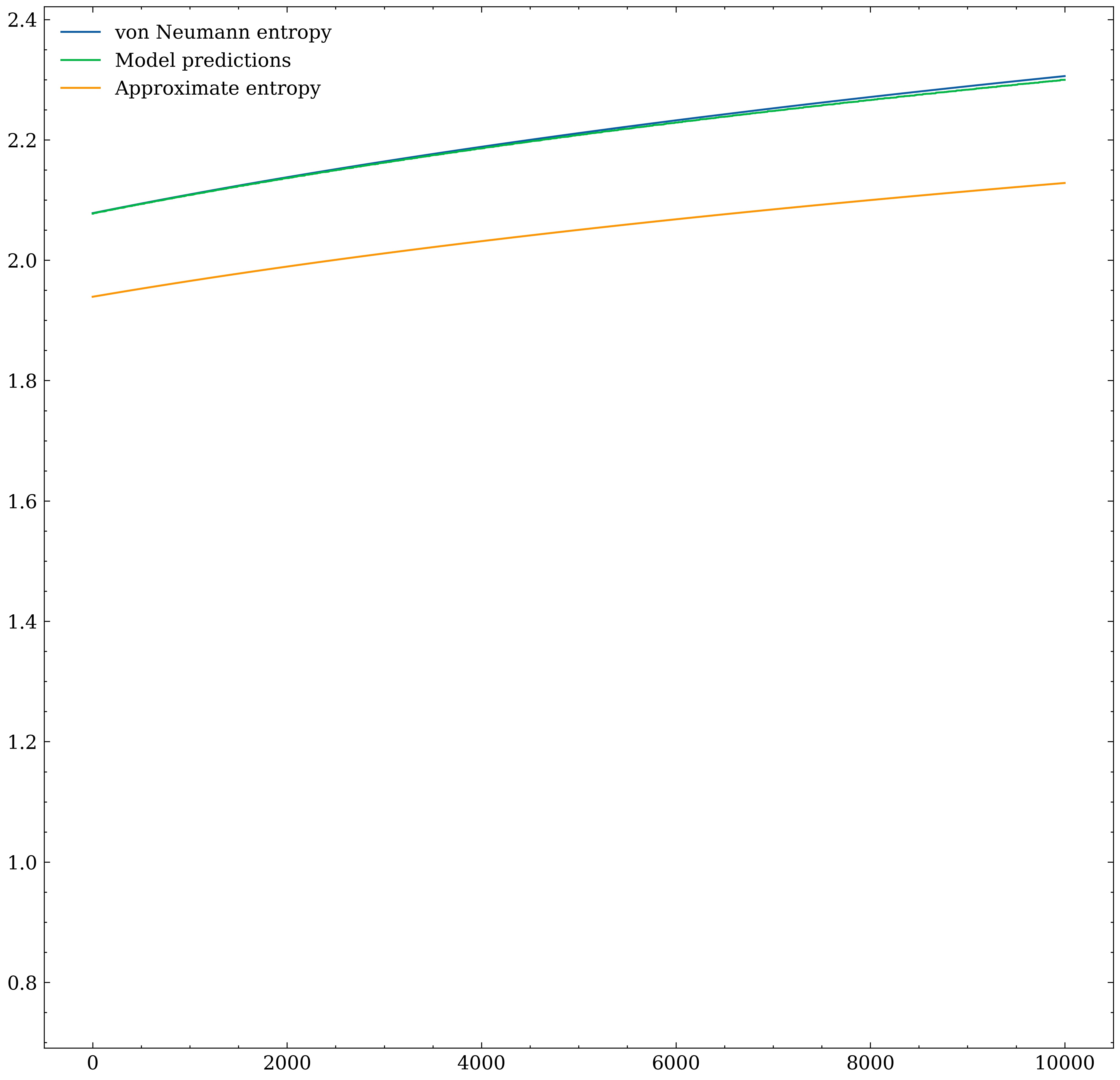}
\caption{We plot the predictions from the model with the analytic von Neumann entropy computed by \er{sin} for the 1000 test datasets (left) from the training-validation-test split and the additional 10000 test datasets (right), with the same scale on both figures. The correct von Neumann entropy overlaps with the model's predictions precisely. We have also included the approximate entropy by summing over $k=50$ terms in the generating function. }
\label{1Entropy}
\end{figure}
\FloatBarrier

Figure~\ref{1Loss} illustrates that the process outlined in the previous subsection effectively minimizes the relative errors in predicting the test data to a very small extent. Moreover, the model's effectiveness is further confirmed by its ability to achieve similarly small relative errors when predicting the additional test datasets. The accuracy of the model's predictions for the two test datasets significantly surpasses the approximate entropy obtained by summing the first $50$ terms of the generating function, as can be seen in Figure~\ref{1Entropy}. We emphasize that in order for the generating function to achieve the same accuracy as the deep neural networks, we generally need to sum $k \geq 400$ from \er{gen1} \cite{DHoker:2020bcv}. This applies to all the following examples.

In this example, the von Neumann entropy is a simple logarithmic function, making it relatively straightforward for the deep learning models to decipher. However, we will now move on to a more challenging example. \\

\noindent \bf{Single interval at finite temperature and length}

We extend the single interval case to finite temperature and length, where $\Tr{\rho^n_A}$ becomes a complicated function of the inverse temperature $\beta=T^{-1}$ and the length $\ell$. The analytic expression of the R\'enyi entropies was first derived in  \cite{Azeyanagi:2007bj} for a two-dimensional free Dirac fermion on a circle from bosonization. We can impose periodic boundary conditions that correspond to finite size and finite temperature. For simplicity, we set the total spatial size $L$ to $1$, and use $\ell$ to denote the interval length. In this case we have \cite{Azeyanagi:2007bj}
\be \la{TL}
\Tr \rho_A^n= \prod_{k=-\frac{n-1}{2}}^{\frac{n-1}{2}} \bigg|\frac{2 \pi \epsilon \eta(\tau)^3}{\theta_{1}(\ell | \tau)} \bigg|^{\frac{2 k^2}{n^2}} \frac{|\theta_{\nu}(\frac{k \ell}{n}| \tau)|^2}{|\theta_{\nu}(0|\tau)|^2},
\ee
where $\epsilon$ is a UV cutoff. We study the case of $\nu = 3$, which is the Neveu-Schwarz (NS-NS) sector. We then have the following Dedekind eta function $\eta (\tau)$ and the Jacobi theta functions $\theta_1 (z| \tau)$ and $\theta_3(z| \tau)$
\be
\eta(\tau) \equiv q^{\frac{1}{24}}\prod_{n=1}^\infty (1-q^n),
\ee
\be
\theta_{1} (z|\tau) \equiv \sum_{n=-\infty}^{n=\infty} (-1)^{n-\frac{1}{2}} e^{(n+\frac{1}{2})^2 i \pi \tau} e^{(2n+1) \pi i z} \,,\qqu
\theta_{3} (z|\tau) \equiv \sum_{n=-\infty}^{n=\infty} e^{n^2 i \pi \tau} e^{2 n \pi i z} \,.
\ee
Previously, the von Neumann entropy after analytically continuing \er{TL} was only known in the high- and low-temperature regimes \cite{Azeyanagi:2007bj}. In fact, only the infinite length or zero temperature pieces are universal. However, the analytic von Neumann entropy for all temperatures was recently worked out by \cite{Blanco:2019cet, Fries:2019acy}, which we present below 
\be\la{see}
S(\rho_A) = \fr{1}{3} \log \fr{\s(\ell)}{\e} + 4i\ell \int_0^\infty dq \fr{\z(iq\ell+1/2+i\b/2)-\z(1/2)-\z(i\b/2)}{e^{2\pi q}-1}.
\ee
Here $\sigma$ and $\zeta$ are the Weierstrass sigma function and zeta function with periods $1$ and $i\b$, respectively. We can see clearly that the analytic expressions for both $\Tr \rho_A^n$ and $S(\rho_A)$ are rather different compared to the previous example.

\begin{figure}[hbt!]
\centering
\includegraphics[width=0.45\textwidth]{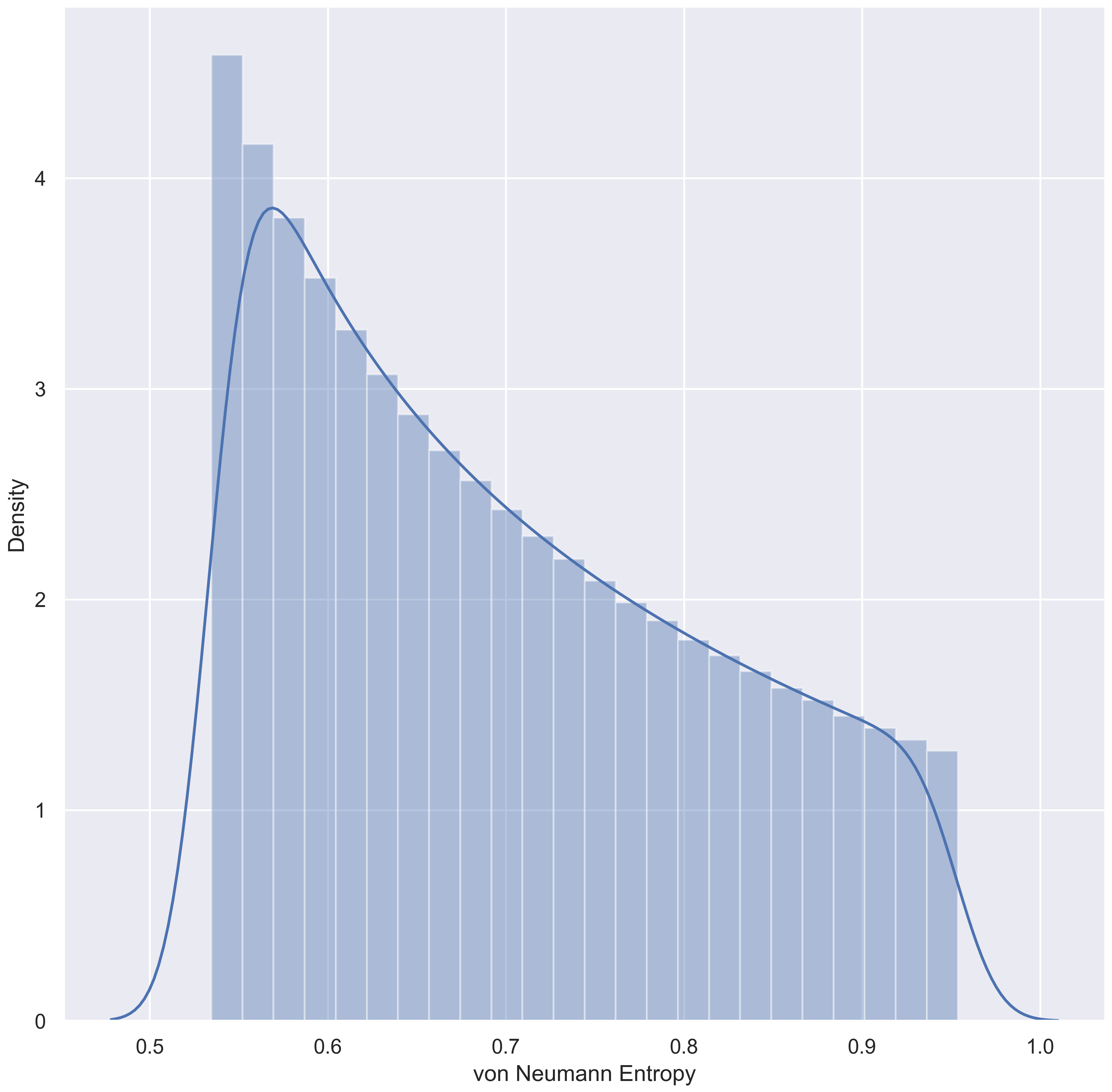}
\includegraphics[width=0.45\textwidth]{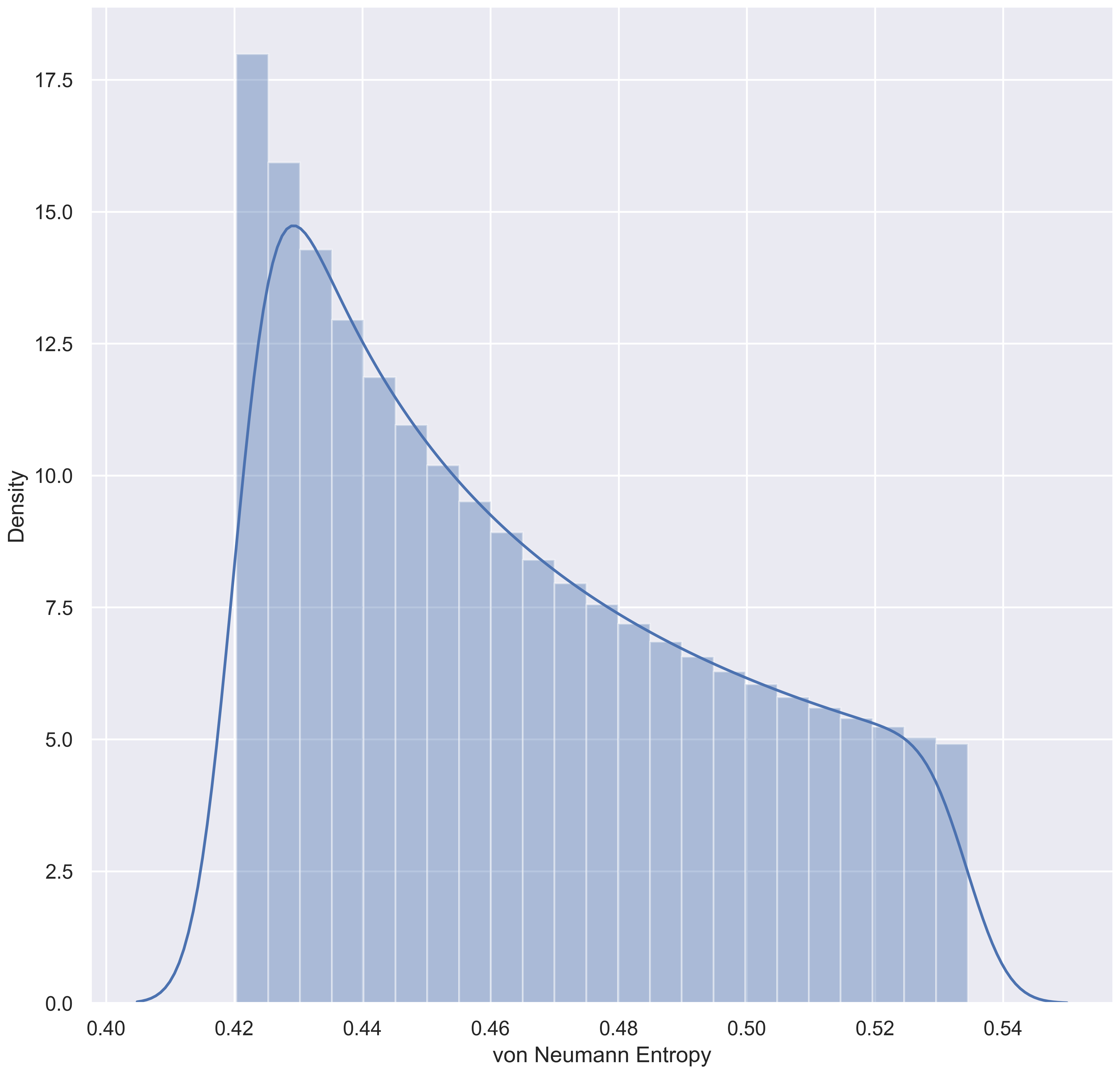}
\caption{The distribution of the two test datasets for
the case of a single interval at finite temperature and length, where we plot density as a function of the
von Neumann entropy computed by \er{see} with varying $\beta$. }
\label{2Data}
\end{figure}

\begin{figure}[hbt!] 
\centering
\includegraphics[height=3.8cm, width=7.9cm]{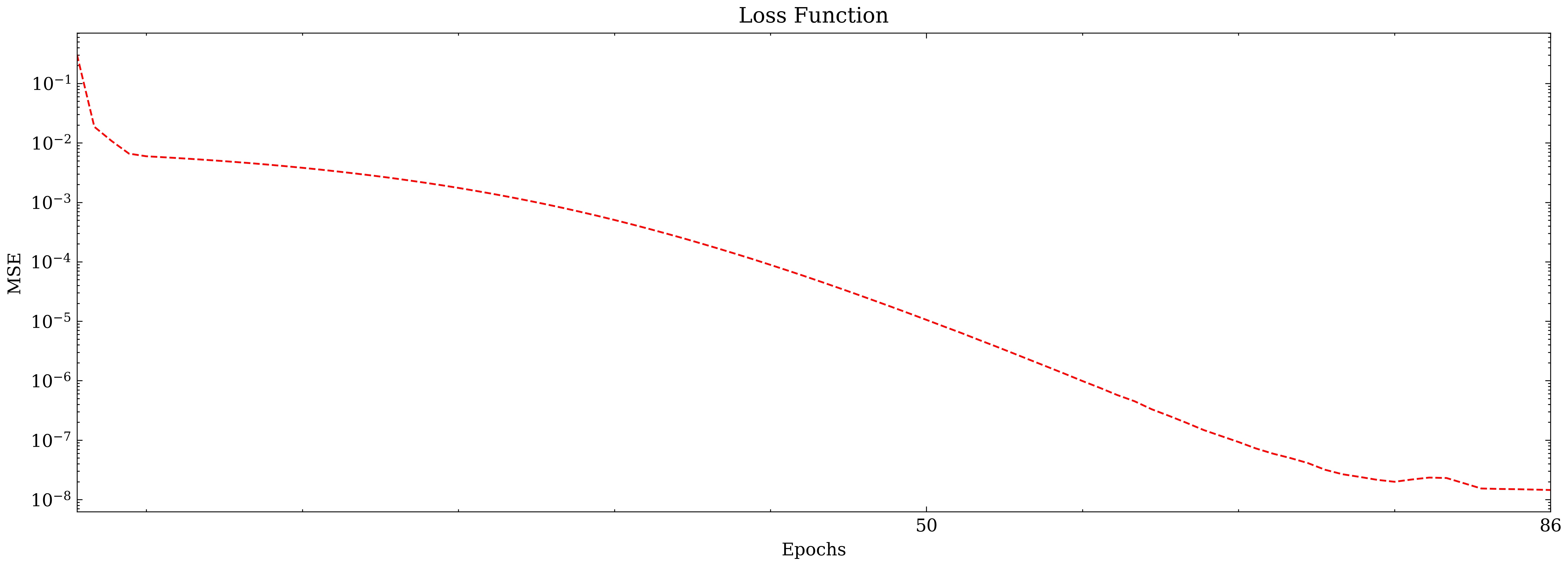}
\includegraphics[width=0.45\textwidth]{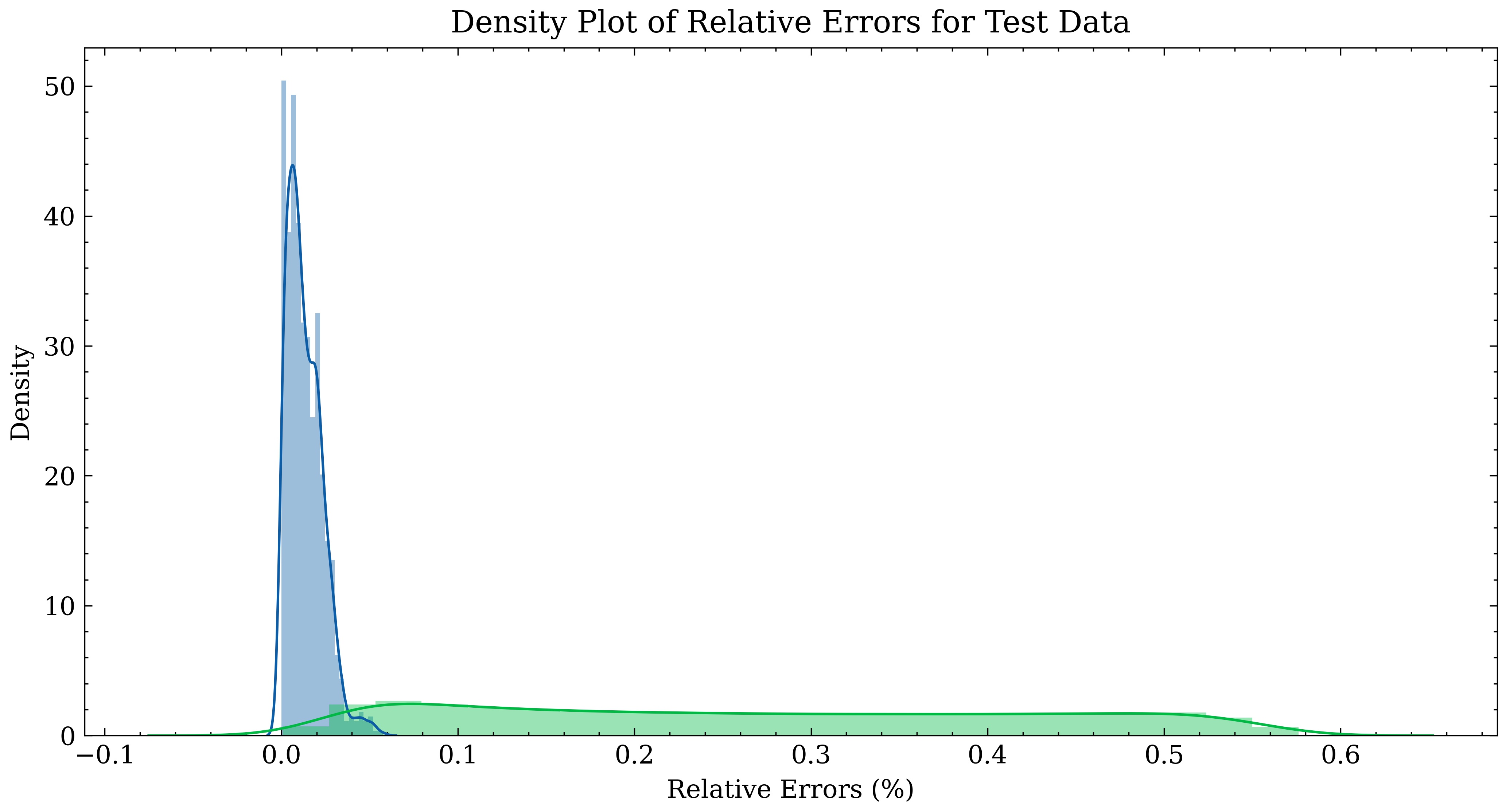}
\caption{Left: The MSE loss function as a function of epochs. The minimum loss close to $10^{-8}$ is achieved at epoch 86 for this instance. Right: The relative errors between the model predictions and targets for the two test datasets, where we have achieved high accuracy with relative errors $\lesssim 0.6 \%$.}
\label{2Loss}
\end{figure}

\begin{figure}[hbt!] 
\centering
\includegraphics[width=0.45\textwidth]{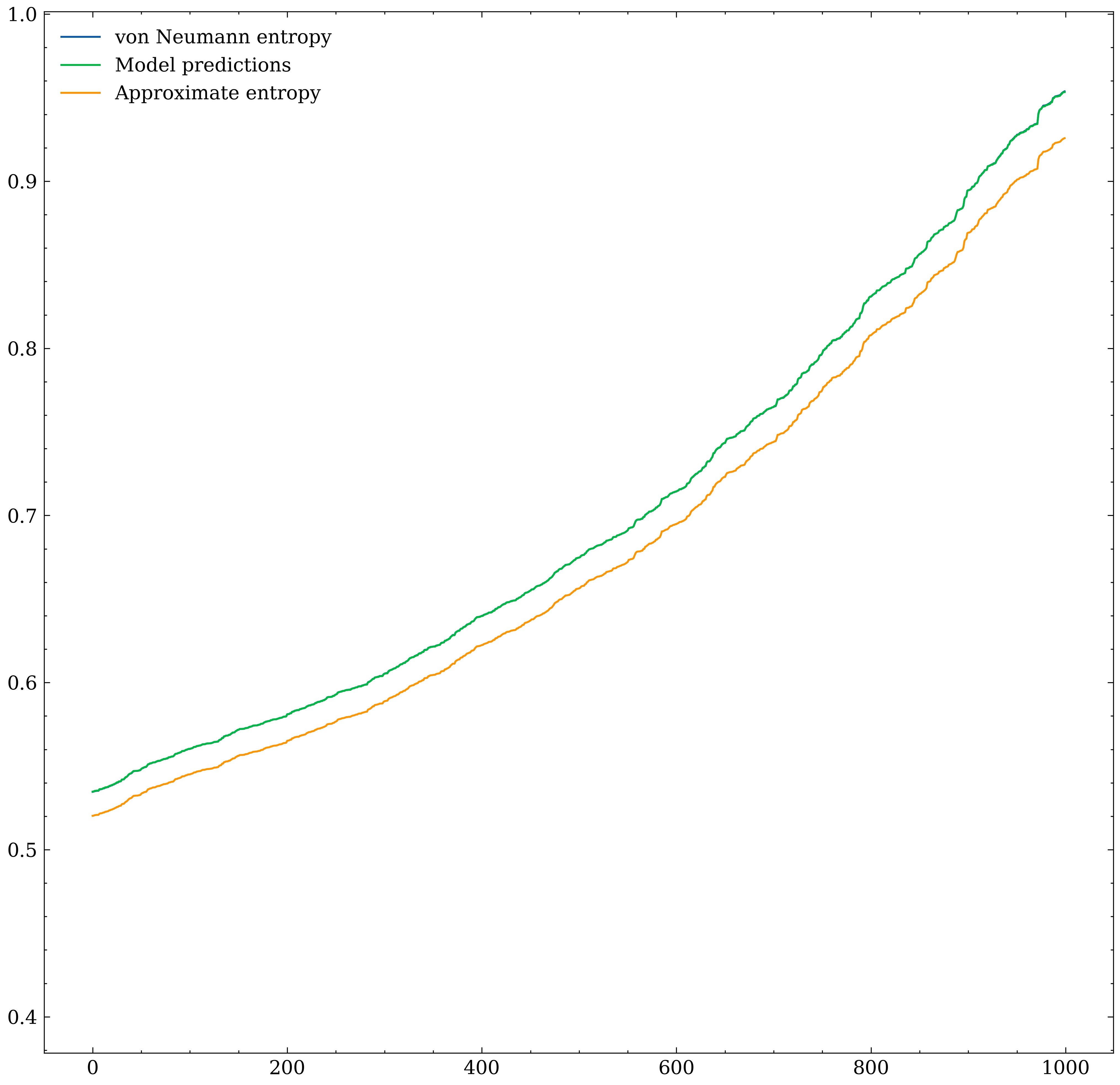}
\includegraphics[width=0.45\textwidth]{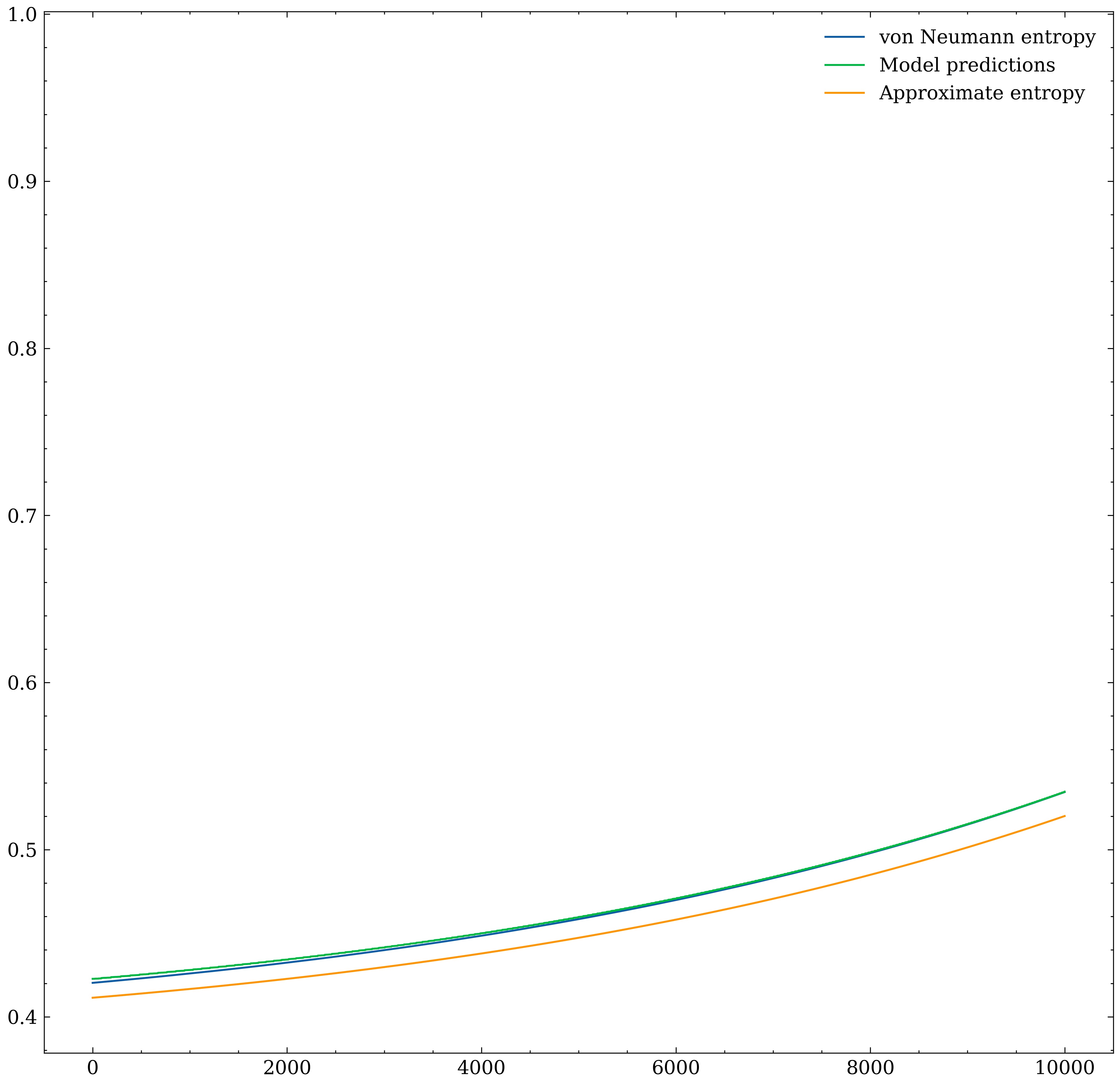}
\caption{We plot the predictions from the model with the analytic von Neumann entropy computed by \er{see} for the two test datasets. Again, the approximate entropy by summing over $k=50$ terms in the generating function is included.}
\label{2Entropy}
\end{figure}

In preparing the datasets, we fixed the interval length $\ell =0.5$ and the UV cutoff $\epsilon=0.1$. We generated 10000 sets of data for train-validation-test split from $\beta =0.5$ to $1.0$, with an increment of $\Delta \beta = 5 \times 10^{-5}$ between each step up to $k=50$ in $G(w;\rho_A)$. Since $\beta$ corresponds to the inverse temperature, this is a natural parameter to vary as the formula \er{see} is valid for all temperatures. To further validate our model, we generated $10000$ additional test datasets for the following physical parameters: $\beta =1.0$ to $1.5$ with $\Delta \beta = 5 \times 10^{-5}$. A density plot of the data with respect to the von Neumann entropy is shown in Figure~\ref{2Data}. As shown in Figure~\ref{2Loss} and Figure~\ref{2Entropy}, our model demonstrates its effectiveness in predicting both test datasets, providing accurate results for this highly non-trivial example.

\FloatBarrier 

\FloatBarrier 

\FloatBarrier

\subsection{Entanglement entropy of two disjoint intervals} \la{s3.3}

We now turn to von Neumann entropy for the union of two intervals on an infinite line. In this case, several analytic expressions can be derived for both R\'enyi and von Neumann entropies. The theory we will consider is a CFT$_2$ for a free boson with central charge $c=1$, and the von Neumann entropy will be distinguished by two parameters, a cross-ratio $x$ and a universal critical exponent $\eta$. The latter is proportional to the square of the compactification radius.

To set up the system, we define the union of the two intervals as $A \cup B$ with $A=[x_1, x_2]$ and $B=[x_3, x_4]$. The cross-ratio is defined to be
\be
x= \frac{x_{12} x_{34}}{x_{13} x_{24}}, \quad x_{ij} = x_i-x_j.
\ee
With the definition, we can write down the generating function for two intervals in a free boson CFT with finite $x$ and $\eta$ \cite{Calabrese:2009ez}
\be \la{ge} 
\Tr(\r^n) =c_n\bigg( \frac{\epsilon^2 x_{13}x_{24}}{x_{12}x_{34}x_{14}x_{23}} \bigg)^{\frac{1}{6}(n-\frac{1}{n})} \mathcal{F}_{n}(x, \eta),
\ee
where $\epsilon$ is a UV cutoff and $c_n$ is a model-dependent coefficient \cite{Calabrese:2009qy} that we set to $c_n=1$ for simplicity. An exact expression for $\mathcal{F}_{n}(x, \eta)$ is given by
\be \la{RS}
\mathcal{F}_{n}(x, \eta)=\frac{\Theta(0 | \eta \Gamma) \Theta (0 | \Gamma/ \eta)}{[\Theta (0 | \Gamma)]^2},
\ee
for integers $n \geq 1$. Here $\Theta(z| \Gamma)$ is the Riemann-Siegel theta function defined as
\be
\Theta(z| \Gamma) \equiv \sum_{m  \in \mathbb{Z}^{n-1}} \exp[i \pi m^{t} \cdot \Gamma \cdot m+2 \pi i m^{t} \cdot z],
\ee
where $\Gamma$ is a $(n-1)\times (n-1)$ matrix with elements
\be
\Gamma_{rs}=\frac{2i}{n} \sum_{k=1}^{n-1} \sin\bigg( \frac{\pi k}{n} \bigg) \beta_{k/n} \cos \bigg[ \frac{2 \pi k}{n}(r-s) \bigg],
\ee
and
\be\la{bfdef}
\beta_{y}=\frac{F_{y}(1-x)}{F_{y}(x)},\qqu
F_{y}(x) \equiv {}_2 F_1(y,1-y;1;x),
\ee
where ${}_2 F_1$ is the hypergeometric function. A property of this example is that \er{RS} is manifestly invariant under $\eta \leftrightarrow 1/\eta $.

The analytic continuation towards the von Neumann entropy is not known, making it impossible to study this example directly with supervised learning. Although the Taylor series of the generating function guarantees convergence towards the true von Neumann entropy for sufficiently large values of $k$ in the partial sum, evaluating the higher-dimensional Riemann-Siegel theta function becomes increasingly difficult. For efforts in this direction, see \cite{deconinck2002computing, frauendiener2017efficient}. However, we will revisit this example in the next section when discussing the sequence model.

However, there are two limiting cases where analytic perturbative expansions are available, and approximate analytic continuations of the von Neumann entropies can be obtained. The first limit corresponds to small values of the cross-ratio $x$, where the von Neumann entropy has been computed analytically up to second order in $x$. The second limit is the decompactification limit, where we take $\eta \rightarrow \infty$. In this limit, there is an approximate expression for the von Neumann entropy. \\

\noindent \bf{Two intervals at small cross-ratio }

Let us consider the following expansion of $\mathcal{F}_{n}(x, \eta)$ at small $x$ for some $\eta \neq 1$
\be \la{smallx}
\mathcal{F}_{n}(x, \eta)=1+ \bigg( \frac{x}{4n^2} \bigg)^\alpha s_{2}(n)+\bigg(\frac{x}{4n^2} \bigg)^{2 \alpha} s_4(n)+ \cdots,
\ee
where we can look at the first order contribution with
\be
s_2(n) \equiv \mathcal N \frac{n}{2} \sum_{j=1}^{n-1} \fr{1}{\[\sin(\pi j/n)\]^{2\a}}.
\ee
The coefficient $\alpha$ for a free boson is given by $\alpha = \text{min}[\eta, 1/\eta]$. $\mathcal{N}$ is the multiplicity of the lowest dimension operators, where for a free boson we have $\mathcal{N}=2$. Up to this order, the analytic von Neumann entropy is given by
\be \la{sma}
S (\rho_{AB}) =\frac{1}{3} \ln \bigg( \frac{x_{12}x_{34}x_{14}x_{23}}{\epsilon^2 x_{13}x_{24}} \bigg)-  \mathcal N  \bigg(\frac{x}{4}\bigg)^{\alpha}\fr{\sqrt{\pi} \G(\a+1)}{4\G\(\a +\fr{3}{2}\)}- \cdots.
\ee
We can set up the numerics by taking $|x_{12}| = |x_{34}|=r$, and the distance between the centers of $A$ and $B$ to be $L$, then the cross-ratio is simply
\be
x=\frac{x_{12} x_{34}}{x_{13}x_{24}}=\frac{r^2}{L^2}.
\ee
Similarly we can express $|x_{14}|=L+r=L(1+\sqrt{x})$ and $|x_{23}|=L-r = L(1-\sqrt{x})$. This would allow us to express everything in terms of $x$ and $L$.

\begin{figure}[hbt!]
\centering
\includegraphics[width=0.45\textwidth]{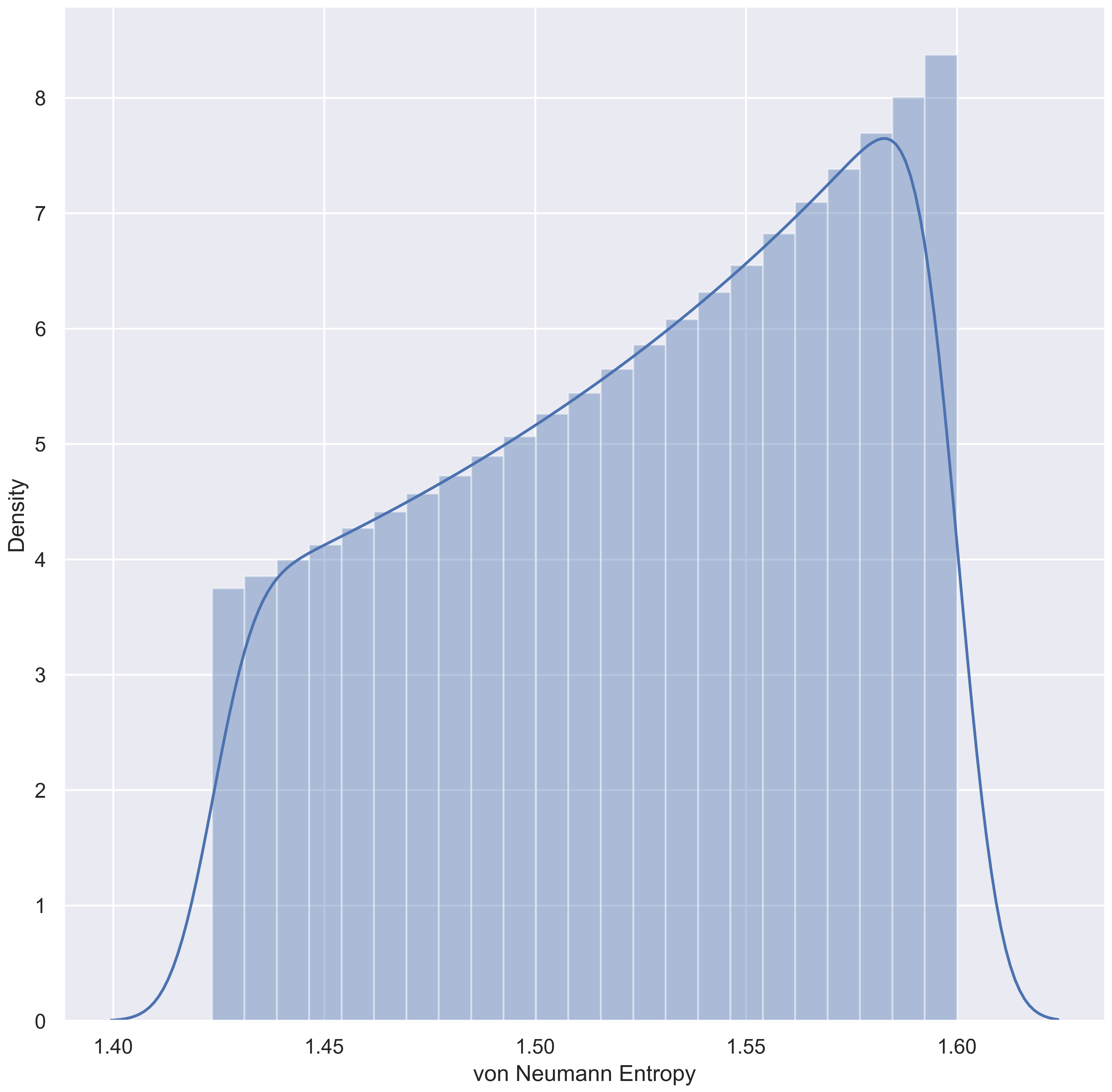}
\includegraphics[width=0.45\textwidth]{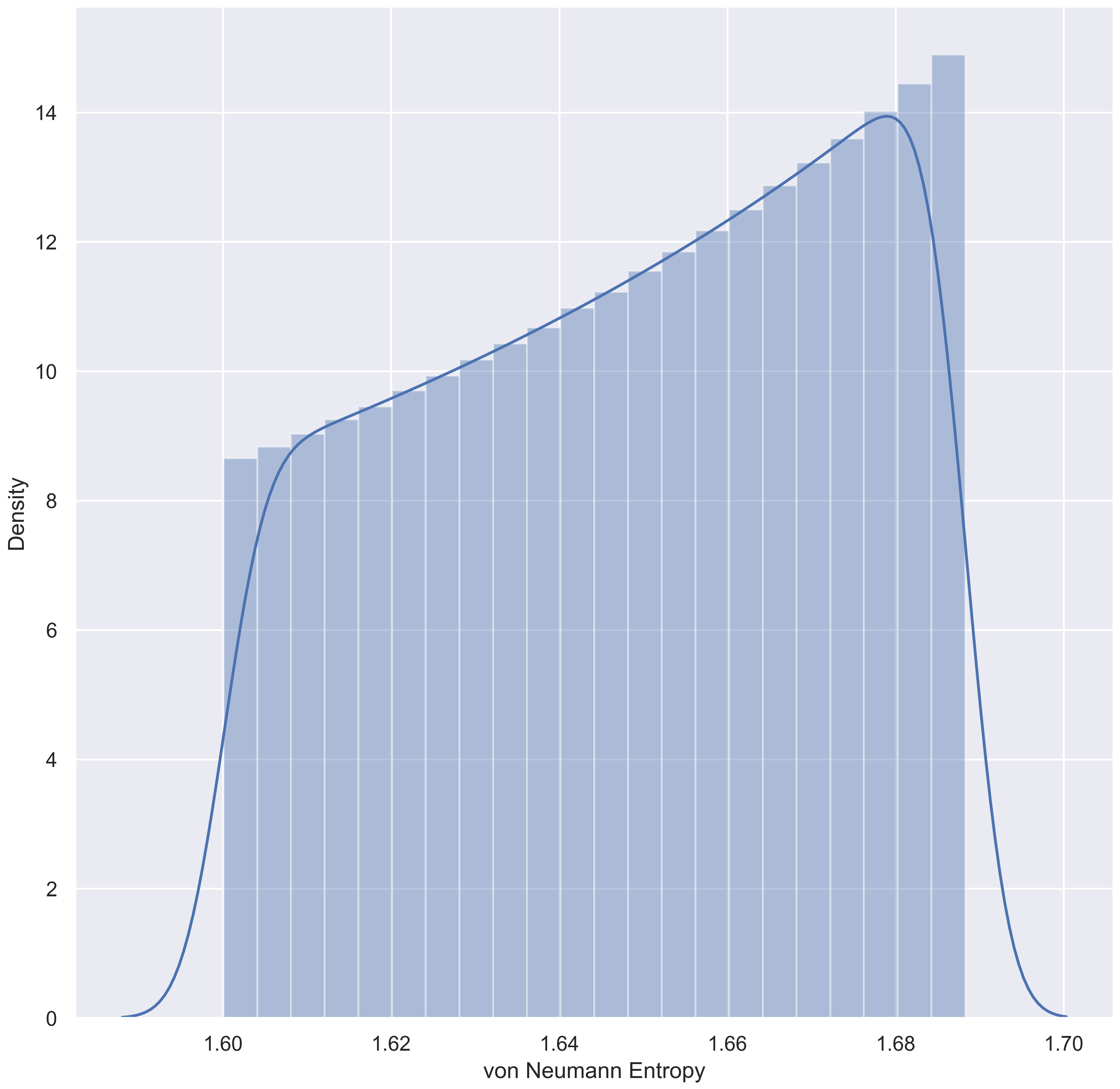}
\caption{The distribution of the two test datasets for
the case of two intervals at small cross-ratio, where we plot density as a function of the
von Neumann entropy computed by \er{sma} with varying $x$.}
\label{3Data}
\end{figure}

\begin{figure}[hbt!] 
\centering
\includegraphics[height=3.8cm, width=7.9cm]{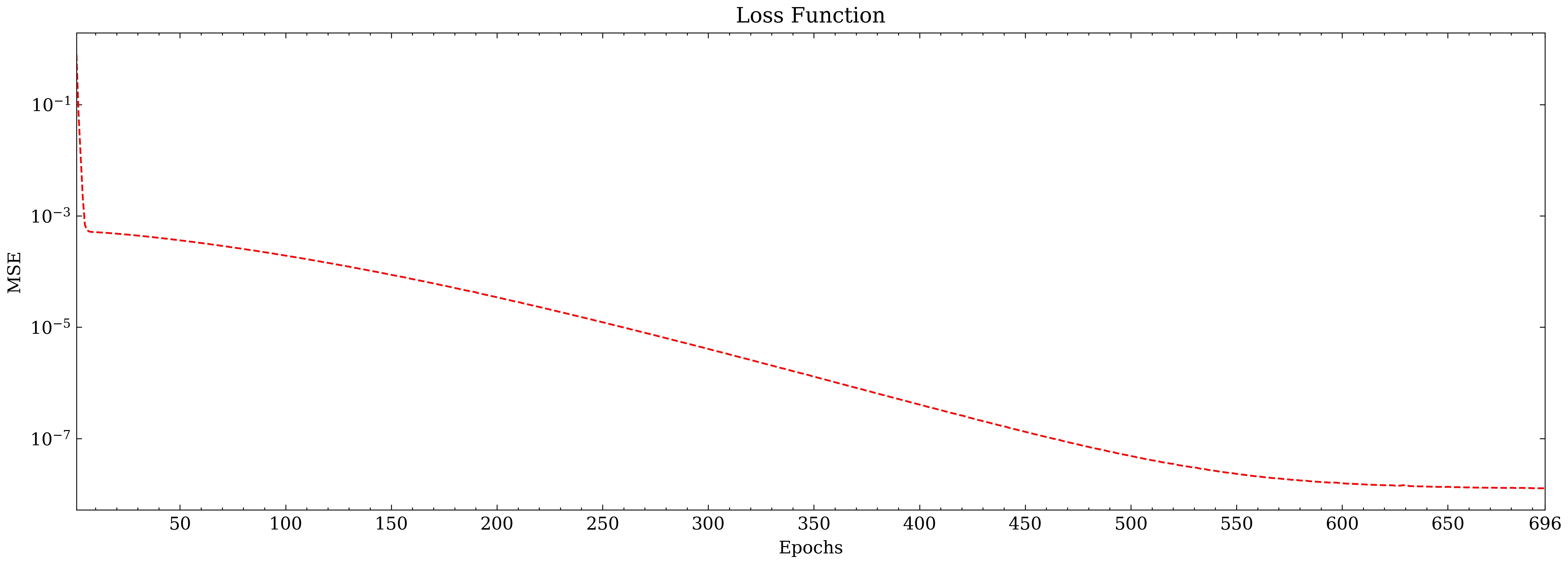}
\includegraphics[width=0.45\textwidth]{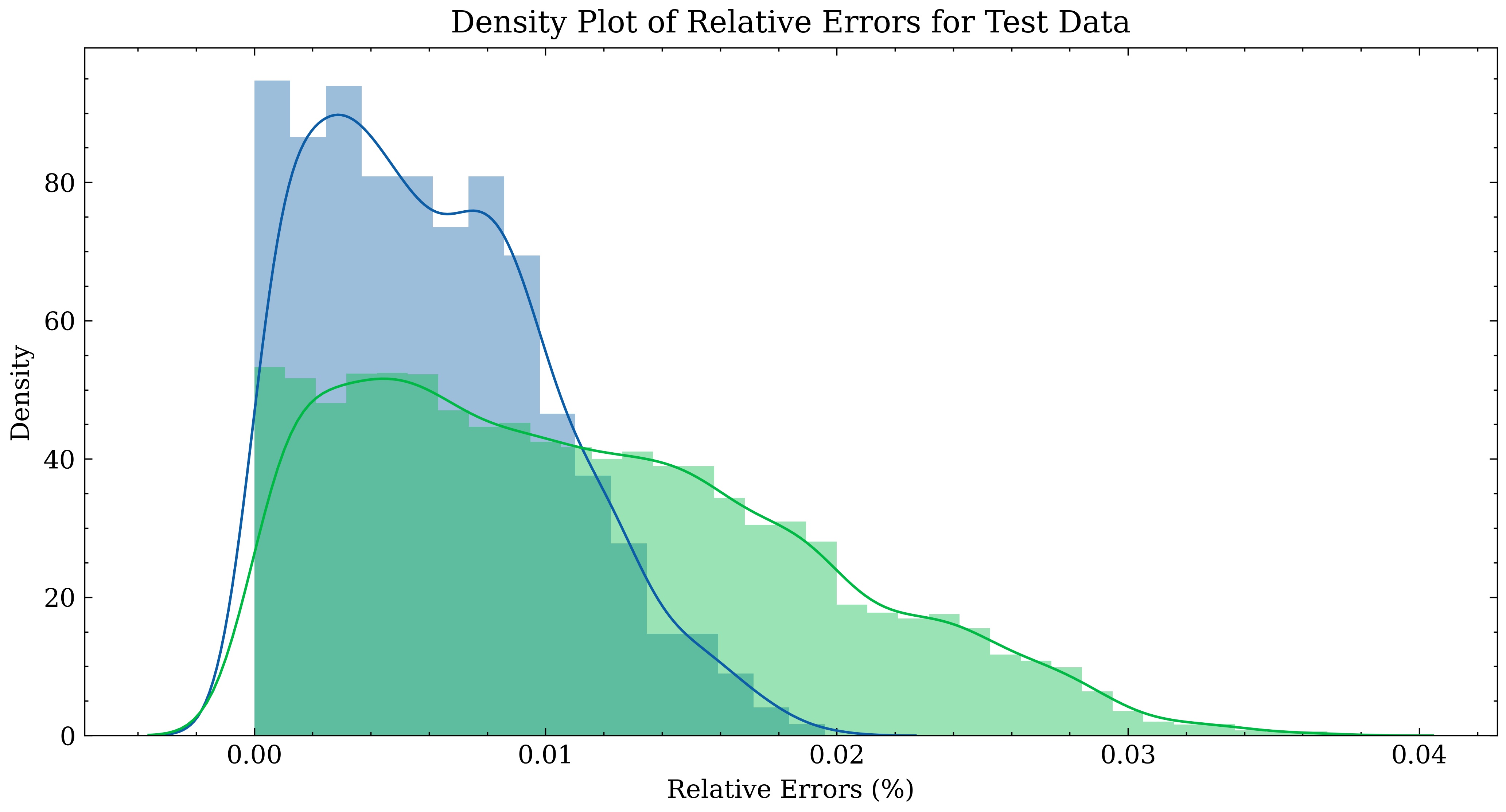}
\caption{Left: The MSE loss function as a function of epochs. The minimum loss close to $10^{-8}$ is achieved at epoch 696 for this instance. Right: The relative errors between the model predictions and targets for the two test datasets, where we have achieved high accuracy with relative errors $\lesssim 0.03 \%$.}
\label{3Loss}
\end{figure}

\begin{figure}[hbt!] 
\centering
\includegraphics[width=0.45\textwidth]{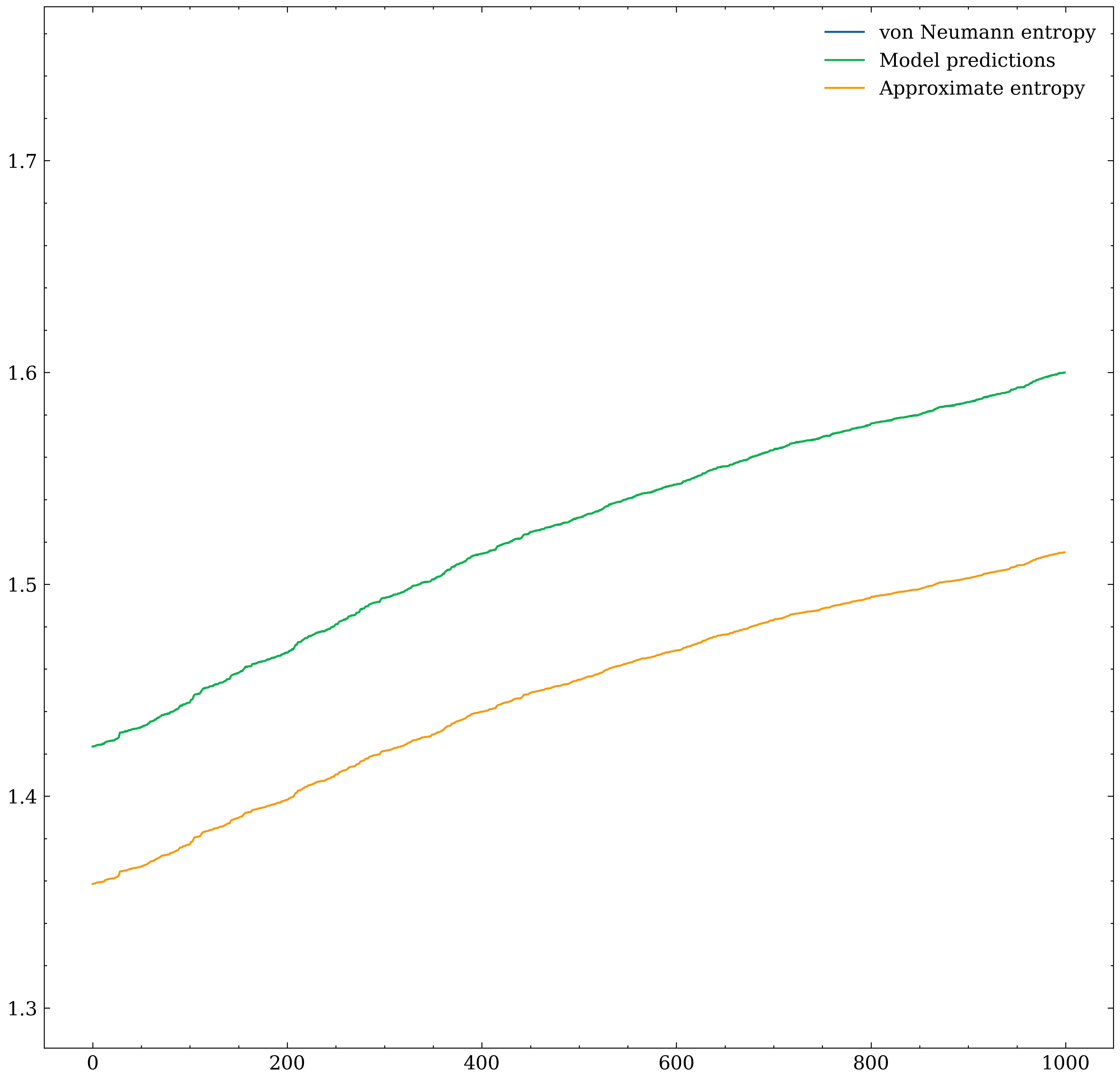}
\includegraphics[width=0.45\textwidth]{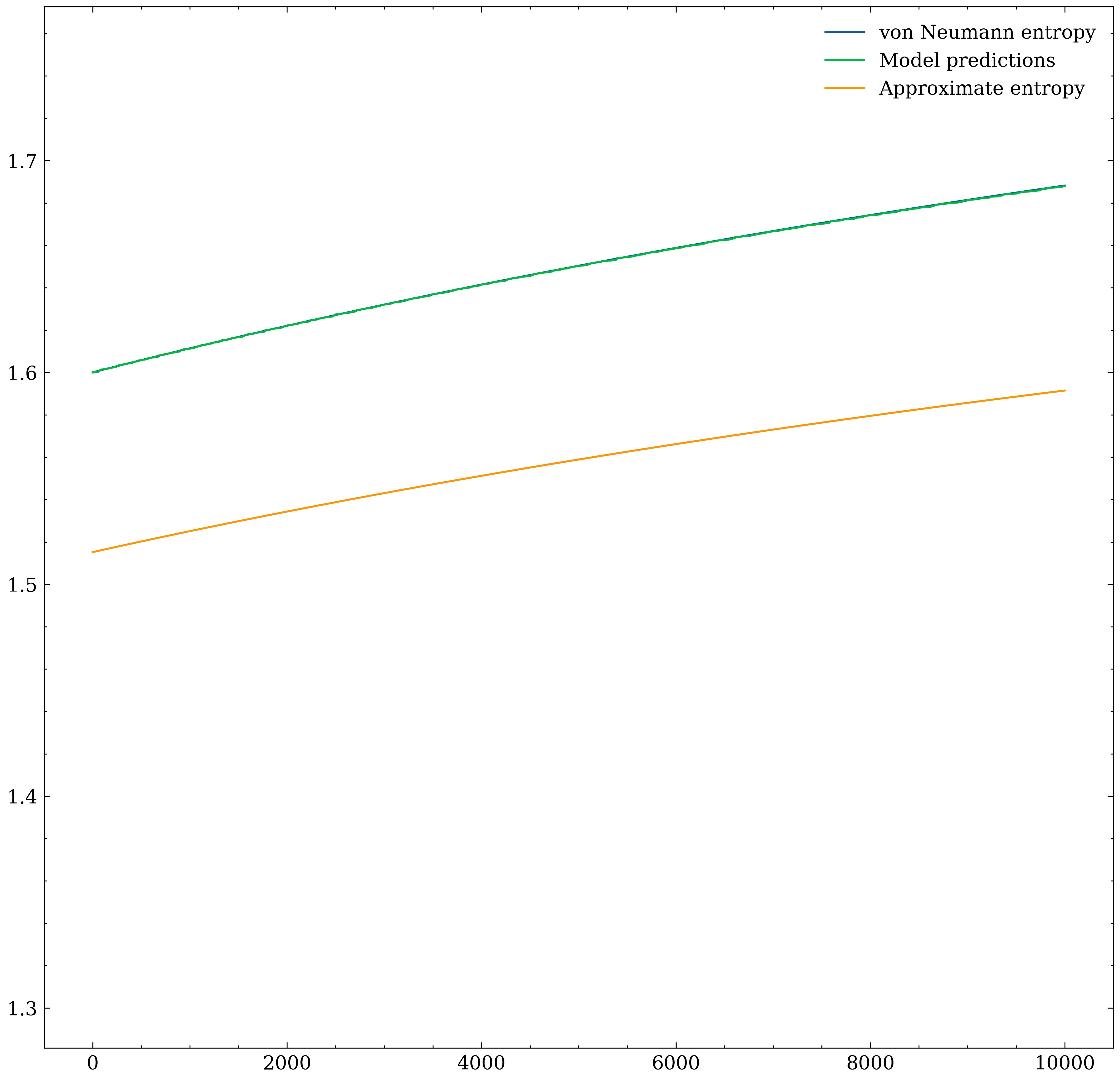}
\caption{We plot the predictions from the model with the analytic von Neumann entropy computed by \er{sma} for the two test datasets. We also include the approximate entropy by summing over $k=50$ terms in the generating function.}
\label{3Entropy}
\end{figure}

For the datasets, we fixed $L=14$, $\alpha =0.5$, and $\epsilon^2=0.1$. We generated $10000$ sets of data for train-validation-test split from $x =0.05$ to $0.1$, with an increment of $\Delta x =5 \times 10^{-6}$ between each step up to $k=50$ in $G(w;\rho_A)$. To further validate our model, we generated $10000$ additional test datasets for the following physical parameters: $x =0.1$ to $0.15$ with $\Delta x =5 \times 10^{-6}$. A density plot of the data with respect to the von Neumann entropy is shown in Figure~\ref{3Data}. We refer to Figure~\ref{3Loss} and Figure~\ref{3Entropy} for a clear demonstration of the learning outcomes.  

The study up to second order in $x$ using the generating function method is available in \cite{DHoker:2020bcv}, as well as through the use of holographic methods \cite{Barrella:2013wja}. Additionally, an analytic continuation toward the von Neumann entropy up to second order in $x$ for general CFT$_2$ can be found in \cite{Perlmutter:2013paa}. Although this is a subleading correction, it can also be approached using our method. \\

\FloatBarrier 

\FloatBarrier 

\FloatBarrier

\noindent \bf{Two intervals in the decompactification limit}

There is a different limit that can be taken other than the small cross-ratio, where an approximate analytic R\'enyi entropies can be obtained. This is called the decompactification limit where we take $ \eta \to \infty$, then for each fixed value of $x$ we have $\mathcal{F}(x, \eta)$ as
\be \la{decom}
\mathcal{F}_{n}(x, \eta)=\bigg[ \frac{\eta^{n-1}}{\prod^{n-1}_{k=1} F_{k/n}(x) F_{k/n}(1-x)} \bigg]^\frac{1}{2},
\ee
where $_2F_1$ is the hypergeometric function. Equation \er{decom} is invariant under $\eta \leftrightarrow 1/\eta$, so we will instead use the result with $\eta \ll 1$
\be 
\mathcal{F}_{n}(x, \eta)=\bigg[ \frac{\eta^{-(n-1)}}{\prod^{n-1}_{k=1} F_{k/n}(x) F_{k/n}(1-x)} \bigg]^\frac{1}{2}.
\ee
In this case, the exact analytic continuation of the von Neumann entropy is not known, but there is an approximate result following the expansion
\be \la{Ap}
S (\rho_{AB}) \simeq S^{W} (\rho_{AB})+\frac{1}{2} \ln \eta - \frac{D'_1(x)+D'_1(1-x)}{2} +\cdots, \quad (\eta \ll 1)
\ee
with $S^{W} (\rho_{AB})$ being the von Neumann entropy computed from the R\'enyi entropies without the special function $\mathcal{F}_n(x, \eta)$ in \er{ge}. Note that
\be
D'_{1}(x)=-\int_{-i \infty}^{i \infty}\frac{dz}{i}\frac{\pi z}{\sin^2 (\pi z)} \ln F_{z}(x).
\ee
This approximate von Neumann entropy has been well tested in previous studies \cite{Calabrese:2009ez, DHoker:2020bcv}, and we will adopt it as the target values in our deep learning models.

\begin{figure}[hbt!] 
\centering
\includegraphics[width=0.45\textwidth]{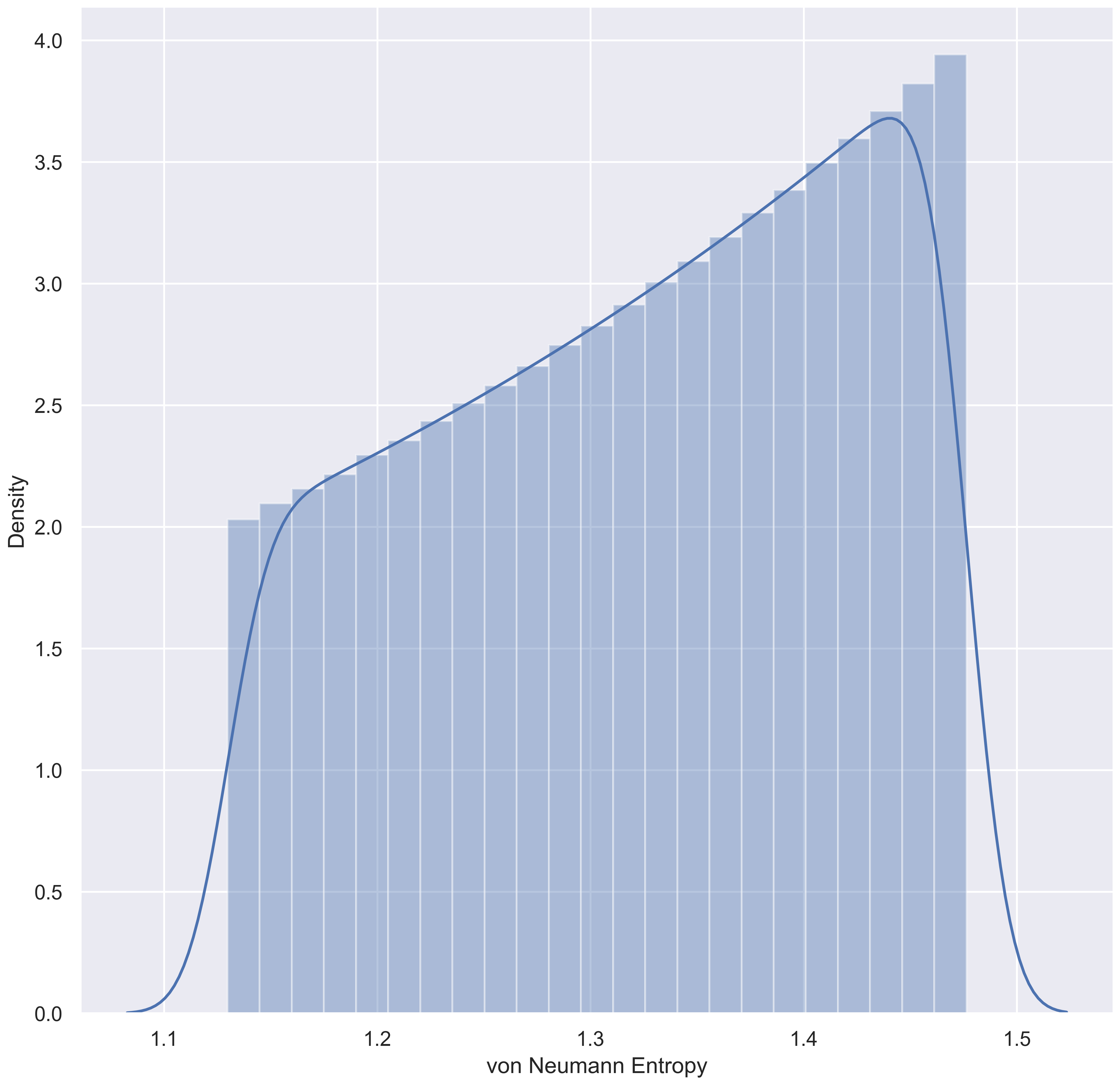}
\includegraphics[width=0.45\textwidth]{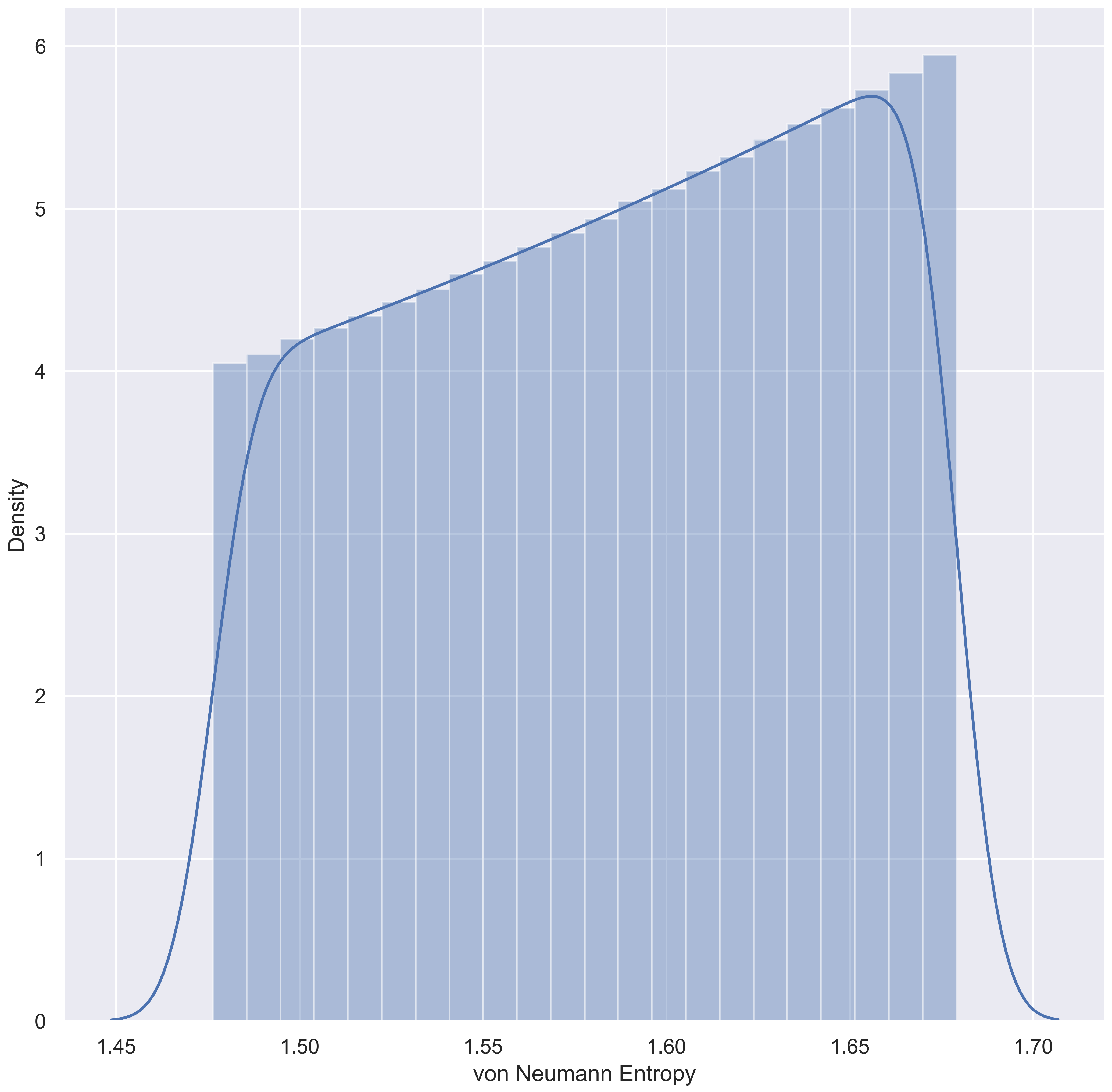}
\caption{ The distribution of the two test datasets for
the case of two intervals in the decompactification limit, where we plot density as a function of the
von Neumann entropy computed by \er{Ap} with varying $\eta$.}
\label{4Data}
\end{figure}

\begin{figure}[hbt!]  
\centering
\includegraphics[height=3.8cm, width=7.9cm]{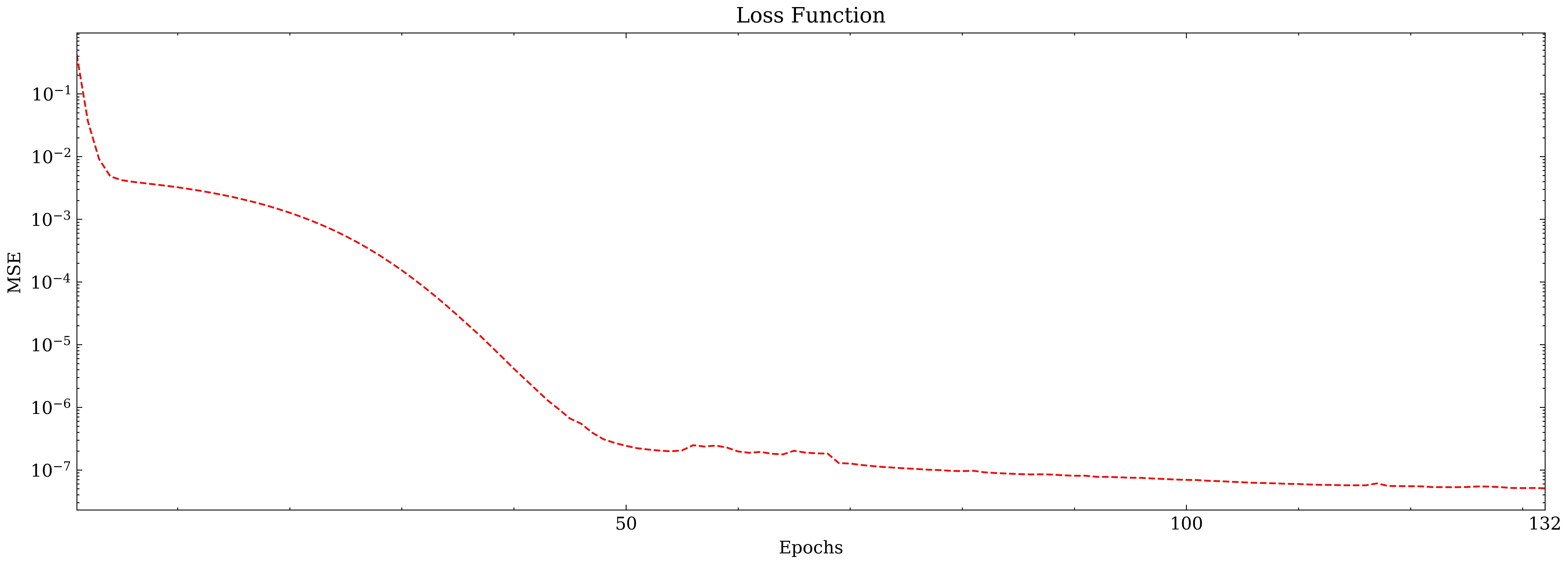}
\includegraphics[width=0.45\textwidth]{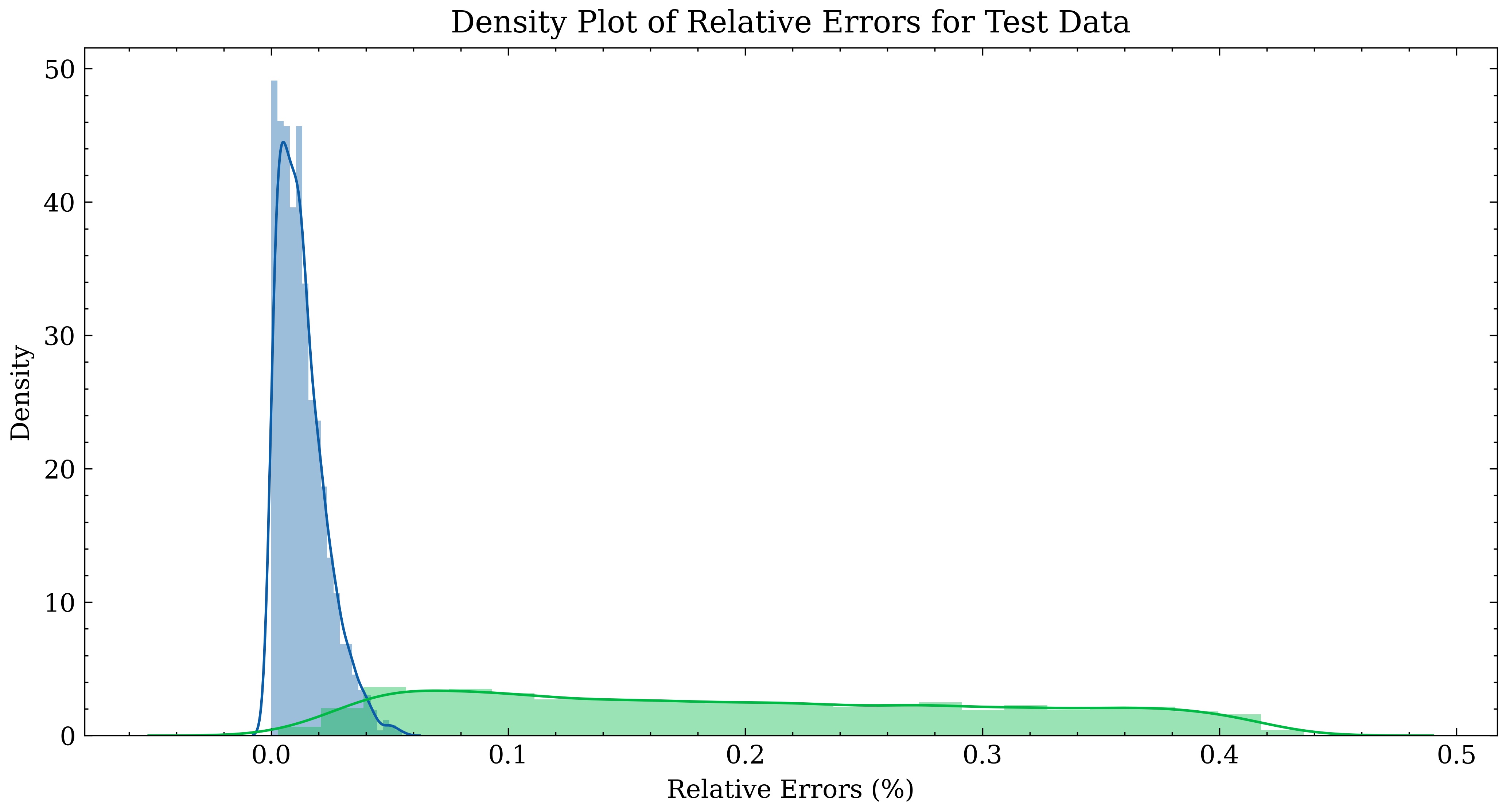}
\caption{Left: The MSE loss function as a function of epochs. The minimum loss at around $10^{-7}$ is achieved at epoch 132 for this instance. Right: The relative errors between the model predictions and targets for the two test datasets, where we have achieved high accuracy with relative errors $\lesssim 0.4 \%$.}
\label{4Loss}
\end{figure}

\begin{figure}[hbt!]  
\centering
\includegraphics[width=0.45\textwidth]{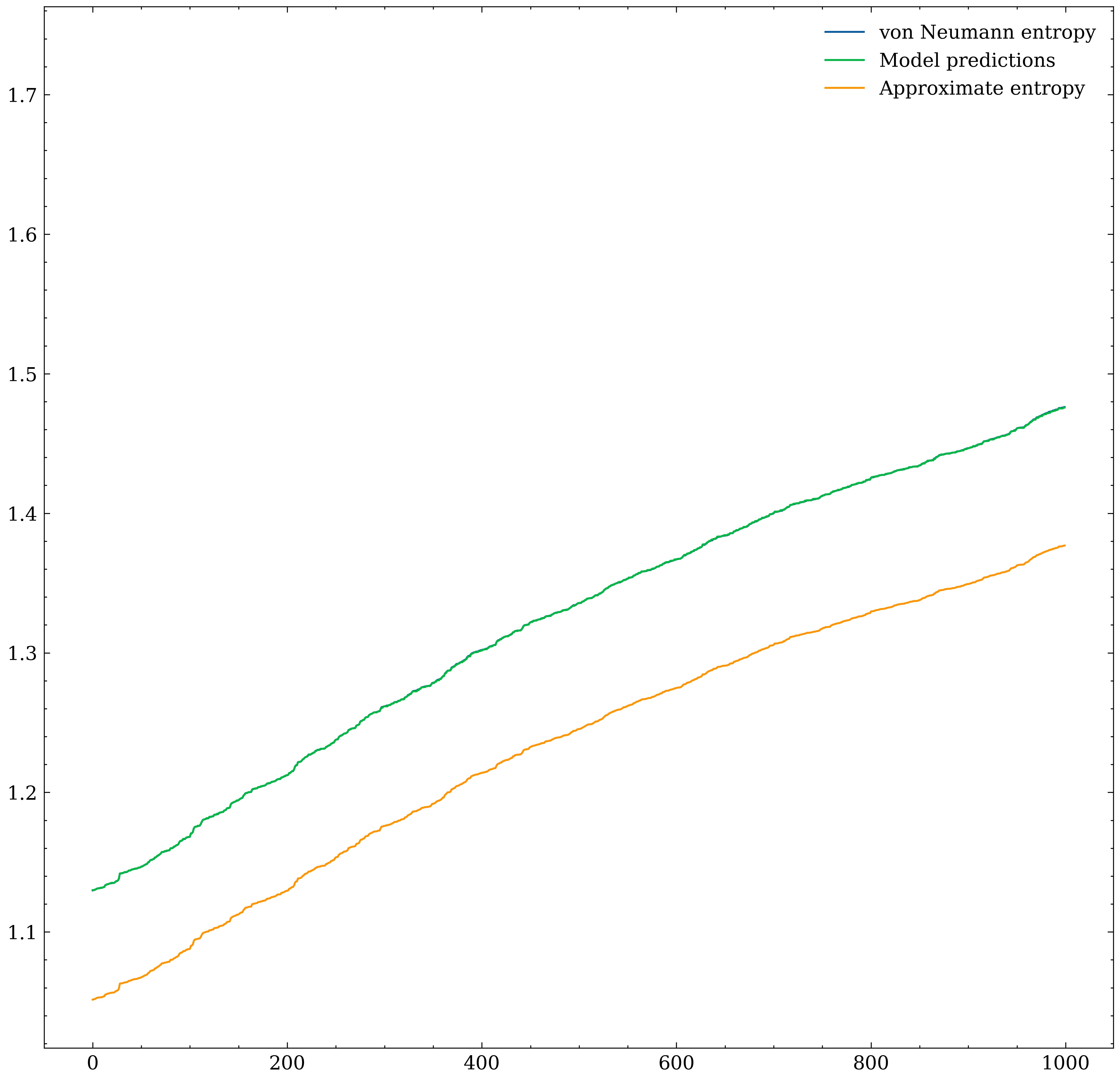}
\includegraphics[width=0.45\textwidth]{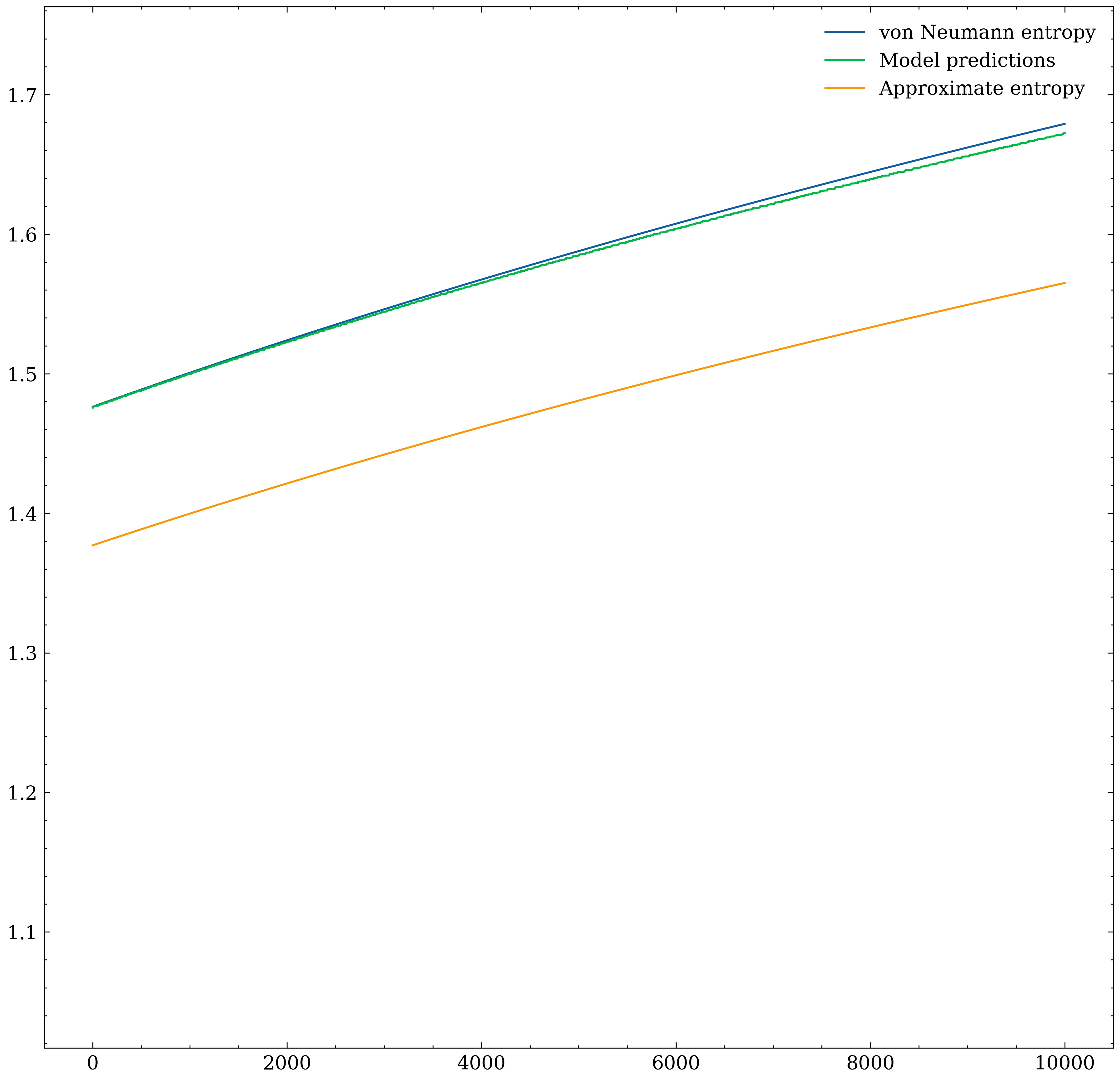}
\caption{We plot the predictions from the model with the analytic von Neumann entropy computed by \er{Ap} for the two test datasets. We also include the approximate entropy by summing over $k=50$ terms in the generating function.}
\label{4Entropy}
\end{figure}

For the datasets, we fixed $L=14$, $x =0.5$ and $\epsilon^2=0.1$. We generated 10000 sets of data for train-validation-test split from $\eta =0.1$ to $0.2$, with an increment of $\Delta \eta = 10^{-5}$ between each step up to $k=50$. To further validate our model, we generated $10000$ additional test datasets for the following physical parameters: $\eta =0.2$ to $0.3$ with $\Delta \eta = 10^{-5}$. A density plot of the data with respect to the von Neumann entropy is shown in Figure~\ref{4Data}. We again refer to Figure~\ref{4Loss} and Figure~\ref{4Entropy} for
a clear demonstration of the learning outcomes.

\FloatBarrier 

\FloatBarrier 

\FloatBarrier

We have seen that deep neural networks, when treated as supervised learning, can achieve accurate predictions for the von Neumann entropy that extends outside the parameter regime in the training phase. However, the potential for deep neural networks may go beyond this.

As we know, the analytic continuation must be worked out on a case-by-case basis (see the examples in \cite{Calabrese:2004eu, Calabrese:2009ez, Calabrese:2009qy, Calabrese:2010he}) and may even depend on the method we use \cite{DHoker:2020bcv}. Finding general patterns in the analytic continuation is still an open question. Although it remains ambitious, the non-linear mapping that the neural networks uncover would allow us to investigate the expressive power of deep neural networks for the analytic continuation problem of the von Neumann entropy.

Our approach also opens up the possibility of using deep neural networks to study cases where analytic continuations are unknown, such as the general two-interval case. Furthermore, it may enable us to investigate other entanglement measures that follow similar patterns or require analytic continuations. We leave these questions as future tasks.

\section{R\'enyi entropies as sequential deep learning} \la{s4}

In this section, we focus on higher R\'enyi entropies using sequential learning models. Studying higher R\'enyi entropies that depend on $\Tr \rho^n_A$ is equivalent to studying the higher-order terms in the Taylor series representation of the generating function \er{gen1}. There are a few major motivations. Firstly, although the generating function can be used to compute higher-order terms, it becomes inefficient for more complex examples.
Additionally, evaluating $\Tr \rho^n_A$ in \er{ge} for the general two-interval case involves the Riemann-Siegel theta function, which poses a challenge in computing higher R\'enyi entropies \cite{DHoker:2020bcv,deconinck2002computing, frauendiener2017efficient}. On the other hand, all higher R\'enyi entropies should be considered independent and cannot be obtained in a linear fashion. They can all be used to predict the von Neumann entropy, but in the Taylor series expansion \er{gen1}, knowing higher R\'enyi entropies is equivalent to knowing a more accurate von Neumann entropy. As we cannot simply extrapolate the series, using a sequential learning approach is a statistically robust way to identify underlying patterns.

\textit{Recurrent neural networks} (RNNs) are a powerful type of neural network for processing sequences due to their "memory" property \cite{Rumelhart1986}. RNNs use internal loops to iterate through sequence elements while keeping a state that contains information about what has been observed so far. This property allows RNNs to identify patterns in a sequence regardless of their position in the sequence. To train an RNN, we initialize an arbitrary state and encode a rank-2 tensor of size (steps, input features), looping over multiple steps. At each step, the networks consider the current state at $k$ with the input, and combine them to obtain the output at $k+1$, which becomes the state for the next iteration.

RNNs incorporate both feedforward networks and \textit{back-propagation through time} (BPTT) \cite{rumelhart1985learning, WERBOS1988339}, with "time" representing the steps $k$ in our case. The networks connect the outputs from a fully connected layer to the inputs of the same layer, referred to as the hidden states. These inputs receive the output values from the previous step, with the number of inputs to a neuron determined by both the number of inputs to the layer and the number of neurons in the layer itself, known as \textit{recurrent connections}. Computing the output involves iteratively feeding the input vector from one step, computing the hidden states, and presenting the input vector for the next step to compute the new hidden states.

RNNs are useful for making predictions based on sequential data, or "sequential regression," as they learn patterns from past steps to predict the most probable values for the next step.

\subsection{Model architectures and training strategies} \la{s4.1}

In this subsection, we discuss the methodology of treating the R\'enyi entropies (the Taylor series of the generating function) as sequence models. \\

\textbf{Data preparation}

To simulate the scenario where $k_{\text{max}}$ in the series cannot be efficiently computed, we generate $N=10000$ datasets for different physical parameters, with each dataset having a maximum of $k_{\text{max}}=50$ steps in the series.  We also shuffle the $N$ datasets since samples of close physical parameters will have most of their values in common. Among the $N$ datasets, we only take a fraction $p<N$ for the train-validation-test split. The other fraction $q=N-p$ will all be used as test data for the trained model. This serves as a critical examination of the sequence models we find. The ideal scenario is that we only need small $p$ datasets while achieving accurate performance for the $q$ datasets.

Due to the rather small number of steps available, we are entitled to adopt the SimpleRNN structure in TensorFlow-Keas\footnote{SimpleRNN suffers from the vanishing gradient problem when learning long dependencies \cite{279181}. Even using ReLU, which does not cause a vanishing gradient, back-propagation through time with weight sharing can still lead to a vanishing gradient across different steps. However, since the length of the sequence is small due to the limited maximum steps available in our case, we have found that SimpleRNN generally performs better than its variants.} instead of the more complicated ones such as LSTM or GRU networks \cite{hochreiter1997long, cho-etal-2014-properties}.

We also need to be careful about the train-validation-test splitting process. In this type of problem, it is important to use validation and test data that is more recent than the training data. This is because the objective is to predict the next value given the past steps, and the data splitting should reflect this fact.  Furthermore, by giving more weight to recent data, it is possible to mitigate the vanishing gradient (memory loss) problem that can occur early in the BPTT. In this work, the first $60\%$ of the steps ($k=1 \sim 30$) are used for training, the middle $20\%$ ($k= 31 \sim 40$) for validation, and the last $20\%$ ($k= 41 \sim 50$) for testing.

\begin{figure}[hbt!]
\centering
\includegraphics[width=0.97\textwidth]{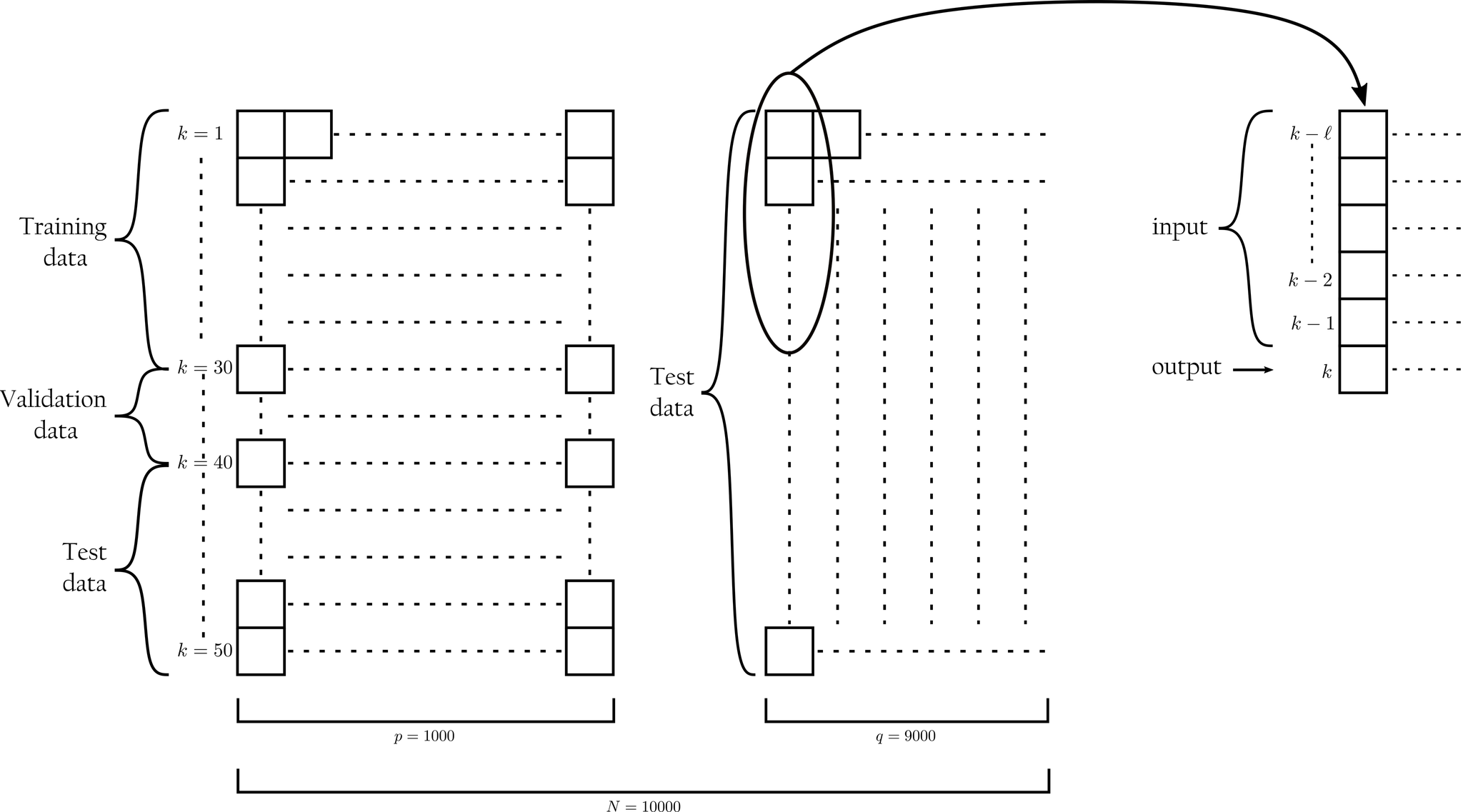}
\caption{Data preparation process for the sequential models. A total of $N$ datasets are separated into two parts: the $p$ datasets are for the initial train-validation-test split, while the $q$ datasets are treated purely as test datasets. The zoomed-in figure on the right hand side illustrates how a single example sequence is generated, where we have used a fixed number of past steps $\ell=5$. Note that for the additional $q$ test datasets, a total of $(\text{steps}-\ell) \times q=405000$ sequences are generated.}
\label{DataPrep}
\end{figure}

We split the datasets in the following way: for a single dataset from each step, we use a fixed number of past steps\footnote{We could also include as many past steps as possible, but we have found it less effective. This can be attributed to our choice of network architectures and the fact that we have rather short maximum steps available.}, specified by $\ell$, to predict the next value. This will create $(\text{steps}-\ell)$ sequences from each dataset, resulting in a total of $(\text{steps}-\ell) \times p$ sequences for the $p$ datasets in the train-validation-test splitting. Using a fixed sequence length $\ell$ allows the network to focus on the most relevant and recent information for predicting the next value, while also simplifying the input size and making it more compatible with our network architectures. We take $p=1000$, $q=9000$, and $\ell=5$. An illustration of our data preparation strategy is shown in Figure~\ref{DataPrep}.

\FloatBarrier

\textbf{Model design}

After the pre-processing of data, we turn to the model design. Throughout the section, we use the ReLU activation function and Adam optimizer with MSE as the loss function.

In KerasTuner, we employ Bayesian optimization by adjusting a few crucial hyperparameters and designs. We summarize them in the following list: 

\begin{itemize}
    \item We introduce one or two SimpleRNN layers, with or without recurrent dropouts. The units of the first layer range from 64 to 256 with a step size of 16. If a second layer is used, the units range from 32 to 128 with a step size of 8. Recurrent dropout is applied with a dropout rate in the range of 0.1 to 0.3 using log sampling.
    \item We take LayerNormalization as a Boolean choice to enhance the training stability, even with shallow networks. The LayerNormalization is added after the SimpleRNN layer if there is only one layer; in between the two layers if there are two SimpleRNN layers.
    \item We allow a Dense layer with units ranging from 16 to 32 and a step size of 8 as an optional regressor after the recurrent layers.
    \item A final dropout with log sampling of a dropout rate in the range of 0.2 to 0.5 is added as a Boolean choice.
    \item In the Adam optimizer, we only adjust the learning rate with log sampling from the range of $10^{-5}$ to $10^{-4}$. All other parameters are taken as the default values in TensorFlow-Keras. We take the AMSGrad \cite{reddi2019convergence} variant of this algorithm as a Boolean choice. 

\end{itemize}

The KerasTuner is deployed for 300 trials with 2 executions per trial. During the process, we monitor the validation loss using EarlyStopping of patience 8. Once the best set of hyperparameters and model architecture are identified based on the validation data, we initialize a new model with the same design, but with both the training and validation data. This new model is trained 30 times while monitoring the training loss using EarlyStopping of patience 10. The final predictions are obtained by averaging the results of the few cases with close yet overall smallest relative errors from the targets. The purpose of taking the average instead of picking the case with minimum loss is to smooth out possible outliers. We set the batch size in both the KerasTuner and the final training to be 2048.

We will also use the trained model to make predictions on the $q$ test data and compare them with the correct values as validation for hitting the benchmark.

\subsection{Examples of the sequential models}

The proposed approach will be demonstrated using two examples. The first example is a simple representative case of a single interval \er{single}; while the second is a more challenging case of the two-interval at decompactification limit \er{Ap}, where the higher-order terms in the generating function cannot be efficiently computed. Additionally, we will briefly comment on the most non-trivial example of the general two-interval case.

\noindent \bf{Single interval}

\begin{figure}[hbt!]
\centering
\includegraphics[height=4.9cm, width=12cm]{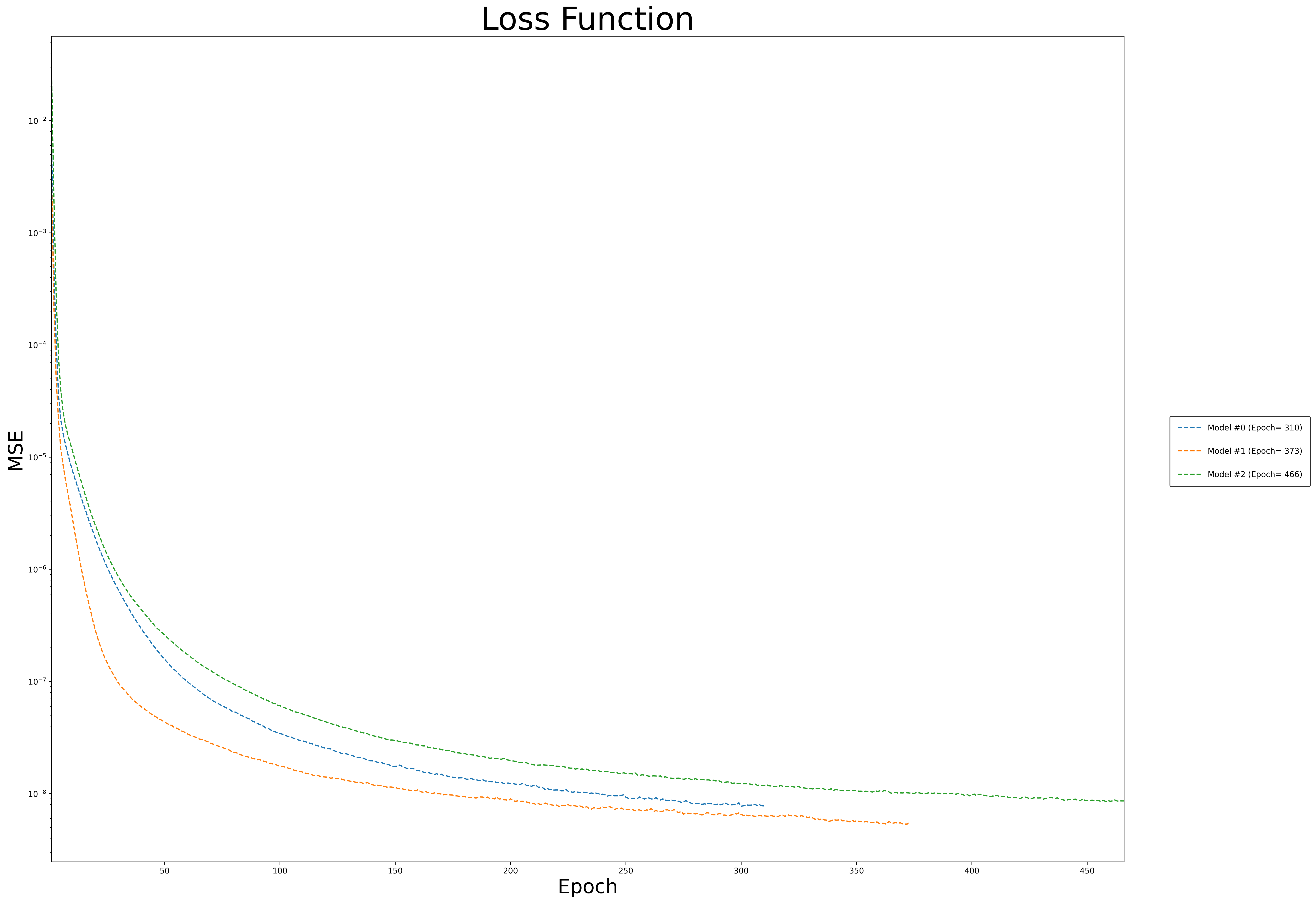}
\includegraphics[width=0.47\textwidth]{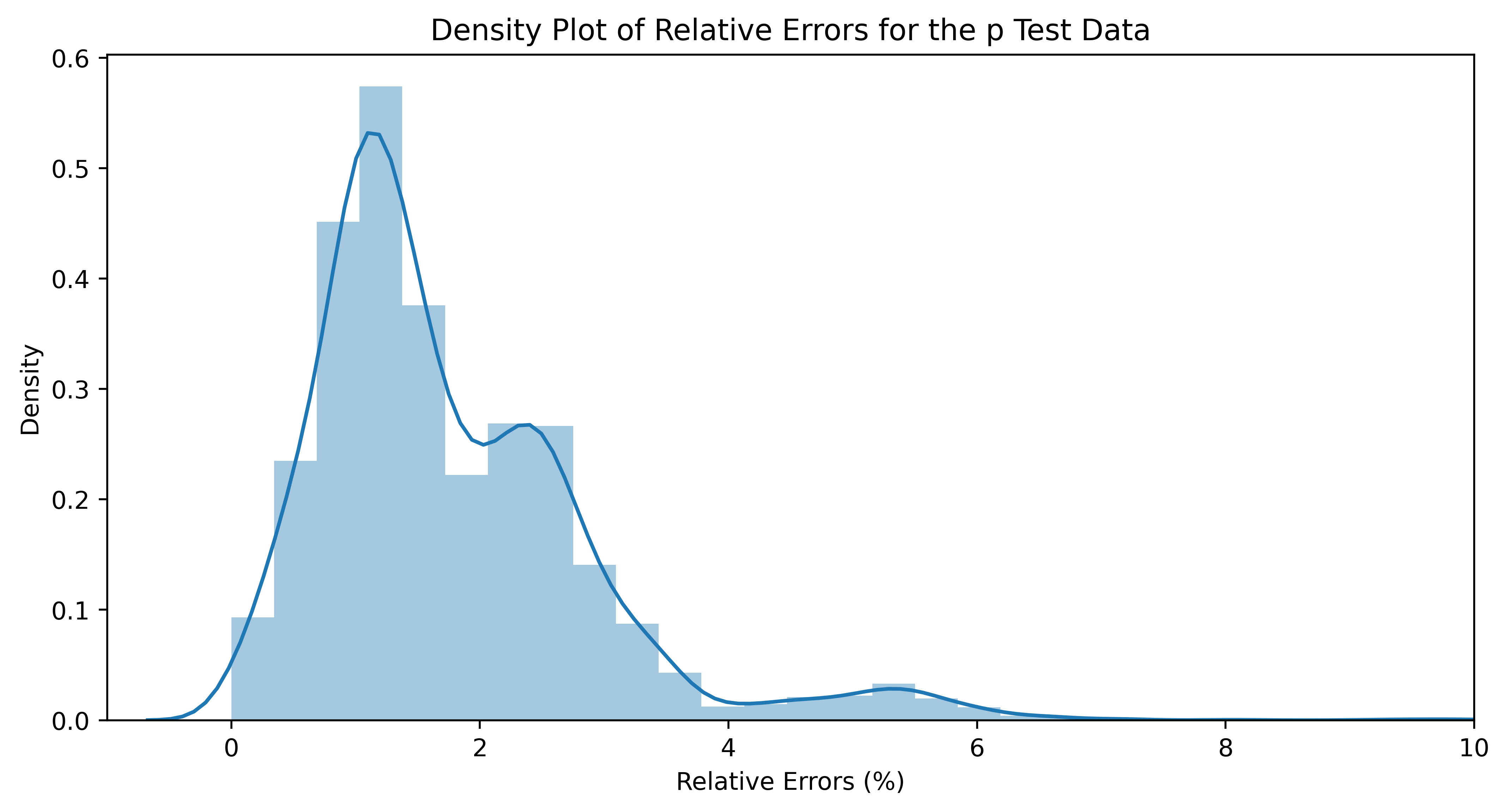}
\includegraphics[width=0.47\textwidth]{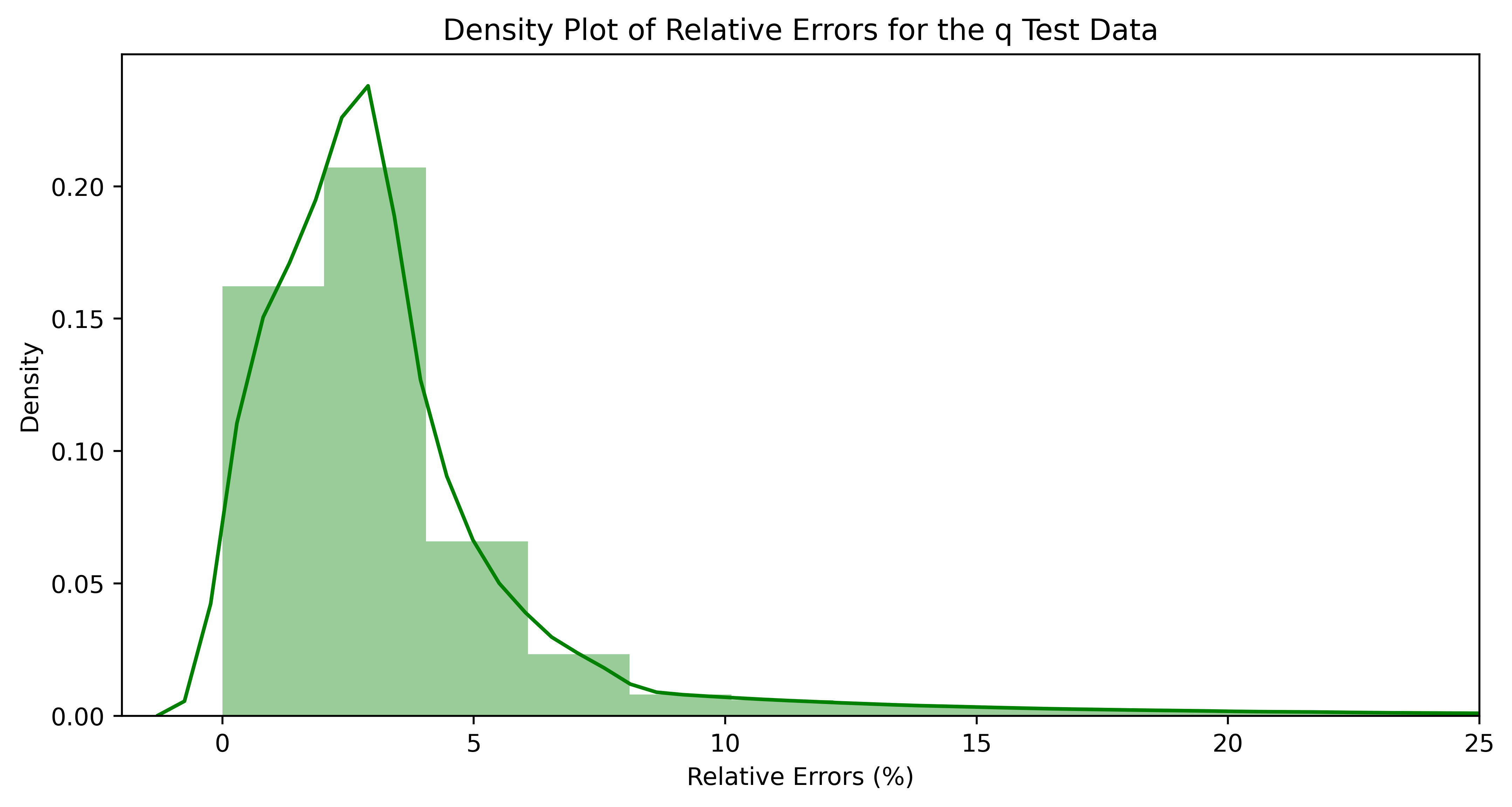}
\caption{Top: The loss function for the best 3 models as a function of epochs. We monitor the loss function with
EarlyStopping, where the epochs of minimum losses at around $10^{-8}$ for different models are specified in the parentheses of the legend. Bottom: The density plots as a function of relative
errors for the two test datasets. The relative errors for the $p$ test datasets are concentrated at around $1 \%$; while for the additional $q$ test datasets, they are concentrated at around $2.5 \%$ with a very small ratio of outliers. }
\label{1Seq}
\end{figure}

In this example, we have used the same $N$ datasets for the single interval as in Sec.~\ref{s3.2}. Following the data splitting strategy we just outlined, it is worth noting that the ratio of training data to the overall dataset is relatively small. We have plotted the losses of the three best-performing models, as well as the density plot of relative errors for the two test datasets in Figure~\ref{1Seq}. Surprisingly, even with a small ratio of training data, we were able to achieve small relative errors on the additional test datasets.

\FloatBarrier

\noindent \bf{Two intervals in the decompactification limit}

\begin{figure}[hbt!] 
\centering
\includegraphics[height=4.9cm, width=12cm]{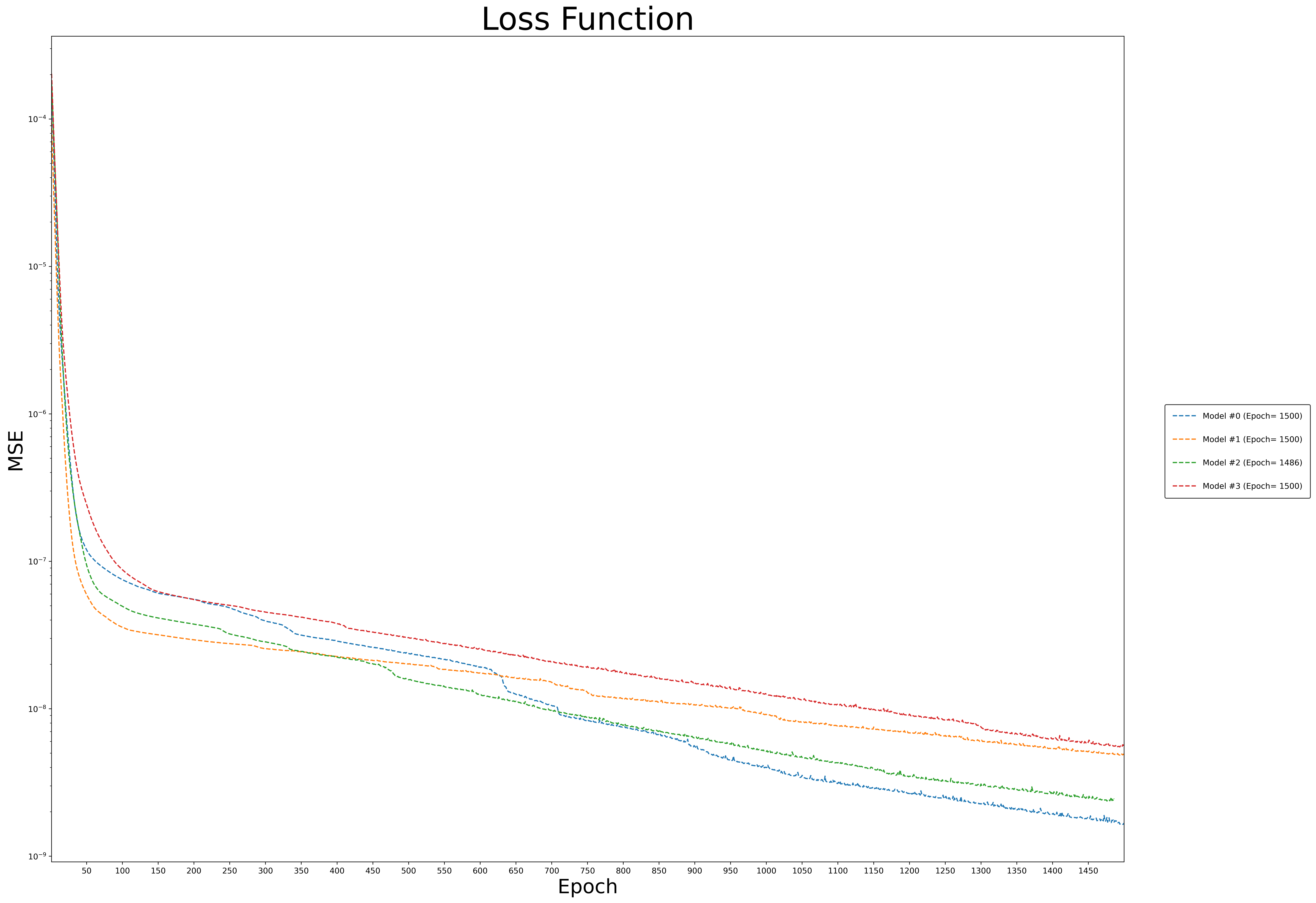}
\includegraphics[width=0.47\textwidth]{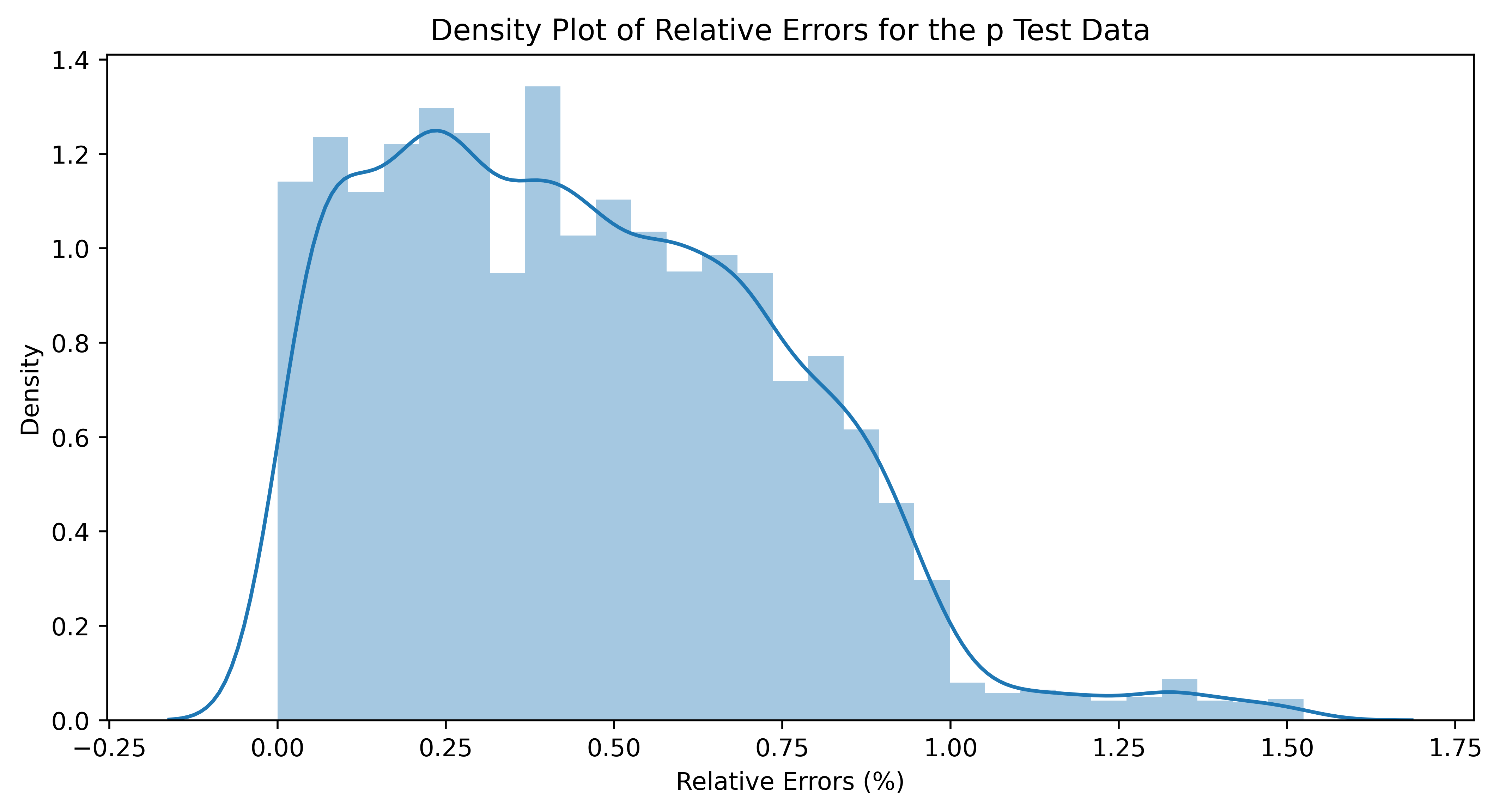}
\includegraphics[width=0.47\textwidth]{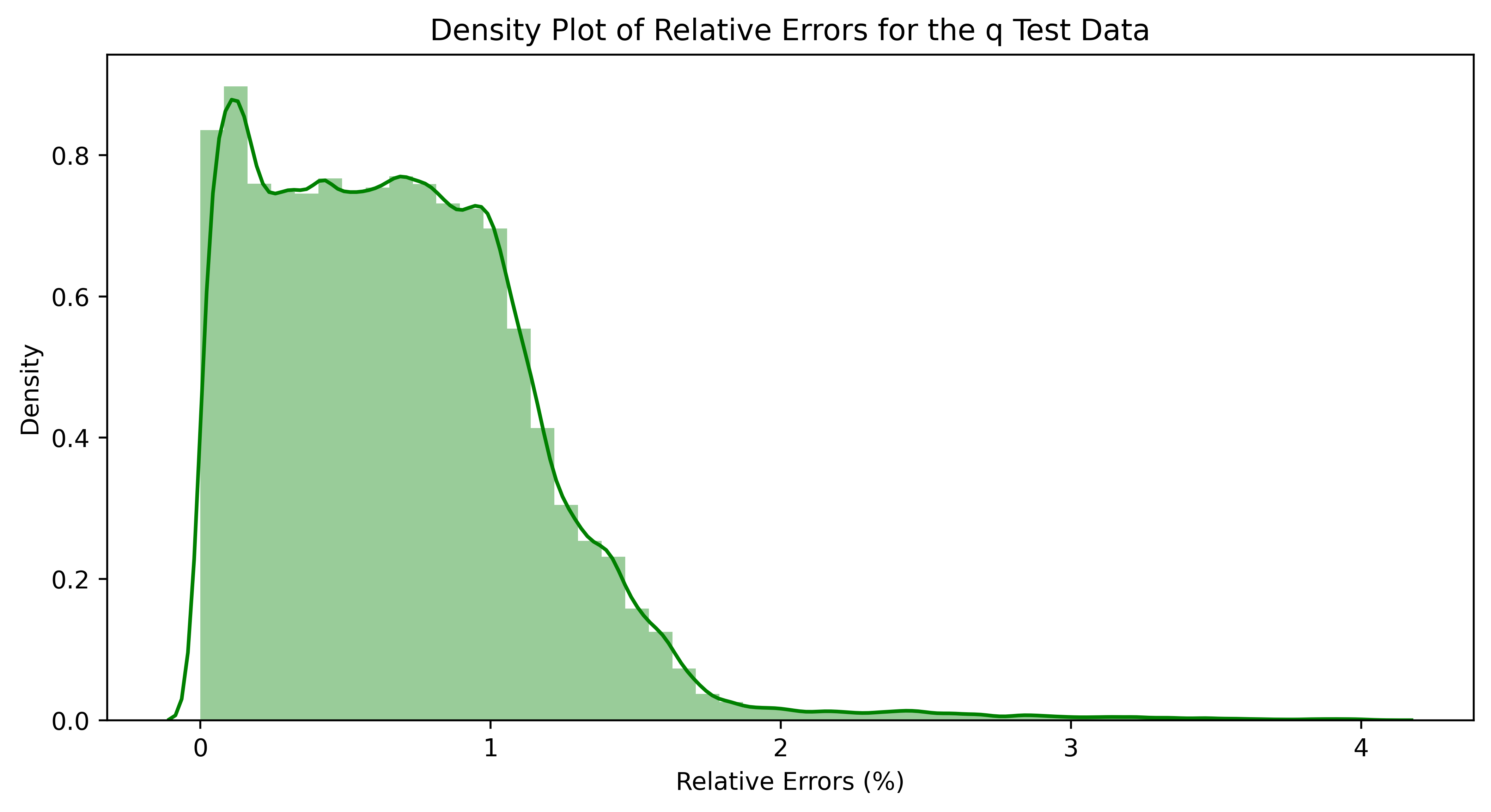}
\caption{Top: The loss function for the best 4 models as functions of epochs. We monitor the loss function with
EarlyStopping.  Bottom: The density plot as a function of relative
errors for the two test datasets. The relative errors for the $p$ test datasets
are well within $\lesssim 1.5 \%$; while for the additional $q$ test datasets, they are well within $\lesssim 2 \%$.}
\label{2Seq}
\end{figure}

Again, we have used the same $N$ datasets for the two intervals in the $\eta \to \infty$ limit as in Sec.~\ref{s3.3}. In Figure~\ref{2Seq}, we have plotted the losses of the four best-performing models and the density plot of relative errors for the two test datasets. In this example, the KerasTuner identified a relatively small learning rate, which led us to truncate the training at a maximum of 1500 epochs since we had achieved the required accuracy. In this case, the predictions are of high accuracy, essentially without outliers.

\FloatBarrier

Let us briefly address the most challenging example discussed in this paper, which is the general two-interval case \er{ge} where the analytic expression for the von Neumann entropy is not available. In this example, only $\Tr \rho^n_A$ is known, and since it involves the Riemann-Siegel theta function, computing the generating function for large $k$ in the partial sum becomes almost infeasible. Therefore, the sequential learning models we have introduced represent the most viable approach for extracting useful information in this case.

Since only $k_{\text{max}} \approx 10$ can be efficiently computed from the generating function in this case, we have much shorter steps for the sequential learning models. We have tested the above procedure with $N=10000$ datasets and $k_{\text{max}} =10$, however, we could only achieve an average of $5\%$ relative errors. Improvements may come from a larger dataset with a longer training time, which we leave as a future task.

In general, sequential learning models offer a potential solution for efficiently computing higher-order terms in the generating function.  To extend our approach to longer sequences beyond the $k_{\text{max}}$ steps, we can treat the problem as self-supervised learning. However, this may require a more delicate model design to prevent error propagation. Nonetheless, exploring longer sequences can provide a more comprehensive understanding of the behavior of von Neumann entropy and its relation to R\'enyi entropies.

\section{Quantum neural networks and von Neumann entropy} \la{s5}

In this section, we explore a similar supervised learning task by treating the quantum circuits as models that map data inputs to predictions, which influences the expressive power of quantum circuits as function approximations.

\subsection{Fourier series from variational quantum machine learning models} \la{s5.1}

We will focus on a specific function class that a quantum neural network can explicitly realize, namely a simple Fourier-type sum \cite{Schuld_2021, gil2020input}.  Before linking it to the von Neumann entropy, we shall first give an overview of the seminal works in \cite{Schuld_2021}.

Consider a general Fourier-type sum in the following form
\be
f_{\theta_i}(\vec{x})=\sum_{\Vec{\omega}\in \Omega}c_{\vec{\omega}}(\theta_i) e^{i \vec{\omega} \cdot \vec{x}},
\ee
with the frequency spectrum specified by $\Omega \subset \mathbb{R}^N$. Note that $c_{\vec{\omega}}(\theta_i)$ are the (complex) Fourier coefficients. We need to come up with a quantum model that can learn the characteristics of the sum by the model's control over the frequency spectrum and the Fourier coefficients. 

\begin{figure}[hbt!]
\centering
\includegraphics[width=0.45\textwidth]{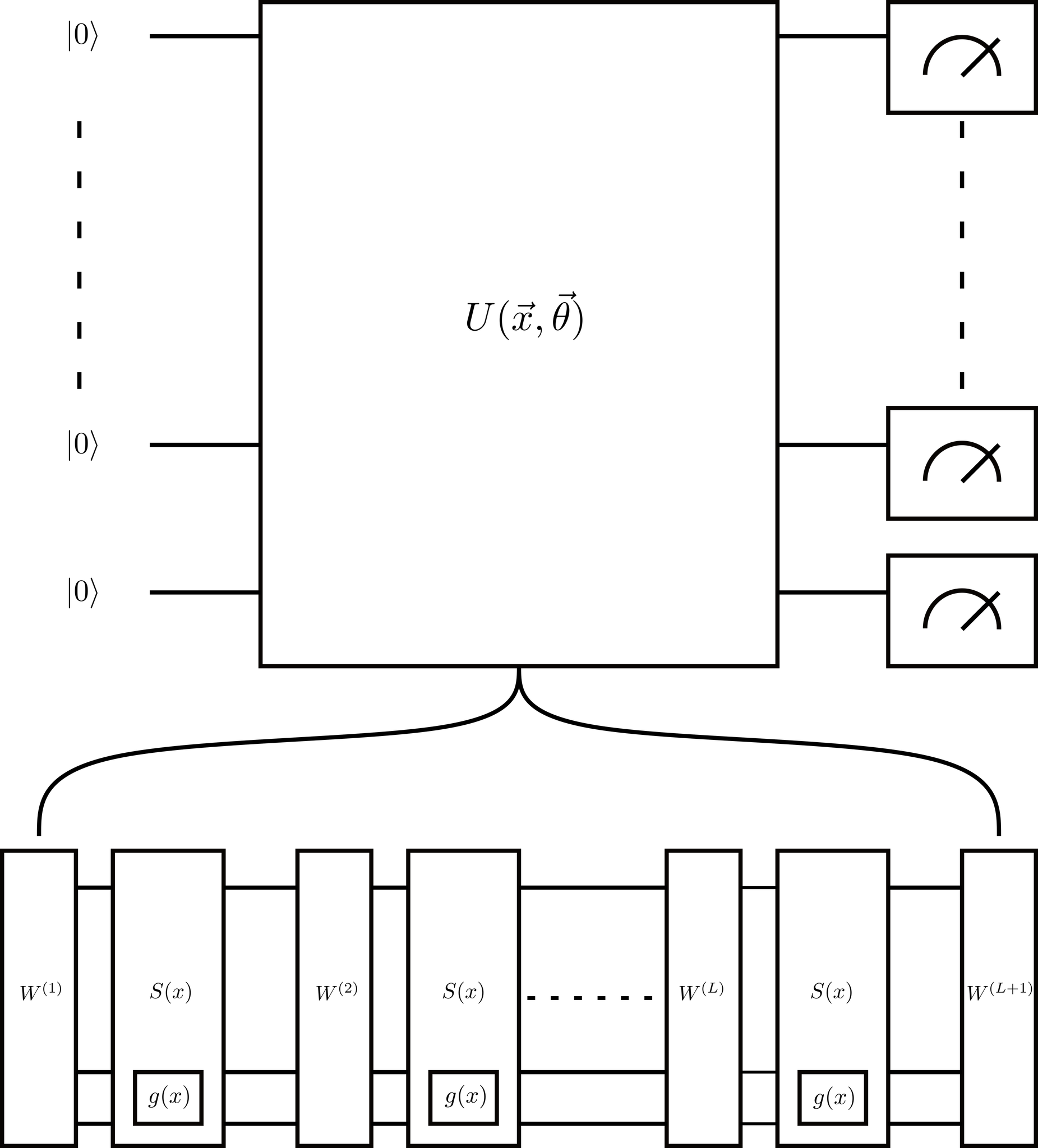}
\caption{Quantum neural networks with repeated data-encoding circuit blocks $S(x)$ (whose gates are of the form $g(x)=e^{-ixH}$) and trainable circuit blocks $W^{(i)}$. The data-encoding circuit blocks determine the available frequency spectrum for $\vec{\omega}$, while the remainder determines the Fourier coefficients $c_{\vec{\omega}}$.}
\label{QMLmodel}
\end{figure}

Now we define the quantum machine learning model as the following expectation value
\be
f_{\theta_i}(x)=\langle 0 | U^\dagger(x, \theta_i) {M} U(x, \theta_i)| 0 \rangle,
\ee
where $|0 \rangle$ is taken to be some initial state of the quantum computer. The ${M}$ will be the physical observable. Note that we have omitted writing the vector symbol and the hat on the operator, which should be clear from the context. The crucial component is $U(x, \theta_i)$, which is a quantum circuit that depends on the data input $x$ and the trainable parameters $\theta_i$ with $L$ layers. Each layer has a data-encoding circuit block $S(x)$, and the trainable circuit block $W(\theta_i)$. Schematically, it has the form 
\be
U(x, \theta_i)=W^{(L+1)}(\theta_i) S(x)W^{(L)}(\theta_i) \cdots W^{(2)}(\theta_i)S(x)W^{(1)}(\theta_i),
\ee
where we refer to Figure~\ref{QMLmodel} for a clear illustration.

\FloatBarrier 

Let us discuss the three major components of the quantum circuit in the following:
\begin{itemize}
    \item The repeated data-encoding circuit block $S(x)$ prepares an initial state that encodes the (one-dimensional) input data $x$ and is not trainable due to the absence of free parameters. It is represented by certain gates that embed classical data into quantum states, with gates of the form $g(x)=e^{-i x H}$, where $H$ is the encoding Hamiltonian that can be any unitary operator. In this work, we use the Pauli X-rotation gate, and the encoding Hamiltonians in $S(x)$ will determine the available frequency spectrum $\Omega$.

    \item The trainable circuit block $W(\theta_i)$ is parametrized by a set of free parameters $\theta_i = (\theta_1, \theta_2, ...)$. There is no special assumption made here and we can take these trainable blocks as arbitrary unitary operations. The trainable parameters will contribute to the coefficients $c_\omega$.

    \item The final piece is the measurement of a physical observable ${M}$ at the output. This observable is general, it could be local for each wire or subset of wires in the circuit.
\end{itemize}

Our goal is to establish that $f(x)$ can be written as a partial Fourier series \cite{Schuld_2021, gil2020input}
\be
f_{\theta_i}(x)=\langle 0 | U^\dagger(x, \theta_i) {M} U(x, \theta_i)| 0 \rangle=\sum_{n \in \Omega}c_n e^{i n x}.
\ee
Note that here for simplicity, we have taken frequencies being integers $\Omega \subset \mathbb{Z}^N$. The training process goes as follows: we sample a quantum model with $U(x, \theta_i)$, and then define the mean square error as the loss function.  To optimize the loss function, we need to tune the free parameters $\theta = (\theta_1, \theta_2, ...)$. The optimization is performed by a classical optimization algorithm that queries the quantum device, where we can treat the quantum process as a black box and only examine the classical data input and the measurement output. The output of the quantum model is the expectation value of a Pauli-Z measurement.

We use the single-qubit Pauli rotation gate as the encoding $g(x)$ \cite{Schuld_2021}. The frequency spectrum $\Omega$ is determined by the encoding Hamiltonians. Two scenarios can be considered to determine the available frequencies: the \textit{data reuploading} \cite{perez2020data} and the \textit{parallel encodings} \cite{rebentrost2014quantum} models. In the former, we repeat $r$ times of a Pauli rotation gate in sequence, which means we act on the same qubit, but with multiple layers $r=L$; whereas in the latter, we perform similar operations in parallel on $r$ different qubits. but with a single layer $L=1$. These models allow quantum circuits to access increasingly rich frequencies, where $\Omega=\{-r, ...,-1,0,1,...,r \}$ with a spectrum of integer-valued frequencies up to degree $r$. This will correspond to the maximum degree of the partial Fourier series we want to compute.

From the discussion above, one can immediately derive the maximum accessible frequencies of such quantum models \cite{Schuld_2021}. But in practice, if the degree of the target function is greater than the number of layers (for example, in the single qubit case), the fit will be much less accurate.\footnote{Certain initial weight samplings may not even converge to a satisfactory solution. This is relevant to the barren plateau problem \cite{mcclean2018barren} generically present in variational quantum circuits with a random initialization, similar to the classical vanishing gradient problem.
} Increasing the value of $L$ typically requires more training epochs to converge at the same learning rate.

This is relevant to a more difficult question of how to control the Fourier coefficients in the training process, given that all the blocks $W^{(i)} (\theta_i)$ and the measurement observable contribute to "every" Fourier coefficient. However, these coefficients are functions of the quantum circuit with limited degrees of freedom. This means that a quantum circuit with a certain structure can only realize a subset of all possible Fourier coefficients, even with enough degrees of freedom. While a systemic understanding is not yet available, a simulation exploring which Fourier coefficients can be realized can be found in \cite{Schuld_2021}.  In fact, it remains an open question whether, for asymptotically large $L$, a single qubit model can approximate any function by constructing arbitrary Fourier coefficients.

\subsection{The generating function as a Fourier series}

Given the framework of the quantum model and its relation to a partial Fourier series, a natural question arises as to whether the entanglement entropy can be realized within this setup. To approach this question, it is meaningful to revisit the generating function for the von Neumann entropy
\be
G(z; \rho_A) \equiv - \Tr \bigg( \rho_A \ln \frac{1-z \rho_A}{1-z} \bigg)= \sum_{k=1}^{\infty} \frac{f(k)}{k} z^k,
\ee
as a manifest Taylor series. The goal is to rewrite the generating function in terms of a partial Fourier series. Therefore, we would be able to determine whether the von Neumann and R\'enyi entropies are the function classes that the quantum neural network can describe. Note that we will only focus on small-scale tests with a low depth or width of the circuit, as the depth or width of the circuit will correspond exactly to the orders that can be approximated in the Fourier series.

But we cannot simply convert either the original generating function or its Taylor series form to a Fourier series. By doing so, it will generally involve special functions in $\rho_A$, for which we will be unable to specify in terms of $\Tr \rho_A^n$. Therefore, it is essential to have an expression of the Fourier series that allows us to compute the corresponding Fourier coefficients at different orders using $\Tr \rho_A^n$, for which we know the analytic form from CFTs.

This can indeed be achieved, see Appendix~\ref{sA} for a detailed derivation. The Fourier series representation of the generating function on an interval $[w_1, w_2]$ with period $T=w_2-w_1$ is given by 
\bea \la{Fourier1}
G(w; \rho)= \frac{a_0}{2} &+&\sum_{n=1}^\infty \bigg\{ \sum_{m=0}^\infty \frac{\tilde{f}(m)}{m} C_{cos}(n,m) \cos{\bigg(\frac{2 \pi n w}{T} \bigg)}
\nn\\
&+& \sum_{m=0}^\infty \frac{\tilde{f}(m)}{m} C_{sin}(n,m) \sin{\bigg(\frac{2 \pi n w}{T} \bigg)} \bigg\},
\eea
where $C_{cos}$ and $C_{sin}$ are some special functions defined as
\bea
C_{cos}(n,m)&=&\frac{2}{(m+1) T}\bigg[ {}_p F_q \bigg( \frac{m+1}{2};\frac{1}{2},\frac{m+3}{2};-\frac{n^2 \pi^2 t^2_2}{T^2} \bigg)t^{m+1}_2
\nn\\
&\quad&-{}_p F_q \bigg( \frac{m+1}{2};\frac{1}{2},\frac{m+3}{2};-\frac{n^2 \pi^2 t^2_2}{T^2} \bigg) t^{m+1}_1 \bigg],
\eea
\bea
C_{sin}(n,m)&=&\frac{4 n \pi}{(m+2)T^2}\bigg[ {}_p F_q \bigg(\frac{m+2}{2};\frac{3}{2}, \frac{m+4}{2};-\frac{n^2 \pi^2 t^2_2}{T^2} \bigg) t^{m+2}_2
\nn\\
&\quad& -{}_p F_q \bigg(\frac{m+2}{2};\frac{3}{2}, \frac{m+4}{2};-\frac{n^2 \pi^2 t^2_1}{T^2} \bigg) t^{m+2}_1 \bigg],
\eea
with ${}_p F_q$ being the generalized hypergeometric function. Note also that
\be
\tilde{f}(m) \equiv \sum_{k=0}^m \frac{(-1)^{2m-k+1} m!}{k! (m-k)!} \Tr{ (\rho_A^{k+1})}.
\ee
Similarly, the zeroth order Fourier coefficient is given by
\be
a_0 = \sum_{m=0}^\infty  \frac{\tilde{f}(m)}{m} C_{cos}(0,m)=\sum_{m=0}^\infty  \frac{\tilde{f}(m)}{m} \frac{2 (w_2^{m+1}-w_1^{m+1})}{(m+1)T}.
\ee
Note that summing to $m=10$ suffices our purpose, while the summation in $n$ corresponds to the degree of the Fourier series.
Note that the complex-valued Fourier coefficients $c_n$ to be used in our simulation can be easily reconstructed from the expression. Therefore, the only required input for evaluating the Fourier series is $\tilde{f}(m)$, with $\Tr \rho_A^{k+1}$ explicitly given. This is exactly what we anticipated and allows for a straightforward comparison with the Taylor series form. 
 
\begin{figure}[hbt!]
\centering
\includegraphics[width=0.45\textwidth]{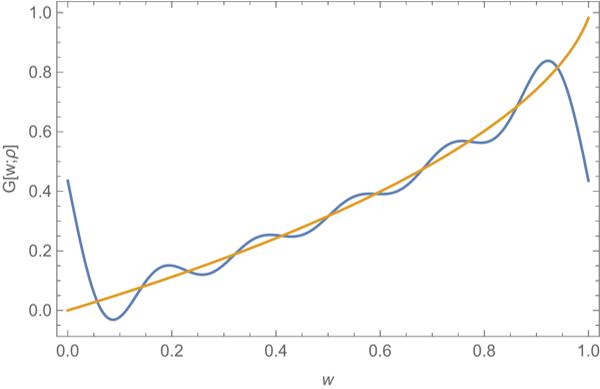}
\caption{Gibbs phenomenon for the Fourier series near the end point for $w \to 1$. We take the single interval example where the yellow curve represents the generating function as a Taylor series, and the blue curve is the Fourier series approximation of the generating function.}
\label{Fourierfig}
\end{figure}

Note the interval for the Fourier series is not arbitrary. We will take the interval $[w_1, w_2]$ to be $[-1, 1]$, which is the maximum interval where the Fourier series \er{Fourier1} is convergent. Furthermore, we expect that as $w \to 1$ from \er{Fourier1}, we arrive at the von Neumann entropy, that is
\be
S(\rho_A)=\lim_{w \to 1} G(w; \rho_A).
\ee
However, as we can see in Figure~\ref{Fourierfig}, there is a rapid oscillation near the end points of the interval for the Fourier series. The occurrence of such "jump discontiunity" is a generic feature for the approximation of discontinuous or non-periodic functions using Fourier series known as the \textit{Gibbs phenomenon}. This phenomenon poses a serious problem in recovering accurate values of the von Neumann entropy because we are taking the limit to the boundary point $w \to 1$. We will return to this issue in Section~\ref{RecovervN}.

\FloatBarrier

\subsection{The expressivity of the quantum models on the entanglement entropy}

In this subsection, we will demonstrate the expressivity of the quantum models of the partial Fourier series with examples from CFTs. We will focus on two specific examples: a single interval and two intervals at small cross-ratio $x$. While these examples suffice for our purpose, it is worth noting that once the Fourier series representation is derived using the expression in \er{Fourier1}, all examples with a known analytic form of $\Tr \rho_A^n$ can be studied.

\begin{figure}[hbt!]
\centering
\includegraphics[width=0.75\textwidth]{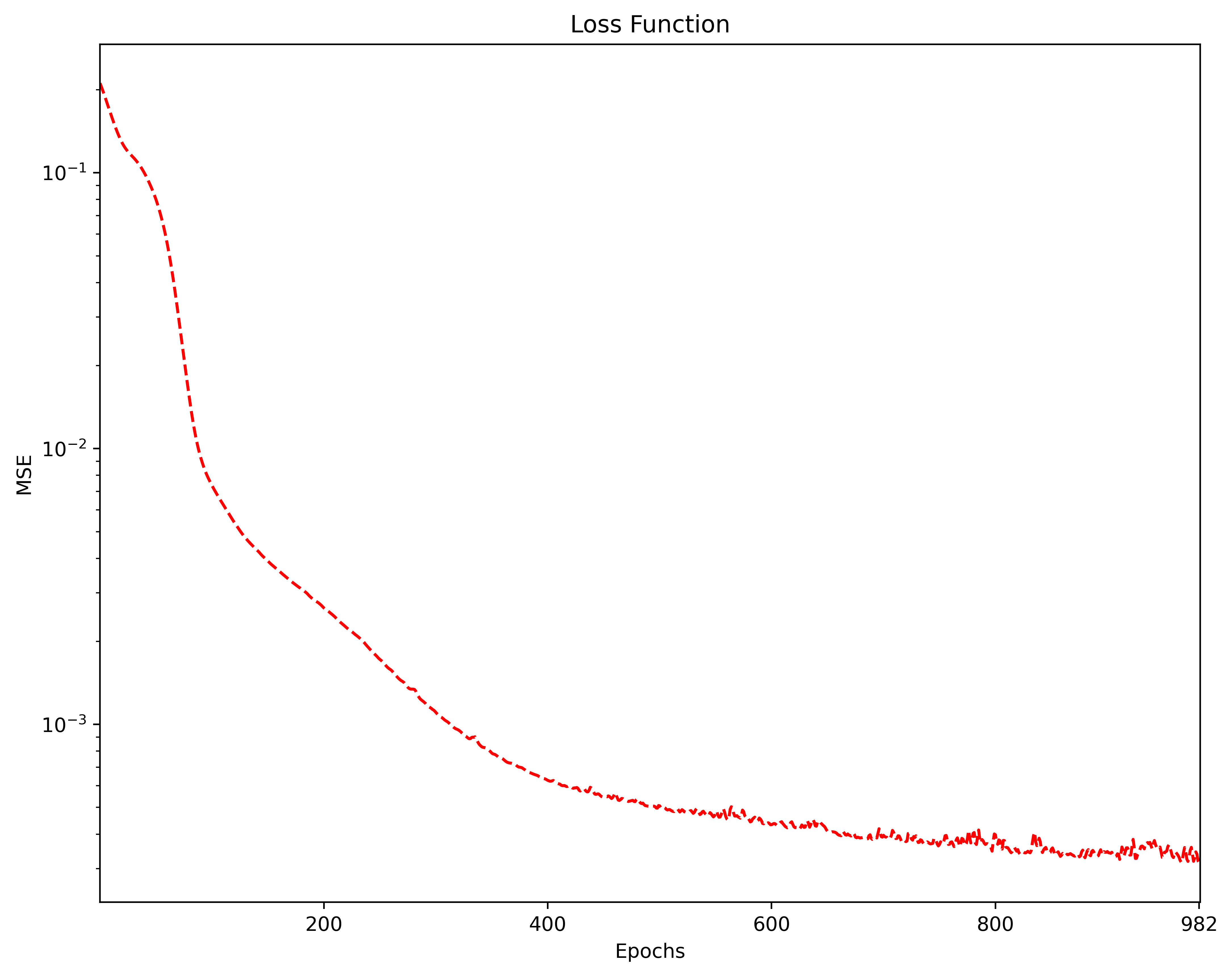}
\includegraphics[width=0.45\textwidth]{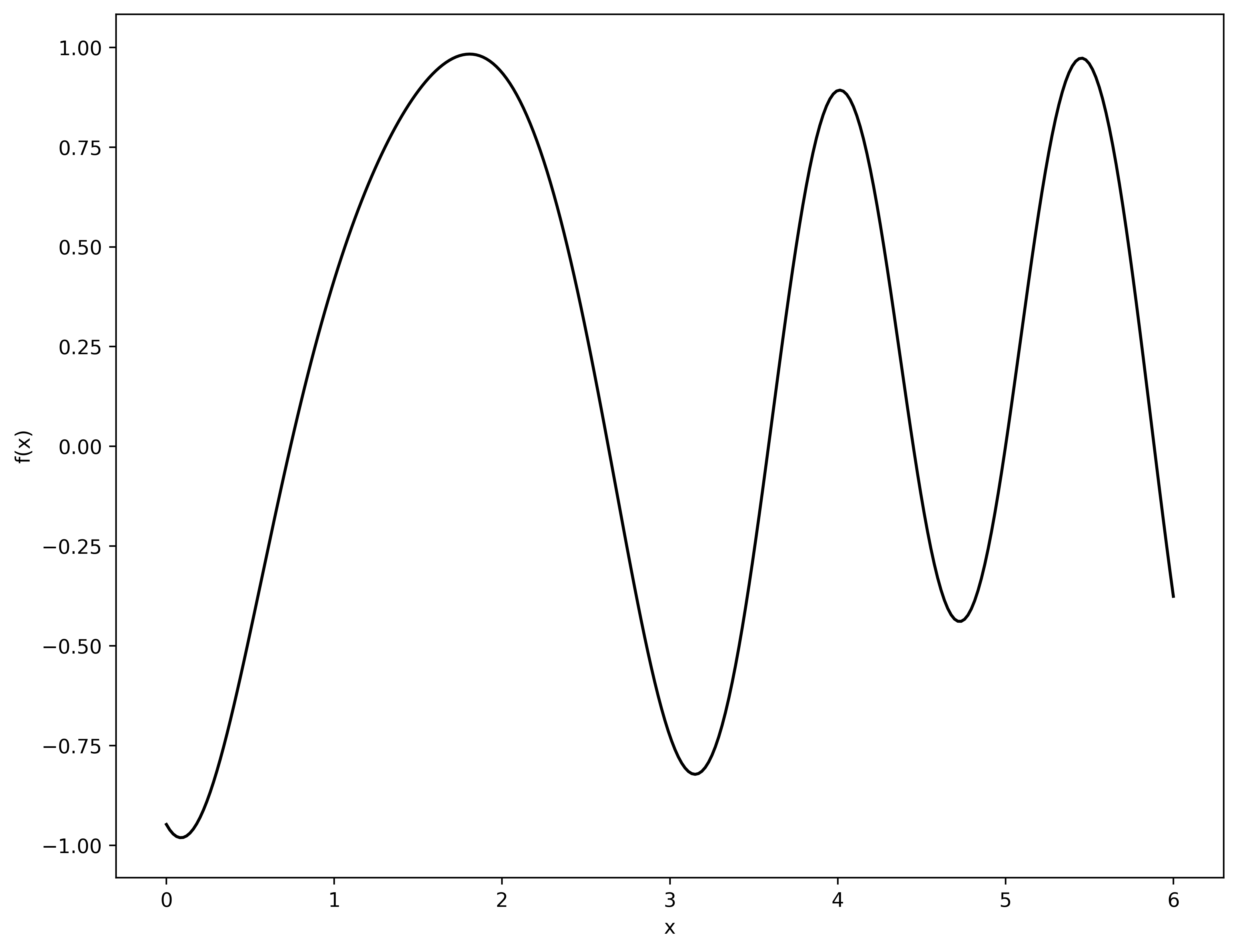}
\includegraphics[width=0.44\textwidth]{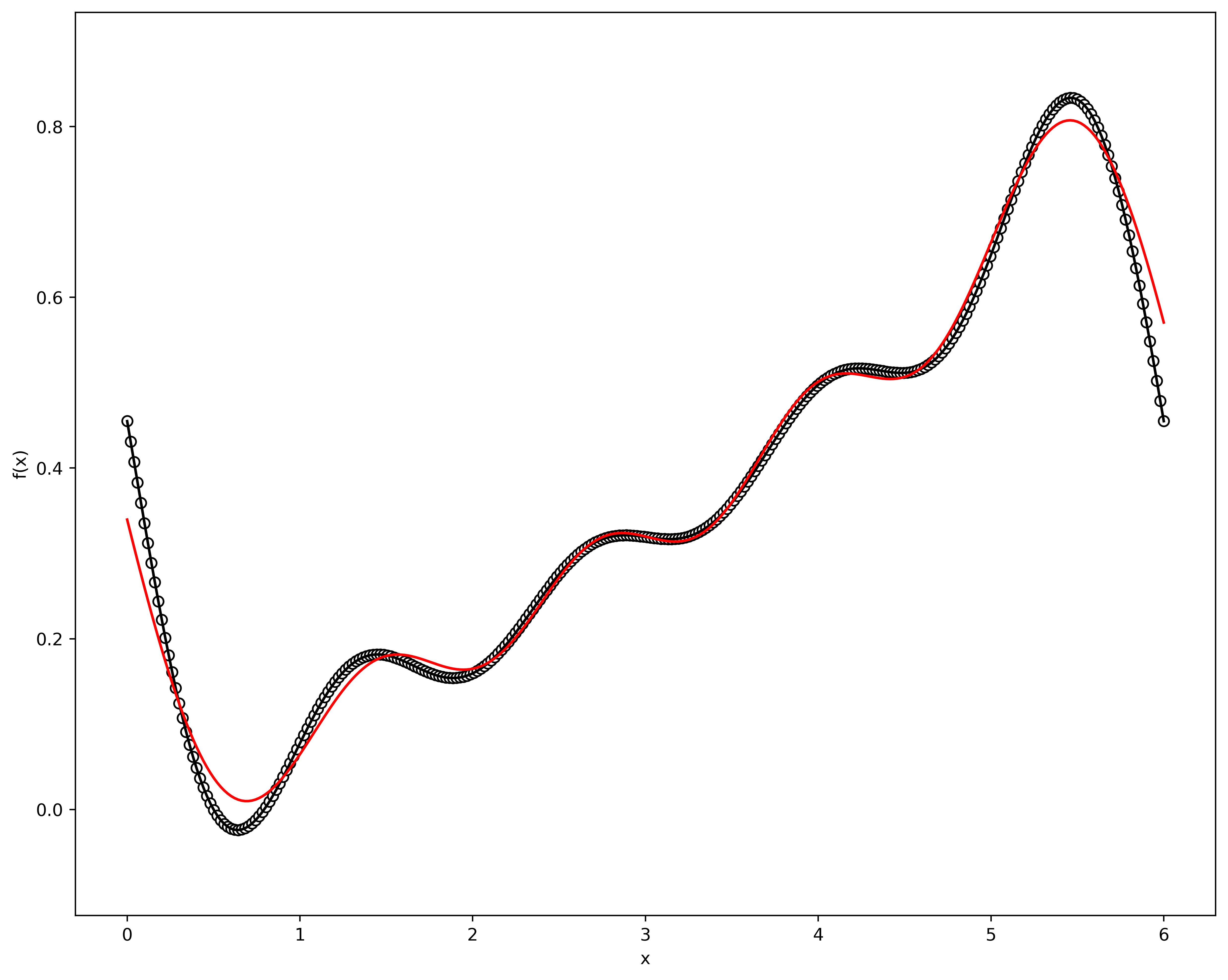}
\caption{A random serial quantum model trained with data samples to fit the target function of the single interval case. Top: the MSE loss function as a function of epochs, where the minimum loss is achieved at epoch 982. Bottom left: a random initialization of the serial quantum model with $r=6$ sequential repetitions of Pauli encoding gates. Bottom right: the circles represent the $300$ data samples of the single interval Fourier series with $\ell=2$ and $\epsilon=0.1$ for \er{sin}. The red curve represents the quantum model after training.}
\label{QMLSS}
\end{figure}

\begin{figure}[hbt!]
\centering
\includegraphics[width=0.75\textwidth]{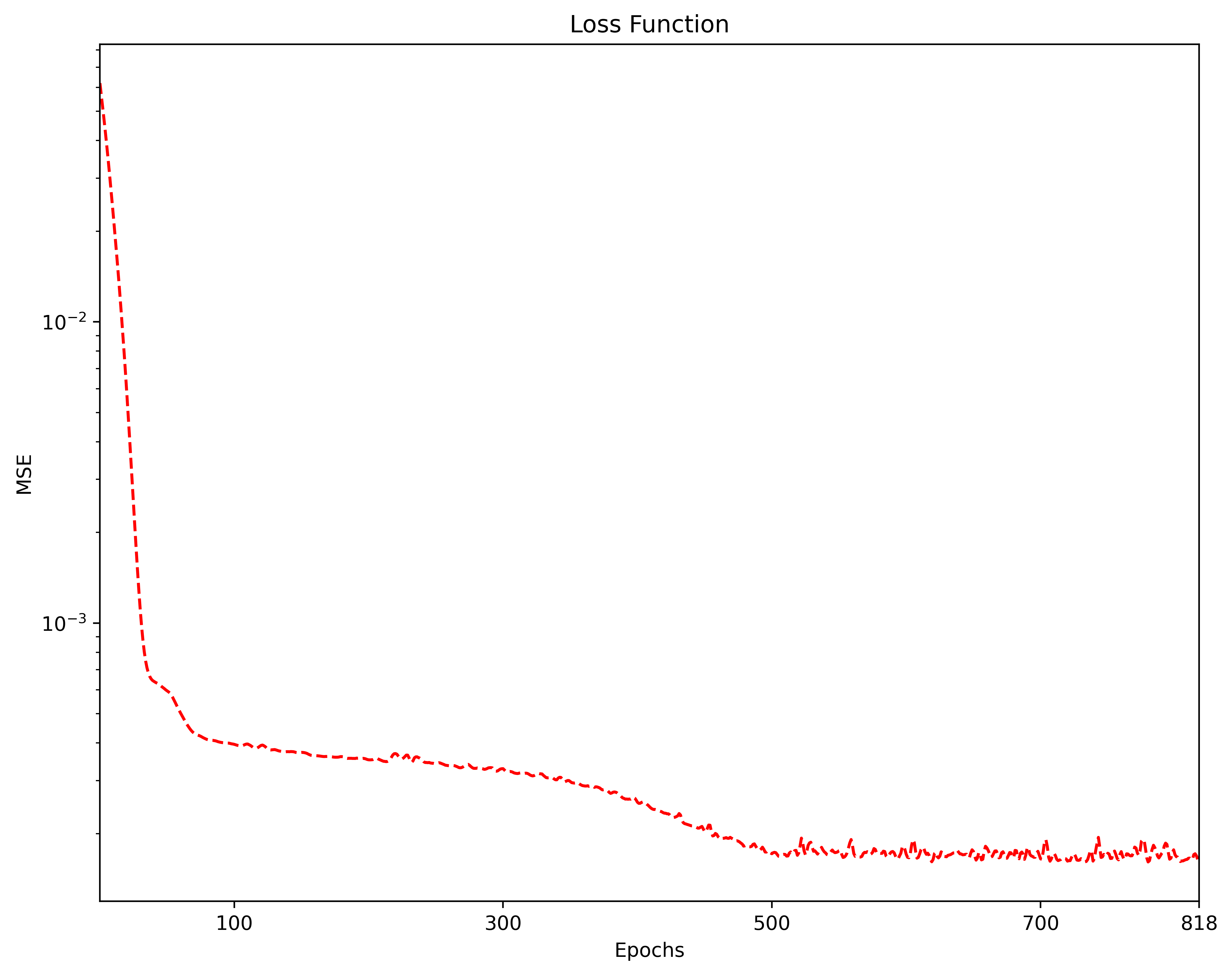}
\includegraphics[width=0.45\textwidth]{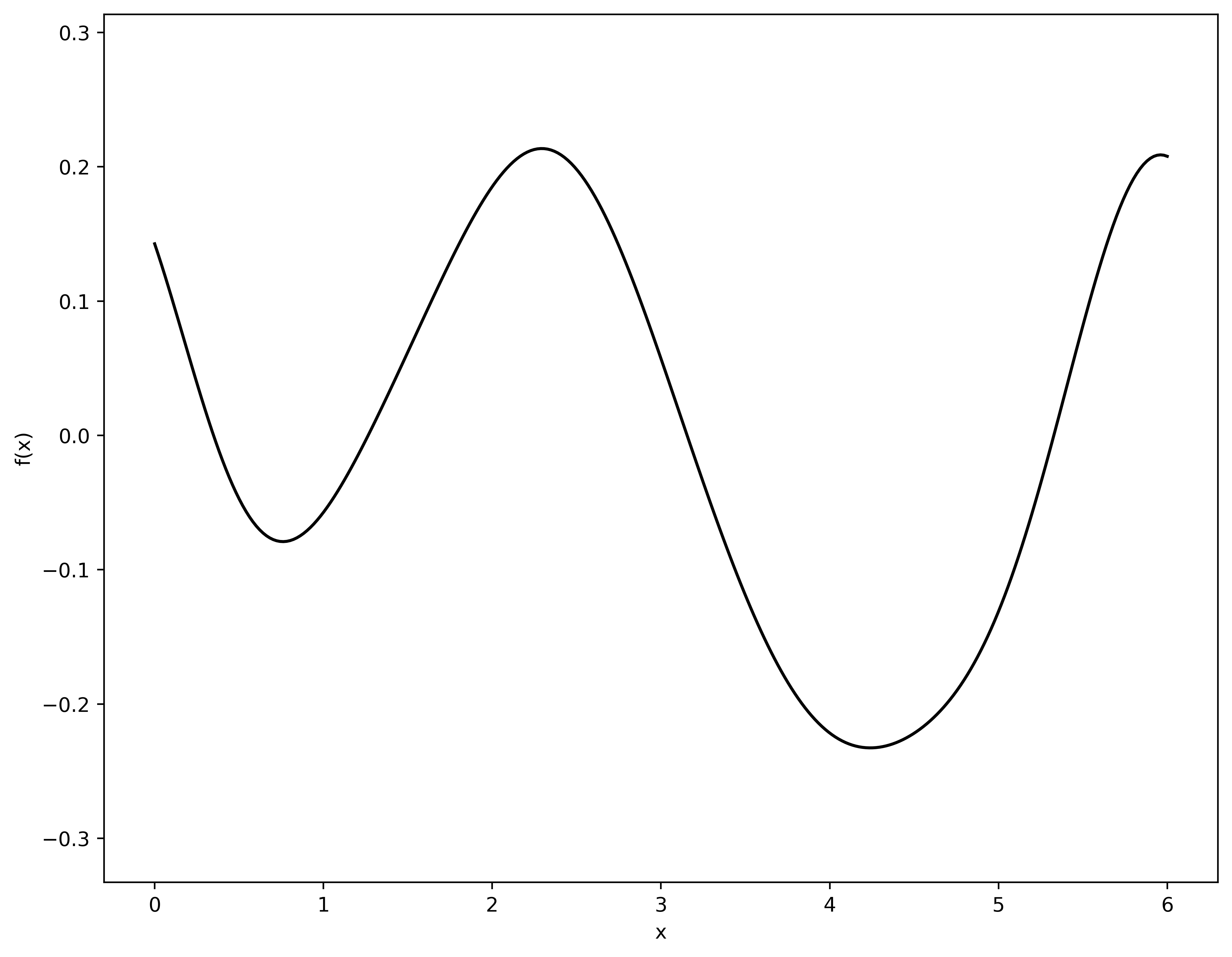}
\includegraphics[width=0.44\textwidth]{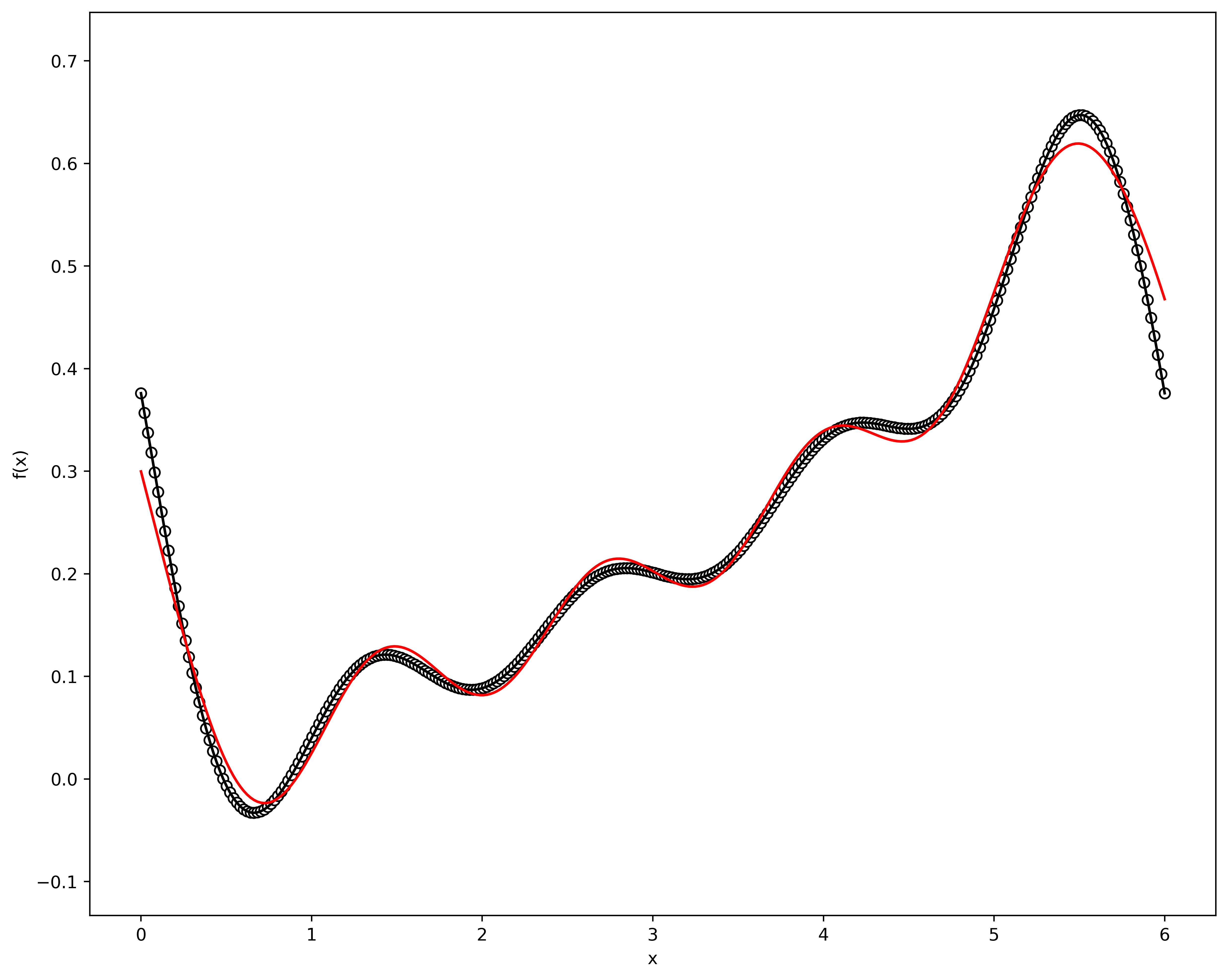}
\caption{A random parallel quantum model for the two-interval case. Top: the loss function achieves minimum loss at epoch 818. Bottom: a random initialization of the quantum model with $r=5$ parallel repetitions of Pauli encoding gates that has achieved a good fit. }
\label{QMLTP}
\end{figure}

The demonstration is performed using Pennylane \cite{Bergholm:2018cyq}. We have adopted the Adam optimizer with a learning rate $0.005$ and batch size of $100$, where MSE is the loss function. Note that we have chosen a smaller learning rate compared to \cite{Schuld_2021} and monitor with EarlyStopping.  For the two examples we study, we have considered both the serial (data reuploading) and parallel (parallel encodings) models for the training. Note that in the parallel model, we have used the StronglyEntanglingLayers in Pennylane with itself of $3$ user-defined layers. In each case, we start by randomly initializing a quantum model with $300$ sample points to fit the target function
\be
f(x)=\sum_{n=-k}^{n=k} c_n e^{-inx}.
\ee
where the complex-valued Fourier coefficients are calculated from the real coefficients in \er{Fourier1}. We have chosen $k=4$ with prescribed physical parameters in the single- and two-interval examples. Therefore, we will need $r$ in the serial and parallel models to be larger than $k=4$. We have executed multiple trials from each case, where we include the most successful results with maximum relative errors controlled in $\lesssim 3\%$ in Figures~\ref{QMLSS}$\sim$\ref{QMLTP}.

\FloatBarrier 

\begin{figure}[hbt!]
\centering
\includegraphics[width=0.75\textwidth]{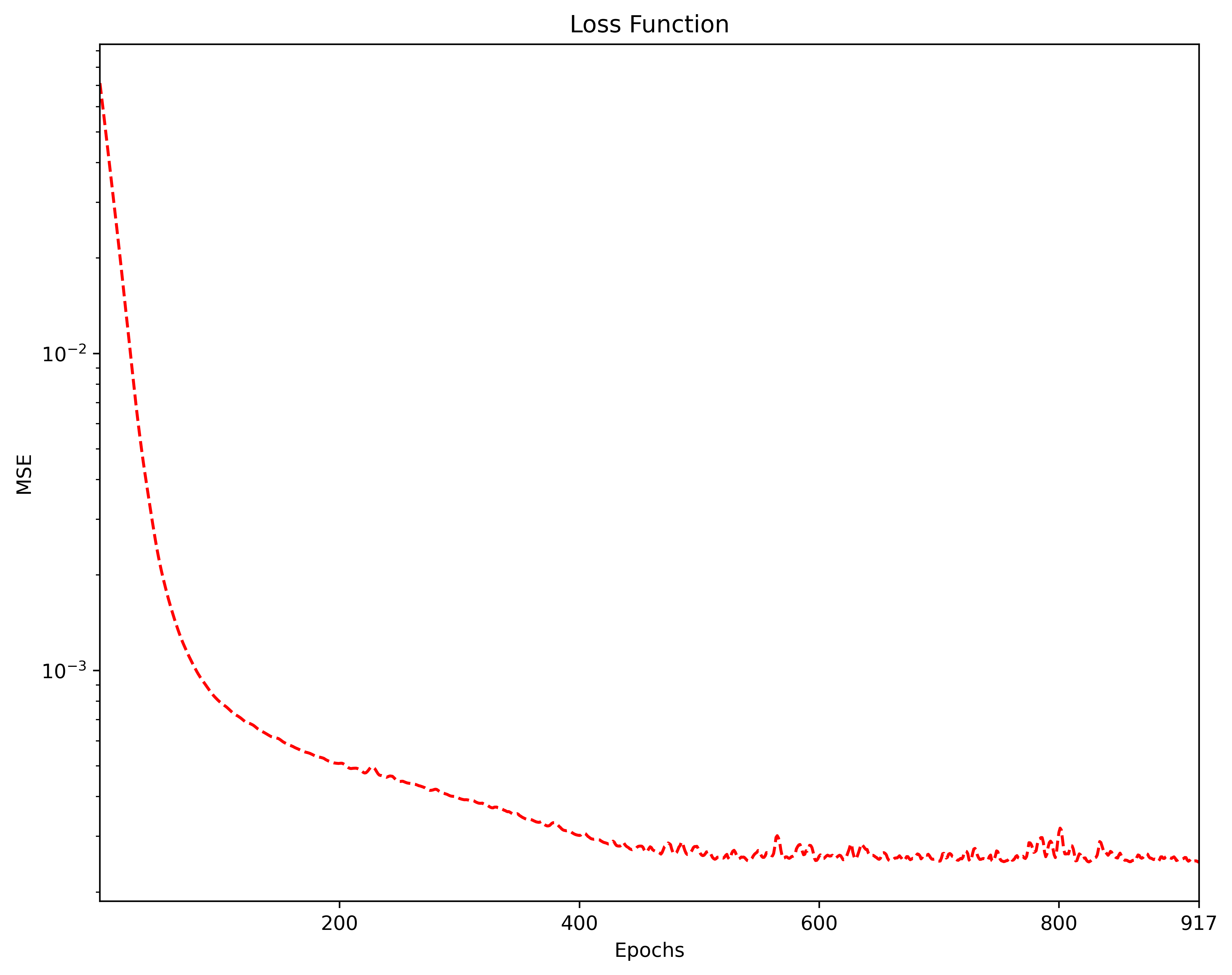}
\includegraphics[width=0.45\textwidth]{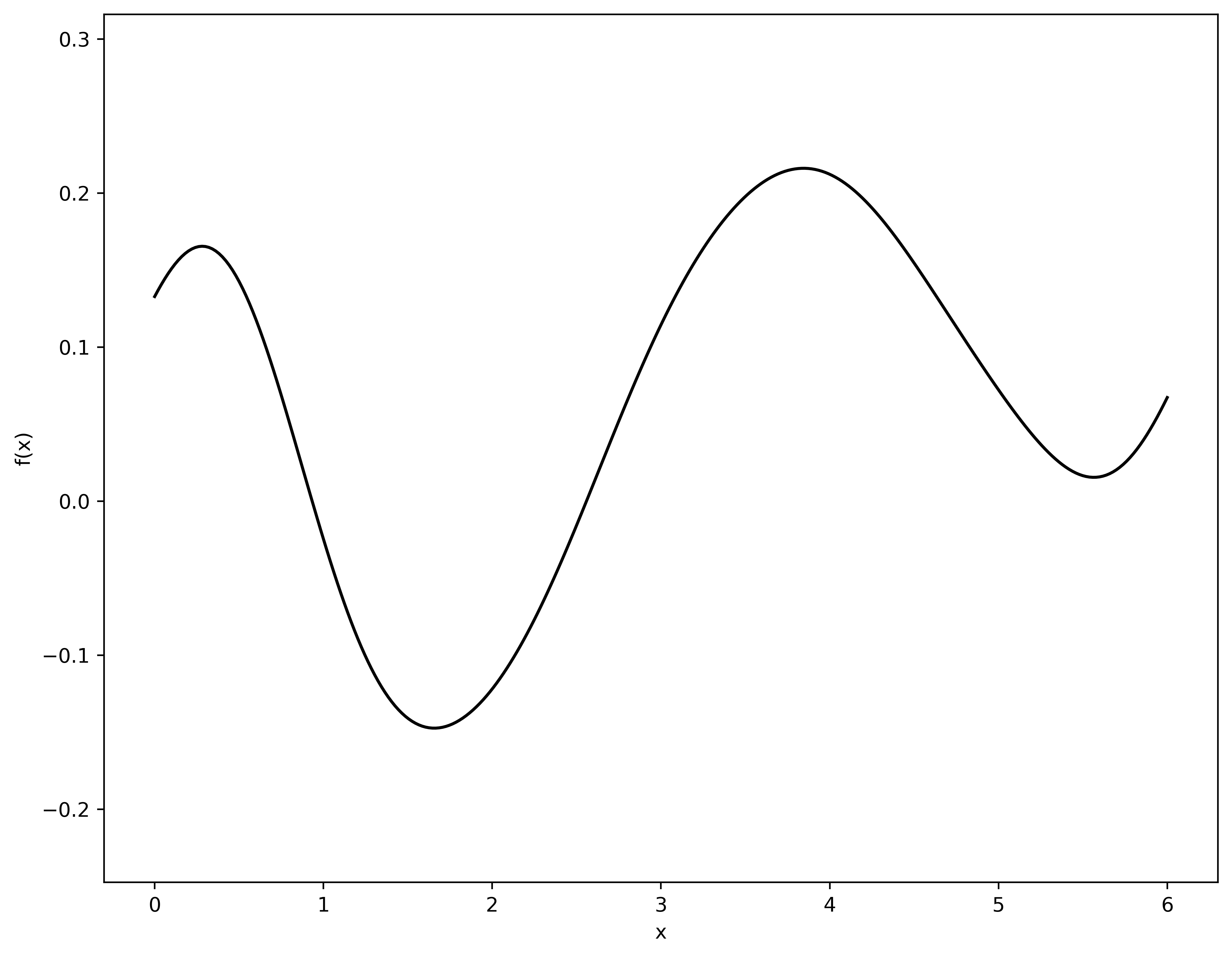}
\includegraphics[width=0.44\textwidth]{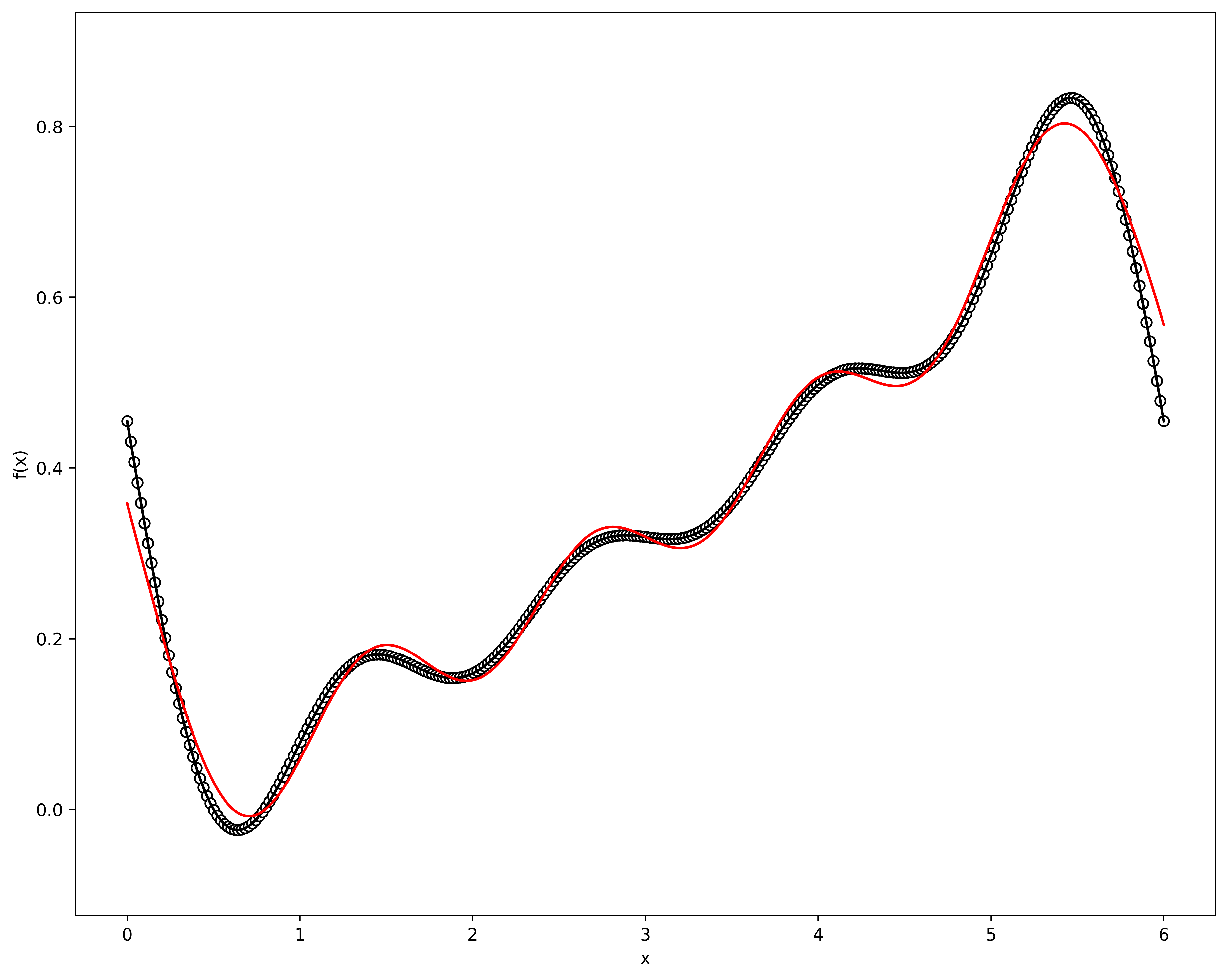}
\caption{A random parallel quantum model for the single interval case. Top: the loss function achieves minimum loss at epoch 917. Bottom: a random initialization of the quantum model with $r=5$ parallel repetitions of Pauli encoding gates that has achieved a good fit.}
\label{QMLSP}
\end{figure}
\FloatBarrier 

\begin{figure}[hbt!]
\centering
\includegraphics[width=0.75\textwidth]{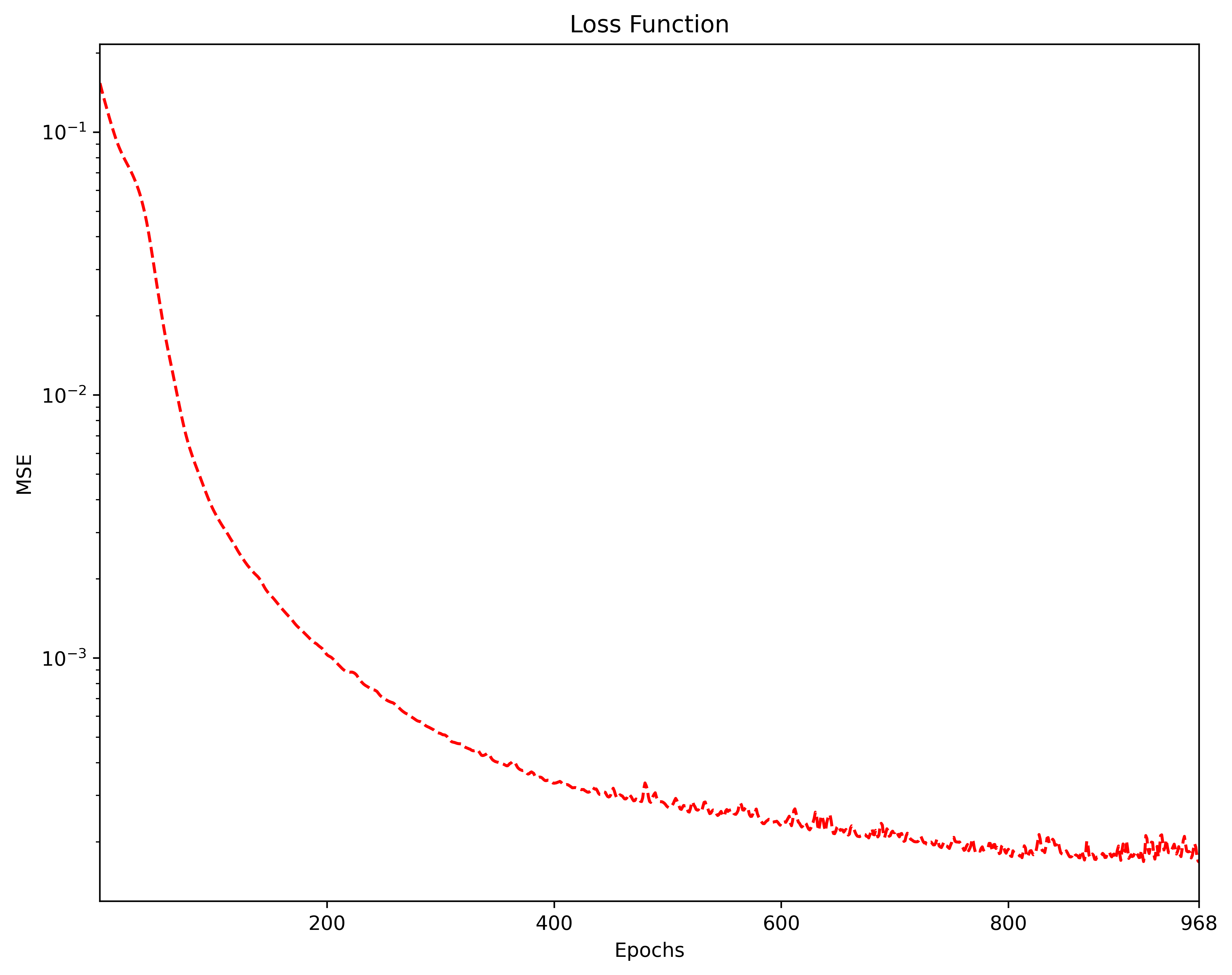}
\includegraphics[width=0.45\textwidth]{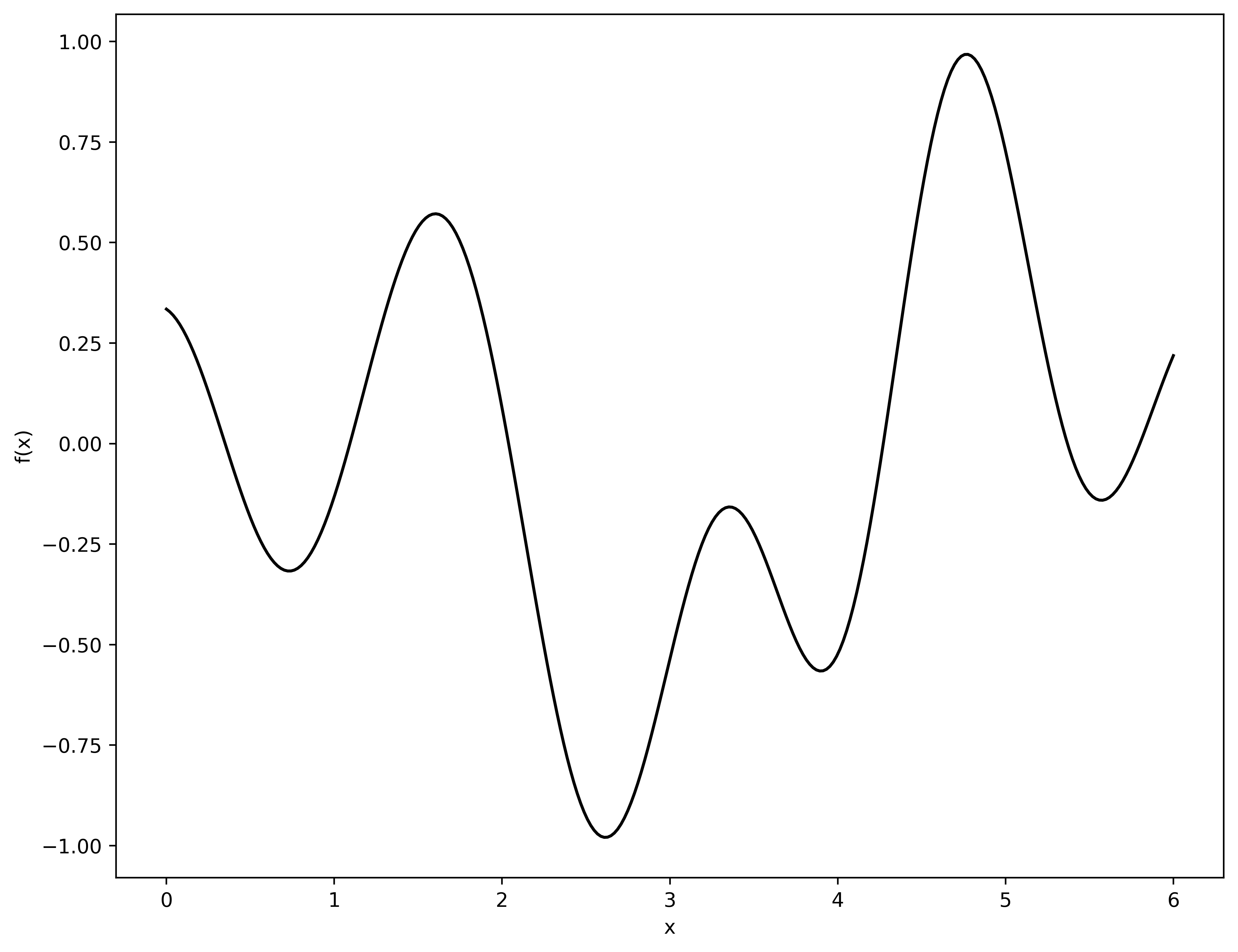}
\includegraphics[width=0.44\textwidth]{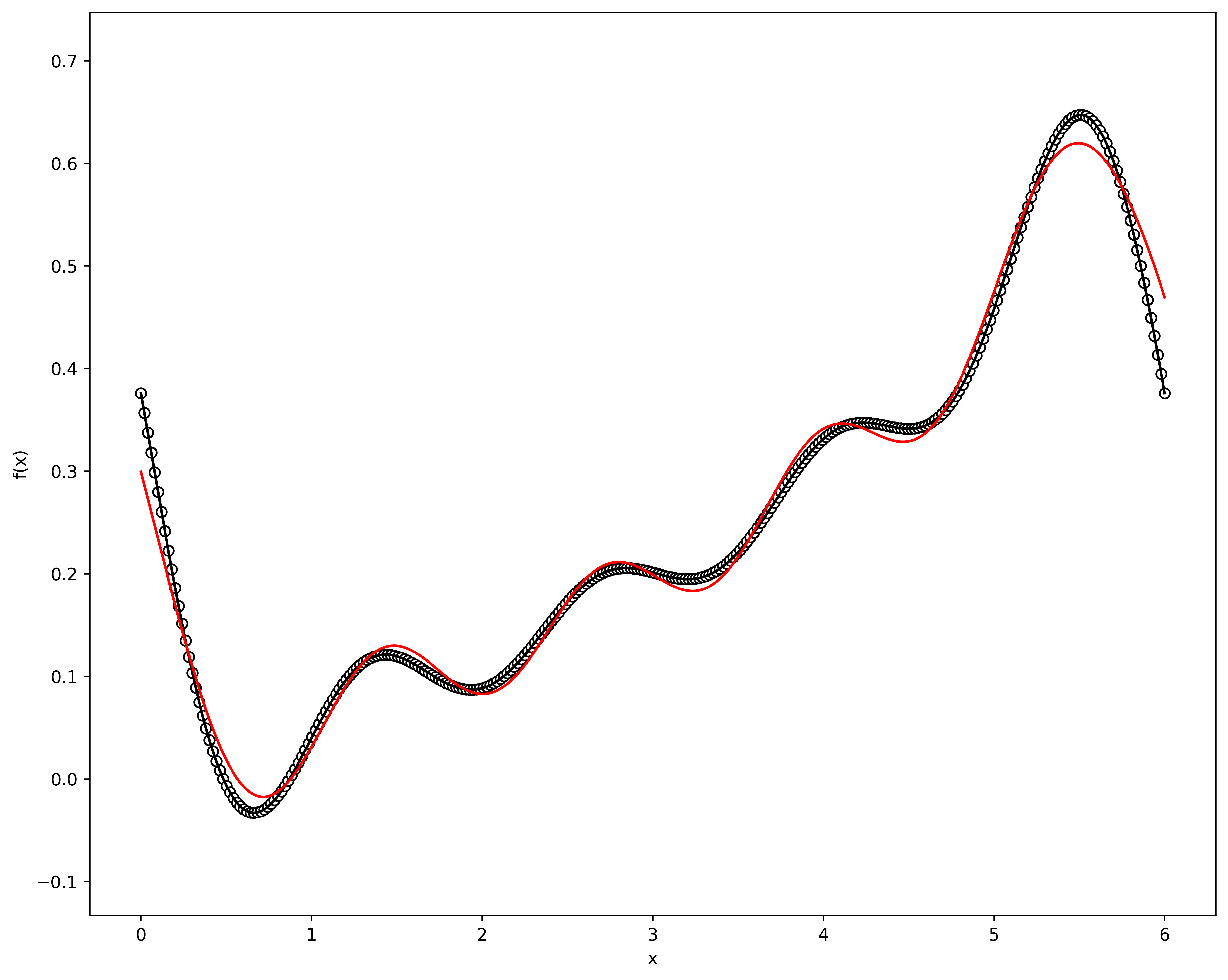}
\caption{A random serial quantum model trained with data samples to fit the target function of the two-interval system with a small cross-ratio. Top: the loss function achieves minimum loss at epoch 968. Bottom left: a random initialization of the serial quantum model of $r=6$ sequential repetitions of Pauli encoding gates. Bottom right: the circles represent the $300$ data samples of the two-interval Fourier series with $x=0.05$, $\alpha=0.1$, and $\epsilon=0.1$ for \er{sma}. The red curve represents the quantum model after training.}
\label{QMLTS}
\end{figure}
\FloatBarrier 

\FloatBarrier

As observed from Figures~\ref{QMLSS}$\sim$\ref{QMLTP}, a rescaling of the data is necessary to achieve precise matching between the quantum models and the Fourier spectrum of our examples. This rescaling is possible because the global phase is unobservable \cite{Schuld_2021}, which introduces an ambiguity in the data-encoding. Consider our quantum model
\be
f_\theta (x)= \langle 0| U^\dagger(x,\theta) M U(x,\theta) | 0 \rangle=\sum_{\omega \in \Omega} c_\omega (\theta) e^{i \omega x},
\ee
where we consider the case of a single qubit $L=1$, then
\be
U(x)=W^{(2)} g(x) W^{(1)}.
\ee
Note that the frequency spectrum $\Omega$ is determined by the eigenvalues of the data-encoding Hamiltonians, which is given by the operator
\be
g(x)=e^{-i x H}.
\ee
$H$ has two eigenvalues $(\lambda_1,\lambda_2)$, but we can rescale the energy spectrum to $(-\gamma, \gamma)$ as the global phase is unobservable (e.g. for Pauli rotations, we have $\gamma=\frac{1}{2}$). We can absorb $\gamma$ from the eigenvalues of $H$ into the data input by re-scaling with
\be
\tilde{x}=\gamma x.
\ee
Therefore, we can assume the eigenvalues of $H$ to be some other values. Specifically, we have chosen $\gamma=6$ in the training, where the interval in $x$ is stretched from $[0,1]$ to $[0,6]$, as can be seen in Figures~\ref{QMLSS}$\sim$\ref{QMLTP}.
    
We should emphasize that we are not re-scaling the original target data, but instead, we are re-scaling how the data is encoded. Effectively, we are re-scaling the frequency of the quantum model itself. The intriguing part is that the global phase shift of the operator acting on a quantum state cannot be observed, yet it affects the expressive power of the quantum model. This can be understood as a pre-processing of the data, which is argued to extend the function classes of the quantum model that can represent \cite{Schuld_2021}.
    
This suggests that one may consider treating the re-scaling parameter $\gamma$ as a trainable parameter \cite{perez2020data}. This would turn the scaling into an adaptive "frequency matching" process, potentially increasing the expressivity of the quantum model. Here we only treat $\gamma$ as a tunable hyperparameter. The scaling does not need to match with the data, but finding an appropriate scaling parameter is crucial for model training.

\subsection{Recovering the von Neumann entropy}
\la{RecovervN}

\begin{figure}[hbt!] 
\centering
\includegraphics[width=0.60\textwidth]{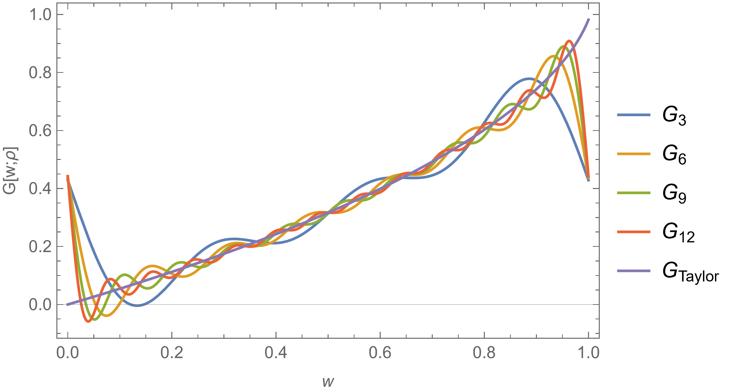}
\caption{We have plotted the single interval example with $L=2$ and $\epsilon=0.1$ for \er{sin}. Here the legends $G_N$ refer to the Fourier series of the generating function to degree $N$, by summing up to $m=10$ in \er{Fourier1}. $G_{\text{Taylor}}$ refers to the Taylor series form \er{gen1} of the generating function by summing up to $k=100$. }
\label{Fourier2}
\end{figure}

So far, we have managed to rewrite the generating function into a partial Fourier series $f_N(w)$ of degree $N$, defined on the interval $w \in [-1,1]$. By leveraging variational quantum circuits, we have been able to reproduce the Fourier coefficients of the series accurately. In principle, with appropriate data-encoding and re-scaling strategies, increasing the depth or width of the quantum models would enable us to capture the series to any arbitrary degree $N$. Thus, the expressivity of the R\'enyi entropies can be established in terms of quantum models. However, a crucial problem remains, that is, we need to recover the von Neumann entropy under the limit $w \to 1$
\be
\lim_{w \to 1} G(w ;\rho_A) =S (\rho_A),
\ee
where the limiting point is exactly at the boundary of the interval that we are approximating. However, as we can see clearly from Figure~\ref{Fourier2}, taking such a limit na\"ively gives a very inaccurate value compared to the true von Neumann entropy. This effect does not diminish even by increasing $N$ to achieve a better approximation of the series when compared to its Taylor series form, as shown in Figure~\ref{Fourier2}. This is because the Fourier series approximation is always oscillatory at the endpoints, a general feature known as the \textit{Gibbs phenomenon} for the Fourier series when approximating discontinuous or non-periodic functions. 

\FloatBarrier

\textit{A priori}, a partial Fourier series of a function $f(x)$ is a very accurate way to reconstruct the point values of $f(x)$, as long as $f(x)$ is smooth and periodic. Furthermore, if $f(x)$ is analytic and periodic, then the partial Fourier series $f_N$ would converge to $f(x)$ exponentially fast with increasing $N$. However, $f_N(x)$ in general is not an accurate approximation of $f(x)$ if $f(x)$ is either discontinuous or non-periodic. Not only the convergence is slow, there is an overshoot near the boundary of the interval. There are many different ways to understand this phenomenon. Broadly speaking, the difficulty lies in the fact that we are trying to obtain accurate local information from the global properties of the Fourier coefficients defined via an integral over the interval, which seems to be inherently impossible.

Mathematically, the occurrence of the Gibbs phenomenon can be easily understood in terms of the oscillatory nature of the Dirichlet kernel, which arises when the Fourier series is written as a convolution. Explicitly, the Fourier partial sum can be written as 
\be
s_n(x)=\frac{1}{\pi} \int_{- \pi}^{\pi} f(\xi) D_n(\xi-x) d \xi,
\ee
where the Dirichlet kernel $D_n(x)$ is given by
\be
D_n(x)=\frac{\sin{(n+\frac{1}{2})x}}{2 \sin{\frac{x}{2}}}.
\ee
This function oscillates between positive and negative values. The behavior is therefore responsible for the appearance of the Gibbs phenomenon near the jump discontinuities of the Fourier series at the boundary.

Therefore, our problem can be accurately framed as follows: given the $2N+1$ Fourier coefficients $\hat{f}_k$ of our generating function \er{Fourier1} for $-N \leq k \leq N$, with the generating function defined in the interval $ w \in [-1,1 ]$, we need to reconstruct the point value of the function at the limit $w \to 1$. The point value of the generating function at this limit exactly corresponds to the von Neumann entropy. Especially, we need the reconstruction to converge exponentially fast with $N$ to the correct point value of the generating function, that is
\be
\lim_{ w \to 1} |G(w;\rho_A)-f_N(w) | \leq e^{- \alpha N}, \quad \alpha > 0.
\ee
This is for the purpose of having a realistic application of the quantum model, where  currently the degree $N$ we can approximate for the partial Fourier series is limited by the depth or the width of the quantum circuits.

We are in need of an operation that can diminish the oscillations, or even better, to completely remove them. Several filtering methods have been developed to ameliorate the oscillations, including the non-negative and decaying Fej\'er kernel, which smooths out the Fourier series over the entire interval, or the introduction of Lanczos $\sigma$ factor, which locally reduces the oscillations near the boundary. For a comprehensive discussion on the Gibbs phenomenon and these filtering methods, see \cite{Jerri_1998}. However, we emphasize that none of these methods are satisfying, as they still cannot recover accurate point values of the function $f(x)$ near the boundary.

Therefore, we need a more effective method to remove the Gibbs phenomenon completely. Here we will adopt a powerful method by re-expanding the partial Fourier series into a basis of Gegenbauer polynomials.\footnote{Note that other methods exist based on periodically extending the function to give an accurate representation within the domain of interest, which involves reconstructing the function based on Chebyshev polynomials \cite{huybrechs2010fourier}. However, we do not explore this method in this work.} This is a method developed in the 1990s by a series of seminal works \cite{gottlieb1992gibbs, gottlieb1994resolution, gottlieb1996gibbsIII, gottlieb1995gibbsIV, gottlieb1995gibbsV, gottlieb1997gibbs}, we also refer to \cite{gelb2007resolution, gottlieb2011review} for more recent reviews. 

The Gegenbauer expansion method allows for accurate representation, within exponential accuracy, by only summing a few terms from the Fourier coefficients. Given an analytic and non-periodic function $f(x)$ on the interval $[-1,1]$ (or a sub-interval $[a,b] \subset [-1,1]$) with the Fourier coefficients
\be
\hat{f}_k=\frac{1}{2}\int_{-1}^1 f(x) e^{-i k \pi x} dx,
\ee
and the partial Fourier series
\be
f_N(x)=\sum_{k=-N}^N \hat{f}_ke^{i k \pi x}.
\ee
The following Gegenbauer expansion represents the original function we want to approximate with the Fourier information
\be
S_{N,M}(x)=\sum_{n=0}^M g^\lambda_{n,N} C^\lambda_n(x),
\ee
where $g^\lambda_{n,N}$ is the Gegenbauer expansion coefficients and $C^\lambda_n(x)$ are the Gegenbauer polynomials.\footnote{The Gegenbauer expansion coefficients $g^\lambda_{n,N}$ are defined with the partial Fourier series $f_N(x)$ as
\be
g^\lambda_{n,N}=\frac{1}{h^\lambda_n}\int_{-1}^{1} (1-x^2)^{\lambda-\frac{1}{2}}f_N(x) C^\lambda_n(x)dx, \quad 0 \leq n \leq M.
\ee
For $\lambda \geq 0$, the Gegenbauer polynomial of degree $n$ is defined to satisfy
\be
\int^1_{-1} (1-x^2)^{\lambda-\frac{1}{2}} C^\lambda_k (x) C^\lambda_n (x) dx =0, \quad k \neq n.
\ee
We refer to Appendix.~\ref{sB} for a more detailed account on the properties of the Gegenbauer expansion.} Note that we have the following integral formula for computing $g^\lambda_{n,N}$
\be
\frac{1}{h^\lambda_n}\int_{-1}^{1} (1-x^2)^{\lambda-\frac{1}{2}}e^{i n \pi x} C^\lambda_n(x)dx=\Gamma(\lambda) \bigg(\frac{2}{\pi k} \bigg)^\lambda i^n (n+\lambda)  J_{n+\lambda}(\pi k),
\ee
then
\be
g^\lambda_{n,N}= \delta_{0,n} \hat{f}(0) + \Gamma(\lambda)i^n (n+\lambda) \sum_{k=-N, k \neq 0}^N J_{n+\lambda} (\pi k) \bigg(\frac{2}{\pi k} \bigg)^\lambda \hat{f}_k,
\ee
where we only need the Fourier coefficients $\hat{f}_k$. 
    
\begin{figure}[hbt!] 
\centering
\includegraphics[width=0.60\textwidth]{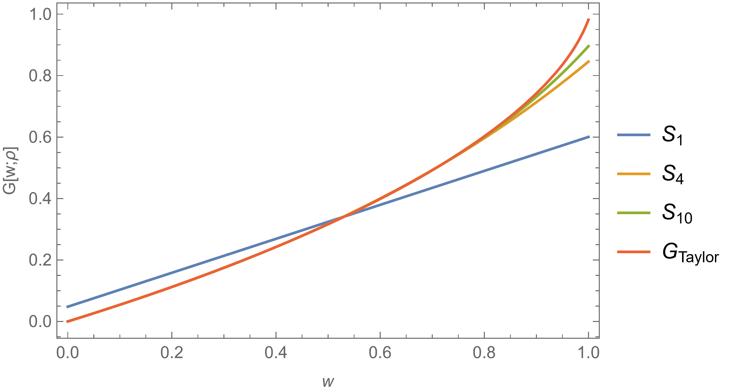}
\caption{Gegenbauer expansion constructed from the Fourier information. Here $S_M$ refers to the Gegenbauer polynomials of order $M$. Note that we set $\beta \epsilon=0.25$, then $\lambda=M=0.25 N$. Therefore, in order to construct the polynomials of order $M$, we need the information of the Fourier coefficients to order $N=4M$. }
\label{Fourier3}
\end{figure}

In fact, the Gegenbauer expansion is a two-parameter family of functions, characterized by $\lambda$ and $M$. It has been shown that by setting $\lambda=M =\beta \epsilon N$ where $\epsilon=(b-a)/2$ and $\beta<\frac{2 \pi e}{27}$ for the Fourier case, the expansion can achieve exponential accuracy with $N$.  Note that $M$ will determine the degrees of the Gegenbauer polynomials, and as such, we should allow the degrees of the original Fourier series to grow with $M$. For a clear demonstration of how the Gegenbauer expansion approaches the generating function from the Fourier data, see Figure~\ref{Fourier3}. We will eventually be able to reconstruct the point value of the von Neumann entropy near $w \to 1$ with increasing order in the expansion. A more precise statement regarding the exponential accuracy can be found in Appendix~\ref{sB}. This method is indeed a process of reconstructing local information from global information with exponential accuracy, thereby effectively removing the Gibbs phenomenon.

\FloatBarrier

\section{Discussion} \la{s6}

In this paper, we have considered a novel approach of using classical and quantum neural networks to study the analytic continuation of von Neumann entropy from R\'enyi entropies. We approach the analytic continuation problem in a way suitable to deep learning techniques by rewriting $\Tr \rho^n_A$ in the R\'enyi entropies in terms of a generating function that manifests as a Taylor series \er{gen1}. We show that our deep learning models achieve this goal with a limited number of R\'enyi entropies.

Instead of using a static model design for the classical neural networks, we adopt the KerasTuner in finding the optimal model architecture and hyperparameters. There are two supervised learning scenarios: predicting the von Neumann entropy given the knowledge of R\'enyi entropies using densely connected neural networks, and treating higher R\'enyi entropies as sequential deep learning using RNNs. In both cases, we have achieved high accuracy in predicting the corresponding targets.

For the quantum neural networks, we frame a similar supervised learning problem as a mapping from inputs to predictions. This allows us to investigate the expressive power of quantum neural networks as function approximators, particularly for the von Neumann entropy. We study quantum models that can explicitly realize the generating function as a partial Fourier series. However, the Gibbs overshooting hinders the recovery of an accurate point value for the von Neumann entropy. To resolve this issue, we re-expand the series in terms of Gegenbauer polynomials, which leads to exponential convergence and improved accuracy.

Several relevant issues and potential improvements arise from our approach:
\begin{itemize}

    \item It is crucial to choose the appropriate architectures before employing KerasTuner, for instances, densely connected layers in Sec.~\ref{s3} and RNNs in Sec.~\ref{s4}. Because these architectures are built for certain tasks \textit{a priori}. KerasTuner only serves as an effective method to determine the optimal complexity and hyperparameters for model training. However, since the examples from CFT$_2$ have different analytic structures for both the von Neumann and R\'enyi entropies, it would be interesting to explore how the different hyperparameters correlate with each example.

    \item Despite being efficient, the parameter spaces we sketched in Sec.~\ref{s3.1} and Sec.~\ref{s4.1} that the KerasTuner searches are not guaranteed to contain the optimal setting, and there could be better approaches.
    \item We can generate datasets by fixing different physical parameters, such as temperature for \er{see} or cross-ratio $x$ for \er{sma}. While we have considered the natural parameters to vary, exploring different parameters may offer more representational power. It is possible to find a Dense model that provides feasible predictions in all parameter ranges, but may require an ensemble of models.

    \item Regularization methods, such as K-fold validation, can potentially reduce the model size or datasets while maintaining the same performance. It would be valuable to determine the minimum datasets required or whether models with low complexity still have the same representational power for learning entanglement entropy.

    \item On the other hand, training the model with more data and resources is the most effective approach to improve the model's performance. One can also scale up the search process in the KerasTuner or use ensemble methods to combine the models found by it.

    \item For the quantum neural networks, note that our approach does not guarantee convergence to the correct Fourier coefficients, as we outlined in Sec.~\ref{s5.1}. It may be beneficial to investigate various pre-processing or data-encoding strategies to improve the approximation of the partial Fourier series with a high degree $r$.

\end{itemize}

There are also future directions that are worth exploring that we shall comment on briefly: 
\begin{itemize}
    \item \textbf{Mutual information:} We can extend our study to mutual information for two disjoint intervals $A$ and $B$, which is an entanglement measure related to the von Neumann entropy defined as 
    \be
    I (A:B) \equiv S (\rho_A)+S (\rho_B)-S(\rho_{A \cup B}).
    \ee
    In particular, there is a conjectured form of the generating function in \cite{DHoker:2020bcv}, with $\Tr \rho^n_A$ being replaced by $\Tr \rho^n_A \Tr \rho^n_B/ \Tr \rho^n_{A \cup B}$. It is worth exploring the expressivity of classical and quantum neural networks using this generating function, particularly as mutual information allows eliminating the UV-divergence and can be compared with some realistic simulations, such as spin-chain models \cite{Furukawa:2008uk}.

    \item \textbf{Self-supervised learning for higher R\'enyi entropies:} Although we have shown that RNN architecture is effective in the sequence learning problem in Sec.~\ref{s4}, it is worth considering other architectures that could potentially offer better performance. For instance, a time-delay neural network, depthwise separable convolutional neural network, or a Transformer may be appropriate for certain types of data. These architectures may be worth exploring in extending the task of extracting higher R\'enyi entropies as self-supervised learning, particularly for examples where analytic continuation is not available.

    \item \textbf{Other entanglement measures from analytic continuation:} There are other important entanglement measures, say, relative entropy or entanglement negativity that may require analytic continuation and can be studied numerically based on neural networks. We may also consider entanglement entropy or entanglement spectrum that can be simulated in specific models stemming from condensed matter or holographic systems.

    \item \textbf{Expressivity of classical and quantum neural networks:} We have studied the expressivity of classical and neural networks for the von Neumann and R\'enyi entropies, with the generating function as the medium. This may help us in designing good generating functions for other entanglement measures suitable for neural networks. It is also worth understanding whether other entanglement measures are also in the function classes that the quantum neural networks can realize.

\end{itemize}

\acknowledgments
We thank Xi Dong for his encouragement of this work. C-H.W. was supported in part by the U.S. Department of Energy under Grant No. DE-SC0023275, and the Ministry of Education, Taiwan. This material is based upon work supported by the Air Force Office of Scientific Research under award number FA9550-19-1-0360.

\begin{appendix}

\section{Fourier series representation of the generating function}
\label{sA}

Suppose there is a Fourier series representation of the generating function from \er{genz}
\be
G(z; \rho_A)= \sum_{n=-\infty}^{ \infty} c_n e^{i n z}.
\ee
The idea is that we want to compute the Fourier coefficients given only the information about $G(z; \rho)$ or $\Tr{\rho_A^n}$. We can compute the complex-valued Fourier coefficients $c_n$ using real-valued coefficients $a_n$ and $b_n$ for a general period $T$ where
\be
G(z; \rho_A)=\frac{a_0}{2} +\sum_{n=1}^\infty a_n \cos{\bigg(\frac{2 \pi n z}{T} \bigg) + b_n \sin{\bigg(\frac{2 \pi n z}{T} \bigg)}}.
\ee
Note that
\be
a_n = \frac{2}{T} \int_{z_1}^{z_2} G(z; \rho_A) \cos{\bigg(\frac{2 \pi n z}{T} \bigg)} dz,
\ee
\be
b_n=\frac{2}{T} \int_{z_1}^{z_2} G(z; \rho_A) \sin{\bigg(\frac{2 \pi n z}{T} \bigg)} dz,
\ee
where we only need to compute the two Fourier coefficients using the generating function of $\Tr{\rho_A^n}$. However, the above integrals are hard to evaluate in general. Instead, we will show that both $a_n$ and $b_n$ can be written as the following series
\be
a_n = \sum_{m=0}^\infty \frac{G(0;\rho)^{(m)}}{m!} C_{cos}(n,m),
\ee
\be
b_n = \sum_{m=0}^\infty \frac{G(0;\rho)^{(m)}}{m!} C_{sin}(n,m).
\ee
where $C_{cos}(n,m)$ and $C_{sin}(n,m)$ involve certain special functions. The definitions of $G(0;\rho_A)^{(m)}$ starts from the following generating function in terms of $w$ from \er{Taylor}
\be
G(w;\rho_A)=-\Tr{ (\rho_A \ln{[1-w(1-\rho_A)]})},
\ee
where the $m$-th derivative with $w \to 0$
\bea
G(0;\rho_A)^{(m)}&=& -\Tr [(-1)^{m+1} (m-1)! \rho_A (\rho_A-1)^m]
\nn\\
&=&-(m-1)! \sum_{k=0}^m \frac{(-1)^{2m-k+1} m!}{k! (m-k)!} \Tr{ (\rho_A^{k+1})}.
\eea
Note that we have to define for $m=0$ such that
\be
G(0;\rho_A)^{(0)}=-\Tr(\rho_A \ln{1})=0.
\ee

Then we have the Fourier series representation of the generating function on an interval $[w_1, w_2]$ with period $T=w_2-w_1$ given by
\bea
G(w; \rho_A)= \frac{a_0}{2} &+&\sum_{n=1}^\infty \bigg\{ \sum_{m=0}^\infty \frac{\tilde{f}(m)}{m} C_{cos}(n,m) \cos{\bigg(\frac{2 \pi n w}{T} \bigg)}
\nn\\
&+& \sum_{m=0}^\infty \frac{\tilde{f}(m)}{m} C_{sin}(n,m) \sin{\bigg(\frac{2 \pi n w}{T} \bigg)} \bigg\},
\eea
where we have defined
\be
\tilde{f}(m) \equiv -\sum_{k=0}^m \frac{(-1)^{2m-k+1} m!}{k! (m-k)!} \Tr{ (\rho_A^{k+1})}.
\ee
with manifest $\Tr{ \rho_A^{k+1}}$ appearing in the expression.

Now we need to work out $ C_{cos}(n,m)$ and $C_{sin}(n,m)$. First, let us consider in general
\be
a_n= \frac{2}{T} \int_{t_1}^{t_2} f(t) \cos{\bigg(\frac{2 \pi n t}{T} \bigg)} dt,
\ee
where we have written $G(w;\rho_A)$ as $f(t)$ for simplicity.
We can write down the Taylor series of both pieces
\be
f(t)= \sum_{j=0}^\infty \frac{f^{(j)}(0)}{j!} t^j, \quad \cos{\bigg( \frac{2 \pi n t}{T} \bigg)}= \sum_{k=0}^\infty \frac{ (-1)^k}{(2k)!} \bigg( \frac{2 \pi n t}{T} \bigg)^{2k},
\ee
Consider the following function
\be
T_{cos}(t)\equiv f(t) \cos{\bigg(\frac{2 \pi n t}{T} \bigg)}= \bigg[\sum_{j=0}^\infty \frac{f^{(j)}(0)}{j!} t^j \bigg] \bigg[ \sum_{k=0}^\infty \frac{ (-1)^k}{(2k)!} \bigg( \frac{2 \pi n t}{T} \bigg)^{2k} \bigg],
\ee
then let us collect the terms in orders of $t$
\bea
T_{cos}(t)= f(0)+ f^{(1)}(0) t&+&\bigg(\frac{1}{2}f^{(2)}(0)-2 f(0) \bigg( \frac{\pi n}{T} \bigg)^2 \bigg) t^2
\nn\\
&+&\bigg( \frac{1}{6} f^{(3)} f(0)-2 f^{(1)}(0)  \bigg( \frac{\pi n}{T} \bigg)^2 \bigg)t^3
\nn\\
&+&\bigg(\frac{1}{24} f^{(4)}(0)-f^{(2)} \bigg( \frac{\pi n}{T} \bigg)^2+\frac{2}{3} f(0) \bigg( \frac{\pi n}{T} \bigg)^4 \bigg) t^4
\nn\\
&+& \cdots,
\eea
then the integral becomes
\bea
\int_{t_1}^{t_2} T_{cos}(t) dt= f(0) (t_2-t_1) &+& \frac{1}{2} f^{(1)}(0) (t^2_2- t^2_1)
\nn\\
&+&\frac{1}{3} \bigg(\frac{1}{2}f^{(2)}(0)-2 f(0) \bigg( \frac{\pi n}{T} \bigg)^2 \bigg) (t^3_2-t^3_1)
\nn\\
&+& \frac{1}{4}\bigg( \frac{1}{6} f^{(3)} f(0)-2 f^{(1)}(0)  \bigg( \frac{\pi n}{T} \bigg)^2 \bigg) (t^4_2 -t^4_1)
\nn\\
&+& \frac{1}{5} \bigg(\frac{1}{24} f^{(4)}(0)-f^{(2)} \bigg( \frac{\pi n}{T} \bigg)^2+\frac{2}{3} f(0) \bigg( \frac{\pi n}{T} \bigg)^4 \bigg) (t^5_2 -t^5_1)
\nn\\
&+&\cdots.
\eea
Now we want to re-order this expression, where we collect terms in terms of $f^{(m)}(0)$
\bea
\int_{t_1}^{t_2} T_{cos}(t) dt &=& f(0) \bigg( (t_2-t_1)-\frac{2}{3} \bigg( \frac{\pi n}{T} \bigg)^2 (t^3_2 -t^3_1)+\frac{2}{15} \bigg( \frac{\pi n}{T} \bigg)^4 (t^5_2-t^5_1)+ \cdots \bigg)
\nn\\
&+& f^{(1)}(0) \bigg( \frac{1}{2}(t^2_2- t^2_1)-\frac{1}{2}  \bigg( \frac{\pi n}{T} \bigg)^2 (t^4_2 -t^4_1) +\cdots \bigg)
\nn\\
&+&f^{(2)}(0) \bigg( \frac{1}{24} (t^4_2 -t^4_1)+\cdots \bigg) + \cdots.
\eea
After multiplying a factor of $2/T$, this can be written as
\be
a_n= \frac{2}{T} \int_{t_1}^{t_2} T_{cos}(t) dt = \sum_{m=0}^\infty \frac{f^{(m)}(0)}{m!} C_{cos}(n,m),
\ee
where
\bea
C_{cos}(n,m)&=&\sum_{p=0}^\infty \bigg[\frac{(-1)^p 2^{(2p+1)} n^{2p} \pi^{2p} (t_2^{(2p+m+1)}-t_1^{(2p+m+1)})}{(2p+m+1)(2p)! T^{2p+1}} \bigg]
\nn\\
&=&\frac{2}{(m+1) T}\bigg[ {}_p F_q \bigg( \frac{m+1}{2};\frac{1}{2},\frac{m+3}{2};-\frac{n^2 \pi^2 t^2_2}{T^2} \bigg)t^{m+1}_2
\nn\\
&\quad&-{}_p F_q \bigg( \frac{m+1}{2};\frac{1}{2},\frac{m+3}{2};-\frac{n^2 \pi^2 t^2_2}{T^2} \bigg) t^{m+1}_1 \bigg].
\eea

Next, we consider the case for $C_{sin}(n,m)$, where we need to work out
\be
b_n=\frac{2}{T} \int_{t_1}^{t_2} f(t) \sin{\bigg(\frac{2 \pi n t}{T} \bigg)} dt,
\ee
again, we know
\be
\sin{\bigg(\frac{2 \pi n t}{T}} \bigg)= \sum_{k=0}^\infty \frac{(-1)^k}{(2k+1)!} \bigg( \frac{2 \pi n t}{T} \bigg)^{(2k+1)},
\ee
then we define
\be
T_{sin}(t) \equiv f(t) \sin{\bigg(\frac{2 \pi n t}{T} \bigg)}= \bigg[\sum_{j=0}^\infty \frac{f^{(j)}(0)}{j!} t^j \bigg] \bigg[ \sum_{k=0}^\infty \frac{(-1)^k}{(2k+1)!} \bigg( \frac{2 \pi n t}{T} \bigg)^{2k+1} \bigg],
\ee
with the only difference being the denominator $(2k)! \to (2k+1)!$ and the power of $\frac{2 \pi n t}{T}$ becomes $2k+1$. Then 
\bea
C_{sin}(n,m)&=&\sum_{p=0}^\infty \bigg[\frac{(-1)^p 2^{(2p+2)} n^{2p+1} \pi^{2p+1} (t_2^{(2p+m+2)}-t_1^{(2p+m+2)})}{(2p+m+2)(2p+1)! T^{2p+2}} \bigg]
\nn\\
&=&\frac{4 n \pi}{(m+2)T^2}\bigg[ {}_p F_q \bigg(\frac{m+2}{2};\frac{3}{2}, \frac{m+4}{2};-\frac{n^2 \pi^2 t^2_2}{T^2} \bigg) t^{m+2}_2
\nn\\
&\quad& -{}_p F_q \bigg(\frac{m+2}{2};\frac{3}{2}, \frac{m+4}{2};-\frac{n^2 \pi^2 t^2_1}{T^2} \bigg) t^{m+2}_1 \bigg].
\eea

\section{The Gegenbauer polynomials and the Gibbs phenomenon}
\la{sB}

In the appendix, we discuss briefly the definition and properties of the Gegenbauer polynomials used to remove the Gibbs phenomenon in Section~\re{RecovervN}. 

The Gegenbauer polynomials $C^\lambda_n(x)$ of degree $n$ for $\lambda \geq 0$ are defined by the integral
\be
\int_{-1}^{1} (1-x^2)^{\lambda-\frac{1}{2}} C^\lambda_k(x) C^\lambda_n (x) dx=0, \quad k \neq n.
\ee
with the following normalization
\be
C^\lambda_n (1)=\frac{\Gamma(n+2 \lambda)}{n! \Gamma(2 \lambda)}.
\ee
Note the polynomials are not orthonormal, the norm of $C^\lambda_n(x)$ is 
\be
\int_{-1}^{1} (1-x^2)^{\lambda-\frac{1}{2}} (C^\lambda_n(x))^2 dx =h^\lambda_n,
\ee
where
\be
h^\lambda_n=\pi^{\frac{1}{2}}C^\lambda_n(1) \frac{\Gamma(\lambda+\frac{1}{2})}{\Gamma(\lambda) (n+\lambda)}.
\ee

Given a function $f(x)$ defined on the interval $[-1,1]$ (or a sub-interval $[a,b] \subset [-1,1]$), the corresponding Gegenbauer coefficients $\hat{f}^\lambda(l)$ are given by
\be
\hat{f}^\lambda(l)=\frac{1}{h^\lambda_n}\int^1_{-1} (1-x^2)^{\lambda-\frac{1}{2}} f(x) C^\lambda_l (x) dx,
\ee
then the truncated Gegenbauer expansion up to the first $m+1$ terms is
\be
f^\lambda_m(x) = \sum_{l=0}^m \hat{f}^\lambda (l) C^\lambda_l (x).
\ee

Here we will sketch briefly how the Gegenbauer expansion leads to a resolution of the Gibbs phenomenon as we discussed in Section~\re{RecovervN}. In fact, one can prove that there is an exponential convergence between the function $f(x)$ we want to approximate and the $m$-th degree Gegenbauer polynomials. We will only sketch the idea behind the proof, and we refer the readers to the review in \cite{gottlieb1997gibbs} for the details.

One can establish exponential convergence by demonstrating that the errors for the $N$-th Fourier coefficient, expanded into Gegenbauer polynomials, can be made exponentially small. Let us call the $f^m_N (x)$ the expansion of $f_N (x)$ into $m$-th degree Gegenbauer polynomials and $f^m (x)$ the expansion of $f(x)$ into $m$-th degree Gegenbauer polynomials. Then we have the following relation, where the approximation of $f(x)$ by $f^m_N (x)$ is obviously bounded by the error between $f(x)$ and $f^m(x)$ and the error between $f^m(x)$ and $f^m_N (x)$
\be
||f(x)-f^m_N (x)|| \leq || f(x)- f^m(x)|| + || f^m(x)-f^m_N(x)||.
\ee
On the right hand side of the inequality, we call the first norm as the \textit{regularization error}, while the second norm as the \textit{truncation error}. Note that we take the norm to be the maximum norm over the interval $[-1,1]$. To be more precise, we can write the truncation error as
\be
||f^m-f^m_N|| = \max_{-1 \leq x \leq 1} \bigg|\sum_{k=0}^m (\hat{f}^\lambda_k-\hat{g}^\lambda_k) C^\lambda_k(x) \bigg|,
\ee
where we take $\hat{f}^\lambda_k$ to be the unknown Gegenbauer coefficients of the function $f(x)$. If both $\lambda$ and $m$ grow linearly with $N$, this error is shown to be exponentially small.
On the other hand, the regularization error can be written as
\be
||f-f^m||=\max_{-1 \leq 1} \bigg|f(x)-\sum_{k=0}^m \hat{f}^\lambda_k C^\lambda_k (x) \bigg|.
\ee
It can also be shown that this error is exponentially small for $\lambda = \gamma m$ with a positive constant $\gamma$.  Since both the regularization and truncation errors can be made exponentially small with the prescribed conditions, the Gegenbauer expansion achieves uniform exponential accuracy and removes the Gibbs phenomenon from the Fourier data.

\end{appendix}

\bibliographystyle{JHEP}
\bibliography{bibliography}

\end{document}